\documentclass[12pt]{article}

\usepackage{geometry}
\newcommand{\pr}{\mathrm{pr}}

\usepackage[authoryear]{natbib}
\usepackage[resetlabels]{multibib} 
 
\usepackage{Y0ung}
\usepackage[colorlinks=true,
linkcolor=red,
citecolor=blue,
urlcolor=red]{hyperref} 
\usepackage{algorithm} 
\usepackage{algpseudocode}
\usepackage{setspace}
\usepackage[bottom]{footmisc}
\usepackage{graphicx}
\usepackage{multirow}
\usepackage{booktabs}
\newcommand{\blind}{1}
\allowdisplaybreaks
\usepackage{setspace}
\newcommand{\spacingset}[1]{\setstretch{#1}}
\usepackage{authblk}
\usepackage{chngcntr}
\newcites{S}{References}

\begin{document}

\if1\blind
{
\spacingset{1.25}
  \title{\bf \Large Communication-efficient distributed quantile regression via signed voting}
\author[1]{Jinyuan Chang}
\author[2]{Yifan Jiang}
\author[1]{Lin Yang}
\author[3]{Wen-Xin Zhou}

\affil[1]{\it \small Joint Laboratory of Data Science and Business
Intelligence, Institute of Statistical Interdisciplinary Research, Southwestern University of Finance and Economics, Chengdu, China} 
\affil[2]{\it \small 
Department of Biostatistics, Epidemiology and Informatics,
University of Pennsylvania, Philadelphia,   USA
} 
\affil[3]{\it \small College of Business Administration, University of Illinois Chicago, Chicago, IL, USA}

		\setcounter{Maxaffil}{0}
		
		\renewcommand\Affilfont{\itshape\small}
		\date{\vspace{-5ex}}
		\maketitle
	} \fi
	\if0\blind
	{
		\bigskip
		\bigskip
		\bigskip
            \begin{center}
			{
			\Large \bf Communication-efficient distributed quantile regression via signed voting
			}
		\end{center}
		\medskip
	} \fi 
\spacingset{1}
  
\setlength{\baselineskip}{24pt}	
	\begin{abstract}
	Quantile regression provides a flexible framework for characterizing heterogeneous covariate effects across the conditional response distribution. In high-dimensional distributed settings, however, communicating full coefficient or gradient vectors can incur considerable costs. To address this challenge, we propose a screening-then-estimation framework that integrates local feature screening, signed voting, and selective coefficient aggregation. In the screening stage, local machines construct candidate support sets using a convolution-smoothed, $\ell_1$-penalized quantile regression estimator coupled with coordinatewise debiasing based on decorrelated scores. Only the selected coordinate indices and the signs of their corresponding screening statistics are transmitted to a master machine, which applies a signed voting rule to obtain an aggregated screening set. In the estimation stage, each machine transmits its local debiased coefficient estimates restricted to this set, and these estimates are averaged and thresholded to construct the final estimator. We establish estimation-error bounds that reveal how aggregating information across machines can mitigate the loss of estimation accuracy caused by more stringent local screening. We further derive bounds on the expected local and global screening sizes, characterizing how the local screening threshold and the number of machines affect communication cost per machine. In particular, we identify sufficient conditions under which the expected communication cost per machine is $O(s)$, with $s$ denoting the true sparsity level. Numerical experiments and a real-data analysis demonstrate that the proposed method achieves competitive estimation and prediction performance with substantially lower per-machine communication cost  than existing high-dimensional distributed quantile regression methods.
	\end{abstract}

	\noindent%

	\bigskip

		
\section{Introduction}

High-dimensional data are often stored across multiple machines, institutions, or devices because centralized storage and computation are impractical or institutional and data-ownership constraints limit the sharing of observations (\citealp[Section~1.2]{fan2020statistical}; \citealp{gao2022review}). Even when estimation is performed locally, exchanging high-dimensional statistical summaries can incur considerable communication costs, particularly when the number of covariates is comparable to or exceeds the local sample size. This motivates the development of communication-efficient
methods for distributed estimation and inference \citep{lee2017communication, jordan2019communication, fan2023communication,liu2025communication}.

Quantile regression (QR), introduced by \citet{koenker1978}, provides a flexible framework for characterizing covariate effects at different conditional quantiles. Compared with least squares regression, QR can reveal distributional heterogeneity and is less sensitive to extreme responses. These properties make it useful in economics and finance, including the analysis of compensation and consumption expenditure, as well as the measurement and forecasting of market and systemic risk \citep{koenker2001quantile, engle2004caviar, adrian2016covar}. The availability of increasingly rich covariate information has motivated extensive research on sparse high-dimensional QR \citep{belloni2011l1, wang2012quantile, fan2014adaptive, tan2022high}. When these data are distributed across machines, distributed QR methods must address both high-dimensional estimation and the communication required to combine local information. 

Existing distributed QR methods approach this problem through divide-and-conquer estimation, iterative aggregation, and communication-efficient updating. Early work focused primarily on moderate-dimensional settings, including the procedures of \cite{volgushev2019distributed} and \cite{chen2019quantile}, followed by pilot-sampling and one-step methods \citep{pan2022distributed} and communication-constrained aggregation \citep{tan2022communication}. In high-dimensional settings, \citet{chen2020} develop an approximate Newton-type procedure that uses gradient information in place of matrix-valued communication and establish guarantees for estimation and support recovery. \citet{wang2024distributed} further develop this approach through double smoothing, allowing error-covariate dependence while obtaining near-oracle estimation rates and exact support recovery. \citet{tan2022communication} combine double convolution smoothing with $\ell_1$-penalization for sparse distributed QR, while \citet{yang2026} develop a folded-concave regularization approach with a distributed-oracle property. These procedures exchange gradient or other vector-valued summaries, avoiding the transmission of full Hessian or covariance matrices containing $O(p^2)$ entries, where $p$ is the number of covariates. Nevertheless, their communication cost per machine per round remains of order $p$. Throughout this paper, communication is measured in machine words, with each scalar value or coordinate index counted as one word.

 Under stringent communication constraints, even an $O(p)$ communication cost per machine per round can be substantial when $p$ is large. However, in sparse high-dimensional models, only a small subset of covariates is relevant to the response, suggesting the possibility of reducing communication costs by transmitting information associated with fewer covariates. In related work, \citet{li2020distributed} develop a distributed feature screening method based on aggregating componentwise statistics under a minimum marginal signal strength condition. Although the resulting screening set can be small, their method still requires communicating information for all $p$ covariates before screening at the master machine. Thus, dimension reduction after aggregation does not reduce the communication cost of the screening procedure itself. This raises a natural question: {\it Can we exploit sparsity to reduce communication costs while retaining the statistical benefits of distributed estimation?}

One promising strategy is to perform variable screening locally before communication. For sparse linear regression, \citet{fonseca2023distributed} develop a communication-efficient procedure in which each machine selects candidate variables using debiased Lasso statistics and transmits only their indices to the master machine. The master aggregates these local selections through voting to recover the true support, followed by distributed least squares estimation on the recovered support. Motivated by this strategy, we develop a screening-then-estimation framework for high-dimensional distributed QR. Specifically, each local machine transmits the indices of selected covariates and the signs of their screening statistics to the master machine, which applies signed voting to construct a global screening set for subsequent distributed QR estimation. The master then averages the local debiased coefficient estimates over the selected coordinates and applies a final hard-thresholding step. Relatedly, \citet{liu2024majority} develop a differentially private procedure for distributed sign selection based on majority voting of local Lasso estimates. Their underlying majority-vote rule requires more than half of the machines to agree on a nonzero sign, whereas our procedure constructs a global screening set by thresholding the absolute sum of local signed votes.  In certain stringent communication regimes, the estimation guarantees of \citet{fonseca2023distributed} rely on exact support recovery under a minimum signal strength condition of order $\sqrt{(\log p)/n}$, where $n$ is the local sample size. Similarly, the sign recovery consistency result of \citet{liu2024majority} is established under a minimum signal strength condition.  In contrast, our analysis establishes estimation guarantees without requiring exact support recovery or imposing a minimum signal strength condition.  Although \citet{fonseca2023distributed} also introduce signed voting as a variant of ordinary voting, they do
not provide a theoretical analysis of this variant. Their theoretical guarantees are established only for
ordinary voting based on the number of local selections, which does not account for the potential cancellation
between positive and negative signed votes for inactive covariates. As discussed in Remark \ref{remark1} in Section \ref{sec.mainprocedure}, establishing theoretical guarantees for signed voting requires a refined characterization of the difference between the corresponding tail probabilities, rather than their sum.

The main contributions of this paper are twofold. First, we establish estimation guarantees for signed voting in high-dimensional distributed QR. A key challenge is to separate the expected signed votes for active and inactive covariates, even when the active signals are too weak to be reliably detected on individual machines. To address this challenge, we establish a uniform Bahadur representation and coordinatewise distributional approximations for the local debiased QR estimators. These results allow us to characterize the positive and negative selection probabilities and show how signed voting aggregates weak-signal information across machines. The resulting estimation guarantees do not require the local screening sets to satisfy the sure screening property or the aggregated screening set to recover the true support exactly.

Second, we quantify how
the local screening threshold $\rho$ and the number of machines $M$
jointly affect estimation accuracy and communication
efficiency.  More stringent local screening reduces
communication by retaining fewer covariates, but may
also discard relevant covariates with weak signals,
thereby increasing estimation error. Our analysis
shows that aggregation across machines can mitigate
this effect. Let $s$ denote the number of nonzero slope coefficients and $\phi$ the standard normal density. We derive a bound on the squared $\ell_2$ estimation error that includes the following contribution from screening:
\begin{align}\label{screening.error}
    \frac{s}{n}\min\bigg\{
\frac{1}{M(1+\rho)\phi(\rho)},\,1+\rho^2
\bigg\}\,.
\end{align} 
Holding the local sample size $n$ fixed, increasing $M$ can reduce this contribution and thus mitigate the estimation loss associated with more stringent local screening. Moreover, for sufficiently small $\rho$, corresponding to
limited local screening and communication of order $p$, the term \eqref{screening.error} is $O(s/N)$, where $N=Mn$ is the total sample size. To characterize communication efficiency, we further establish bounds on the expected local and global screening sizes. Specifically, let $\cS^{(m)}$ denote the local candidate support set produced by machine $m\in[M]$, and let $\hat{\cS}$ be the aggregated screening set. We identify sufficient conditions on the local screening threshold $\rho$ under which $\max_{m\in[M]}E\{|\cS^{(m)}|\}=O(s)$. For the global screening set, we further establish that, with the voting threshold suitably calibrated to the local screening threshold $\rho$, $E(|\hat{\cS}|)=O(s)$ can be achieved either by aggregating signed votes across sufficiently many machines or, even with a relatively small number of machines, through sufficiently stringent local screening. These results characterize how local screening and signed voting jointly determine communication cost and estimation error.

 
{\it Notation.} For any integer $q\geq2$, write $[q]=\{1,\ldots,q\}$. Let $\mathbf{1}(\cdot)$ denote the indicator function. For any vector $\bu = (u_1,\ldots,u_p)^{\T}\in\RR^p$,   define $\|\bu\|_0 = \sum_{j=1}^p \mathbf{1}(u_j \neq 0)$, $\|{\bu}\|_1 = \sum_{j=1}^p|{u_j}|$, $\|{\bu}\|_2 = (\sum_{j=1}^p u_j^2)^{1/2}$, $\|{\bu}\|_\infty = \max_{j\in[p]}|u_j|$, and  ${\rm supp}(\bu)=\{j\in[p]:u_j \neq 0\}$. For each $j\in[p]$,  write $\bv_{-j}=(v_k)_{k\in [p]\setminus\{j\}}\in\RR^{p-1}$ for any vector $\bv=(v_1,\ldots,v_p)^{\T}\in\RR^p$, and define  
$\bA_{j,-j}=(A_{j,k})_{k\in [p]\setminus\{j\}}\in\RR^{1\times(p-1)}$, $\bA_{-j,j}=(A_{k,j})_{k\in [p]\setminus\{j\}}\in\RR^{p-1}$, and $\bA_{-j,-j}=(A_{k,l})_{k,l\in [p]\setminus\{j\}}\in\RR^{(p-1)\times(p-1)}$ for any matrix $\bA=(A_{k,l})_{k,l\in [p]}$. Denote by $|\cI|$ the cardinality of a countable set $\cI$. For two sequences of positive numbers $\{a_n\}$ and $\{b_n\}$, we write $a_n\lesssim b_n$ or $b_n\gtrsim a_n$ if $\limsup_{n\rightarrow\infty}a_n/b_n\leqslant c $ for some positive constant $c $, write $a_n\asymp b_n$ if $a_n\lesssim b_n$ and $b_n\lesssim a_n$ hold simultaneously, and write $a_n\ll b_n$ or $b_n\gg a_n $ if $\limsup_{n\rightarrow\infty}a_n/b_n=0$. For a symmetric matrix $\bSigma$, let $\lambda_{\min}(\bSigma)$ and
$\lambda_{\max}(\bSigma)$ denote its smallest and largest eigenvalues, respectively. 
Let $\phi(\cdot)$ and $\Phi(\cdot)$ denote the probability density function and cumulative distribution function of the standard normal distribution, respectively, and let $\bar{\Phi}(\cdot)=1-\Phi(\cdot)$.

\section{Methodology}
\label{sec.methodology}

\subsection{Distributed QR in high dimensions}

Consider a distributed system consisting of $M$ local machines connected to a master machine. For ease of exposition, suppose that the observations are evenly distributed across the local machines, with each machine storing $n$ observations and the total sample size given by $N=Mn$.   
Let $\{(y_i,\bx_i)\}_{i\in[N]}$ denote the combined sample, where the observations are independent and identically distributed copies of $(y,\bx)\in\RR\times\RR^p$. We assume that the conditional $\tau$-quantile of $y$ given $\bx$ follows the linear model $F_{y\,|\,\bx}^{-1}(\tau):= \inf\{ a \in \RR: F_{y\,|\,\bx}(a)\geq \tau \}=\bx^{\T}\bbeta^*$, where $F_{y\,|\,\bx}(\cdot)$ denotes the conditional distribution function of $y$ given $\bx$, and $\bbeta^*=(\beta_1^*,\ldots,\beta_p^*)^{\T}$ is the true QR coefficient vector. Equivalently, letting $\varepsilon_i=y_i-\bx_i^{\T}\bbeta^*$ with $\bx_i=(x_{i,1},\ldots,x_{i,p})^{\T}$, the conditional quantile restriction implies $\pr(\varepsilon_{i} < 0\,|\,\bx_i) \leq \tau \leq \pr(\varepsilon_{i}\leq 0\,|\,\bx_i)$. When the conditional distribution of $\varepsilon_i$ given $\bx_i$ is continuous at zero, this restriction reduces to
\begin{align}\label{linear.model}
  y_i = \bx_i^{\T}\bbeta^* + \varepsilon_i\,, ~~  
  \pr(\varepsilon_{i}\leq 0\,|\,\bx_i)=\tau\,,~~i\in[N]\,.
\end{align} 
Here, $x_{i,1}\equiv 1$ so that $\beta_1^*$ represents the intercept. For theoretical convenience, we assume without loss of generality that \(E(x_{i,j})=0\) for each \(j\in\{2,\ldots,p\}\), since centering the covariates at their population means leaves the slope coefficients unchanged and modifies only the intercept. For notational simplicity, we suppress the dependence of $\bbeta^*$ and $\varepsilon_i$ on $\tau$.

We consider the high-dimensional setting in which the slope coefficient vector
$\bbeta_{-1}^*$ is $s$-sparse ($s\geq 1$) and satisfies
$\|\bbeta_{-1}^*\|_\infty=O(1)$, where
$s=|\cS|\ll\min(n,p)$ with
$\cS=\{j\in\{2,\ldots,p\}:\beta_j^*\neq0\}$.
Our goal is to estimate $\bbeta^*$ in a distributed environment subject to communication constraints between the local
and master machines. A full-dimensional aggregation procedure requires each local machine to transmit a $p$-dimensional vector to the master machine, incurring a per-machine communication cost of $O(p)$ per round. When $p$ is large, this cost can be prohibitive. If the true support $\cS$ were known, estimation could be restricted to the $s$ active covariates, reducing the communication cost to $O(s)$ per machine. Motivated by this observation, we develop a screening-then-estimation strategy that exploits sparsity to reduce communication costs without requiring prior knowledge or exact recovery of $\cS$.

Specifically, our screening-then-estimation strategy proceeds in two stages. In the screening stage, each local machine $m\in[M]$ constructs a candidate set $\cS^{(m)}\subseteq[p]$ using locally computed screening statistics and transmits the selected coordinate indices together with the signs of the corresponding statistics to the master machine. The master then aggregates this information through a signed voting procedure to obtain a global screening set $\hat{\cS}\subseteq[p]$. In the estimation stage, the master machine broadcasts $\hat{\cS}$ to all local machines, which then transmit their local debiased coefficient estimates for the selected covariates and the intercept. The master averages these estimates across machines for each selected coordinate and the intercept, and applies a final hard-thresholding step. Thus, the communication costs in the two stages are governed by the sizes of the local candidate sets and the global screening set, respectively. To implement this strategy, Section \ref{sec.debiase} develops a debiasing-based approach for constructing local debiased coefficient estimates and screening statistics in high-dimensional QR. Section \ref{sec.mainprocedure} then presents the complete screening-then-estimation procedure based on these local quantities.

\subsection{Local debiased estimation}
\label{sec.debiase}

For each $m\in[M]$, local machine $m$ stores a subsample of $n$ observations, denoted by $\cD_m=\{(y_i,\bx_i)\}_{i\in\cI_m}$, where $\{\cI_m\}_{m\in[M]}$ forms a partition of $[N]$ with $|\cI_m|=n$. To stabilize estimation and induce sparsity in high dimensions, we combine $\ell_1$-penalized QR \citep{belloni2011l1} with convolution-smoothing \citep{tan2022high}. Let $\rho_\tau(u)=u\{\tau-\mathbf{1}(u<0)\}$ denote the check loss at quantile level $\tau$, and let $K:\mathbb{R}\to[0,\infty)$ be a symmetric kernel that integrates to one. Given  a smoothing bandwidth $h>0$ and a regularization parameter $\lambda>0$, define the local $\ell_1$-penalized smoothed QR estimator by
\begin{align}\label{l1.est}
    \hat{\bbeta}_{h}^{(m)}
    \in
    \arg\min_{\bbeta\in\RR^p}
    \big\{ 
        \hat{Q}_{h}^{(m)}(\bbeta) +  \lambda\|\bbeta\|_1
    \big\} \,  ~\mbox{ with }~ \hat{Q}_h^{(m)}(\bbeta) = \frac{1}{n } \sum_{i\in\cI_m} (\rho_\tau \circ K_h )(y_i-\bx_i^{\T}\bbeta) \,.
\end{align}
Here, $\circ$ denotes convolution and $K_h(u) = h^{-1}K(u/h)$.  Thus, $\hat{Q}_{h}^{(m)}(\bbeta)$ is a convolution-smoothed counterpart of the local empirical quantile loss $\hat{Q}^{(m)}(\bbeta) = {n}^{-1}\sum_{i\in\cI_m} \rho_\tau (y_i-\bx_i^{\T}\bbeta)$. Under some regularity conditions on $K$, 
$\hat{Q}_{h}^{(m)}$ is twice continuously differentiable, with gradient $\nabla\hat{Q}_{h}^{(m)}(\bbeta)={n}^{-1}\sum_{i\in \cI_m}\{\bar{K}_h(\bx_i^{\T}\bbeta-y_i)-\tau\}\bx_i$ and Hessian $\nabla^2 \hat{Q}_{h}^{(m)}(\bbeta)={n}^{-1}\sum_{i\in \cI_m} K_h (\bx_i^{\T}\bbeta-y_i)\bx_i{\bx}_i^{\T}$, where $\bar{K}_h(u)=\int_{-\infty}^{u/h}K(v)\,\mbox{d}v$.  Further details on convolution smoothing are provided in Section~\ref{sec.smoothing} of the supplementary material. In our implementation, the initial estimator is computed using the authors' publicly available code, with the intercept left unpenalized, as discussed in Remark~2 of \cite{tan2022high}.

Write $\bbeta=(\beta_1,\ldots,\beta_p)^{\T}$. For a fixed $j\in[p]$, regard $\beta_j$ as the target parameter and
$\bbeta_{-j} \in \mathbb{R}^{p-1}$ as the nuisance parameter. On local machine $m\in[M]$, the naive score for $\beta_j$ associated with the smoothed loss $\hat{Q}^{(m)}_h(\bbeta)$ is $\{\nabla\hat{Q}^{(m)}_h(\bbeta)\}_j$. In high-dimensional settings, direct use of this score is strongly affected by the regularization bias arising from estimation of $\bbeta_{-j}$. To mitigate this bias, we follow the decorrelated score framework of \citet{ning2017general} and define the local decorrelated score for the $j$th coordinate on machine $m$ as 
$$
{S}_j^{(m)}( \bbeta ) 
=
     \{\nabla\hat{Q}^{(m)}_h( \bbeta ) \}_j  - 
     {\bw}_{(j)}^{\T}
     \{\nabla\hat{Q}^{(m)}_h( \bbeta )\}_{-j} \,,
$$
where $\bw_{(j)}=\{\bH_{-j,-j}^{(h)}\}^{-1}\bH^{(h)}_{-j,j}$, $\bH^{(h)}   = \nabla^2 Q_h(\bbeta^*)$ and $Q_h(\bbeta)=E\{\hat Q_h^{(m)}(\bbeta)
-\hat Q_h^{(m)}(\bbeta^*)\}$. This choice of $\bw_{(j)}$ satisfies $\bH^{(h)}_{j,-j} - \bw_{(j)}^{\T}\bH^{(h)}_{-j,-j}
= \boldsymbol{0}$, and therefore makes the population decorrelated score locally insensitive to changes in the nuisance parameter at the true coefficient vector. A similar orthogonalization idea has also been employed by
\citet{chang2021high} to facilitate inference on parameters
of interest in more general framework. Because $\bw_{(j)}$ is unknown, we can estimate it using a Dantzig-type program \citep{candes2007dantzig}.  Given the initial estimator $\hat{\bbeta}^{(m)}_h=(\hat{\beta}_{h,1}^{(m)},\ldots,\hat{\beta}_{h,p}^{(m)})^{\T}$ and a tuning parameter $\Delta>0$, define
\begin{align}\label{Ch3dantzig}
      \hat{\bw}_{(j)}^{(m)} \in \arg\min_{\bw \in\mathbb{R}^{p-1}} \|{\bw}\|_1 ~~ \mathrm{s.t.}~~ \big \|\{\nabla^2\hat{Q}^{(m)}_h(\hat{\bbeta}^{(m)}_h)\}_{j,-j} -  {\bw}^{\T}\{\nabla^2\hat{Q}^{(m)}_h(\hat{\bbeta}^{(m)}_h)\}_{-j,-j}\big\|_\infty\leq \Delta\,.
\end{align} 
 The resulting estimated decorrelated score is 
\begin{align*} 
    \hat{S}_j^{(m)}(\hat{\bbeta}^{(m)}_h)
    =
    \{\nabla\hat{Q}^{(m)}_h(\hat{\bbeta}^{(m)}_h)\}_j
    -
    \{\hat{\bw}_{(j)}^{(m)}\}^{\T}
    \{\nabla\hat{Q}^{(m)}_h(\hat{\bbeta}^{(m)}_h)\}_{-j}\,.
\end{align*}
Motivated by \cite{ning2017general}, we define the debiased estimator
of $\beta_j^*$ as
\begin{align}\label{eq:debiased-beta-j}
    \tilde{\beta}_j^{(m)}
    =
    \hat{\beta}_{h,j}^{(m)}
    -
    \{\hat{H}_{j|-j}^{(m)}\}^{-1}
    \hat{S}_j^{(m)}(\hat\bbeta^{(m)}_h)\,, 
\end{align} 
where $\hat{H}_{j|-j}^{(m)} = \{\nabla^2\hat{Q}^{(m)}_h(\hat{\bbeta}^{(m)}_h)\}_{j,j} - \{\hat{\bw}_{(j)}^{(m)}\}^{\T} \{\nabla^2\hat{Q}^{(m)}_h(\hat{\bbeta}^{(m)}_h)\}_{-j,j}$. As we will establish in Theorem~\ref{thm.normal} in Section \ref{sec.theory}, under suitable regularity conditions, $\sqrt{n} ( \tilde{\beta}_j^{(m)}-\beta_j^*)\hat{H}^{(m)}_{j|-j}/\hat{\sigma}^{(m)}_{j} \rightarrow N(0,1)$ in distribution for each $j\in[p]$ and $m\in[M]$, where 
\begin{align*} 
    (\hat{\sigma}^{(m)}_{j})^2= \frac{1}{n}\sum_{i\in\cI_m} \big[ \{  \tau-\bar{K}_h( \bx_i^{\T}\hat{\bbeta}^{(m)}_h - y_i) \}  \bx_i^{\T}\hat{\bv}_{(j)}^{(m)} \big]^2\,.
\end{align*}
Here,  
$\hat{\bv}_{(j)}^{(m)}\in\RR^p$ is defined by $\{\hat{\bv}_{(j)}^{(m)}\}_j=1$ and $\{\hat{\bv}_{(j)}^{(m)}\}_{-j}=-\hat{\bw}_{(j)}^{(m)}$. For coordinatewise support screening, define the studentized statistic
\begin{align}\label{statistic}
     \hat{\xi}_j^{(m)} =  \frac{\sqrt{n} \,  \tilde{\beta}_j^{(m)}
            \hat{H}^{(m)}_{j|-j} }{\hat{\sigma}^{(m)}_{j}}\,,~~j\in[p]\,, \, m\in[M] \,.
\end{align}      
We set $\tilde{\beta}_j^{(m)}=0$ if $\hat{H}_{j|-j}^{(m)}=0$,
and $\hat{\xi}_j^{(m)}=0$ if $\hat{\sigma}_j^{(m)}=0$. Throughout, ratios with denominator $\hat\sigma_j^{(m)}$ are defined to be zero on the event ${\hat\sigma_j^{(m)}=0}$. For an inactive slope coordinate $j\in \{2,\ldots,p\}\!\setminus \!\cS$, $\hat{\xi}_j^{(m)}$ is centered around
zero, whereas for an active slope coordinate $j\in\mathcal{S}$ it exhibits a location shift
proportional to $\sqrt{n}\beta_j^*$. This distinction motivates the use of $\{\hat\xi_j^{(m)}\}$ for local variable screening in the screening-then-estimation procedure developed in Section \ref{sec.mainprocedure}.

\subsection{Screening-then-estimation procedure}\label{sec.mainprocedure}

Motivated by the distributed support recovery approach of \cite{fonseca2023distributed}, we first develop a screening procedure consisting of two steps: local screening and global screening. In the local screening step, each machine $m\in[M]$ uses the screening statistics $\{\hat\xi_j^{(m)}\}$ defined in \eqref{statistic} to construct a local candidate set
\begin{align}\label{eq:local.screening}
    \cS^{(m)}
    = \{
    j\in\{2,\ldots,p\}:
    |\hat\xi_j^{(m)}|\geq\rho
    \}\,,
\end{align}
where $\rho\geq 0$ is a prescribed screening threshold that controls the size of the local candidate set. Each local machine $m\in[M]$ then transmits
$\{(j,\operatorname{sign}\{\hat\xi_j^{(m)}\}):j\in\cS^{(m)}\}$
to the master machine. In the global screening step, the master machine computes the signed voting statistic $V_j
    =
    {M}^{-1}\sum_{m=1}^M D_j^{(m)} $ for each $j\in \{2,\ldots,p\}$, where $D_j^{(m)}=\operatorname{sign}\{\hat\xi_j^{(m)}\}\mathbf{1}\{j\in\cS^{(m)}\}$ denotes the signed vote for the $j$th coordinate from machine $m$. The global screening set is defined as 
\begin{align}\label{eq:global.screening}
    \hat{\cS}
    =
    \{
    j\in\{2,\ldots,p\}:
    |V_j|\geq v
    \}\,,
\end{align}
where $v>0$ is a prescribed voting threshold that controls the size of the global screening set.

\begin{remark} \label{remark1}
The ordinary voting scheme primarily studied by {\rm \citet{fonseca2023distributed}} counts the proportion of machines selecting each coordinate: $V_j^{\mathrm{ord}}=M^{-1}\sum_{m=1}^M\mathbf{1}\{j\in\cS^{(m)}\}$. Signed voting additionally uses the signs of the screening statistics, allowing positive and negative votes for inactive coordinates to cancel. Specifically, $E(V_j)=\pr\{\hat\xi_j^{(m)}\geq\rho\}-\pr\{\hat\xi_j^{(m)}\leq-\rho\}$, whereas, for $\rho>0$,  ordinary voting yields $E(V_j^{\mathrm{ord}})=\pr\{\hat\xi_j^{(m)}\geq\rho\}+\pr\{\hat\xi_j^{(m)}\leq-\rho\}$. When $\rho=0$, all coordinates are locally retained, so $E(V_j^{\mathrm{ord}})=1$. For inactive coordinates, such cancellation can  reduce $|{E}(V_j)|$, potentially allowing a smaller voting threshold $v$ to retain weaker active signals while controlling the inclusion of inactive coordinates. Establishing this separation requires precise bounds on the differences between the positive and negative tail probabilities.
Moreover, the voting threshold $v$ should be chosen as a function of the local screening threshold $\rho$, with the primary objective of retaining relevant covariates while controlling the size of the global screening set. When $\rho$ is small, local screening is relatively permissive, and a relatively large $v$ helps exclude irrelevant covariates. In contrast, when $\rho$ is large, local screening becomes more stringent, making inactive coordinates less likely to be selected but also increasing the risk of missing active coordinates at individual machines. A smaller $v$ can then help retain active coordinates that are selected by only a small fraction of machines, thereby mitigating the impact of missed selections at individual machines.
\end{remark}

The communication efficiency of the screening stage relies on controlling the sizes of the local candidate sets.   As established in Proposition \ref{prop.screening.size} in  Section \ref{sec.theory}, for suitably large $\rho$, the expected number of coordinates transmitted by each local machine $m\in[M]$ satisfies $E\{| \cS^{(m)}| \}=O(s)$, indicating that the expected per-machine communication cost of the screening stage can be controlled at the true sparsity level. Moreover, the size of the global screening set $\hat{\cS}$ directly affects the communication cost of the subsequent estimation stage. 
Proposition \ref{prop.screening.size} further shows that, with the voting threshold $v$ suitably calibrated to $\rho$, $E(|\hat{\cS}|)=O(s)$ can be achieved by aggregating signed votes across sufficiently many machines.
Alternatively, when the local screening threshold $\rho$ is
sufficiently large, the same bound can be achieved even with
relatively few machines, as fewer irrelevant variables are
retained at the local screening stage.

Once the global screening set $\hat{\cS}$ is obtained, we average the local debiased coefficient estimates $\{\tilde{\beta}_j^{(m)}\}$ given in \eqref{eq:debiased-beta-j} across machines for each selected coordinate and the intercept.  More precisely, the master machine first broadcasts $\hat{\cS}$ to all local machines. Each local machine $m\in[M]$ then transmits $\{\tilde{\beta}_j^{(m)}: j \in\{1\}\cup\hat{\cS}\}$ to the master machine, which computes   
\begin{align}\label{eq:final.estimator}
    \hat\beta_j
    =
    \begin{cases}
    \displaystyle
    \frac{1}{M}\sum_{m=1}^M
    \tilde\beta_j^{(m)}\,,
    & j\in\{1\}\cup\hat{\cS}\,,\\[8pt]
    0\,,
    & j\in\{2,\ldots,p\}\!\setminus\!\hat{\cS}\,.
    \end{cases}
\end{align} 
Let $\hat{\bbeta}=(\hat{\beta}_1,\ldots,\hat{\beta}_p)^{\T}$ denote the resulting averaged estimator. Finally, we apply a hard-thresholding operator $\mathcal T_\gamma$ to obtain the final estimator $\mathcal T_\gamma(\hat{\bbeta})$, where $\mathcal T_\gamma$ acts coordinatewise according to $\{\mathcal T_\gamma(\hat{\bbeta})\}_j
=
\hat{\beta}_j\mathbf{1}\{|\hat{\beta}_j|\geq\gamma\}$ for each $j\in[p]$, with $\gamma>0$ being a prescribed threshold.  The additional hard-thresholding step, motivated by \cite{battey2018distributed}, controls the estimation error arising from falsely selected coordinates without requiring an explicit bound on the size of the global screening set $\hat{\cS}$.

\section{Statistical theory}
\label{sec.theory}


We first establish a uniform Bahadur representation and coordinatewise normal approximations for the local debiased estimators. We then derive an estimation error bound for the distributed estimator and characterize its communication cost through the expected local and global screening sizes. Throughout this section, we assume that
$\max\{M,n\}\leq p^{\iota}$ for some fixed constant  $\iota >0$. We further impose the following assumptions.

\begin{condition} \label{cond}
(a) The conditional distribution of $\varepsilon=y-\bx^{\T}\bbeta^*$ given $\bx$ admits a density
$f_{\varepsilon\,|\,\bx}$. There exist constants $l_0>0$ and
$0<f_l<f_u<\infty$ such that, almost surely with respect to $\bx$,
$\sup_{u\in\RR}|f'_{\varepsilon\,|\,\bx}(u)|\leq l_0$,
$f_{\varepsilon\,|\,\bx}(0)\geq f_l$, and $\sup_{u\in\RR} f_{\varepsilon\,|\,\bx}(u)\leq f_u$.  (b) There exists a constant $\upsilon_0 \geq 1$ such that $\pr(|\langle \bSigma^{-1/2}\bx, \bu\rangle|\geq \upsilon_0 z)\leq  2 e^{-z}$ for every $\bu \in \mathbb{S}^{p-1}$ and $z\geq 0$, where $\bSigma = E(\bx\bx^{\T})$ satisfies $ 0< \gamma_p :=\lambda_{\min}(\bSigma) \leq \lambda_{\max}(\bSigma) =: \gamma_1 <\infty$, with $\gamma_p$ and $\gamma_1$ uniformly bounded away
from zero and infinity, respectively. (c) Let $\tilde{\bH}=
E\{f_{\varepsilon\,|\,\bx}(0)\bx\bx^{\T}\}$. Assume that $\max_{j\in [p]}\|(\tilde{\bH}^{-1})_{j,-j}\|_0 \leq \tilde{s}$ for some $\tilde{s}\geq 1$. (d) The kernel function $K:\mathbb{R}\to[0,\infty)$ is symmetric about zero and satisfies $\int_{-\infty}^{\infty}K(u)\,\mbox{d}u=1$. For $\ell=1,2$, let $\kappa_\ell
= \int_{-\infty}^{\infty}|u|^\ell K(u)\,\mbox{d}u<\infty$. Moreover, $K$ is twice continuously differentiable and satisfies $\kappa_u = \sup_{u\in\mathbb{R}}K(u)<\infty$, $\kappa_u'
= \sup_{u\in\mathbb{R}}|K'(u)|<\infty$, and $\kappa_u''= \sup_{u\in\mathbb{R}}|K''(u)|<\infty$. 
In addition, $C_K = \int_{-\infty}^{\infty}|K'(u)|\,\mbox{d}u<\infty$, and there exists a constant $\underline{\kappa}>0$ such that $\inf_{|u|\leq 1}K(u)\geq \underline{\kappa}$.
\end{condition}

Condition~1(a) requires the conditional density and its first derivative to be uniformly bounded and the conditional density at zero to be bounded away from zero. Such conditions are standard in the high-dimensional QR literature \citep{belloni2011l1,tan2022high}. Condition~1(b) imposes uniformly sub-exponential tails on the standardized covariates, facilitating the exponential concentration bounds used in our theoretical analysis. The matrix $\tilde{\bH} = E\{ f_{\varepsilon\,|\,\bx}(0)\bx\bx^{\T} \}$
is the Hessian matrix of the unsmoothed population quantile loss $\bbeta\mapsto E\{\rho_{\tau}(y-\bx^{\T}\bbeta)-\rho_{\tau}(y-\bx^{\T}\bbeta^*)\}$ evaluated at the true coefficient $\bbeta^*$. Recall that $\bw_{(j)}=\{\bH_{-j,-j}^{(h)}\}^{-1}\bH^{(h)}_{-j,j}$ is the decorrelating direction associated with the smoothed population loss $Q_h$. Condition~1(c) imposes row sparsity on the bandwidth-free inverse Hessian $\tilde{\bH}^{-1}$, rather than directly on the bandwidth-dependent matrix $\{ {\bH}^{(h)}\}^{-1}$. To see its implication for the decorrelating direction, define $\tilde{\bw}_{(j)}=(\tilde{\bH}_{-j,-j})^{-1}\tilde{\bH}_{-j,j}$. Writing \(\tilde{\boldsymbol\Theta} =(\tilde{\Theta}_{k,l})_{k,l\in[p]}=\tilde{\bH}^{-1}\), the block-inverse identity yields $ \tilde{\bw}_{(j)} = - \tilde{\boldsymbol\Theta}_{-j,j} /
{\tilde{\Theta}_{j,j}}$. Since $\tilde{\boldsymbol\Theta}$ is symmetric,
Condition~1(c) implies
$\|\tilde{\bw}_{(j)}\|_0
=\|\tilde{\boldsymbol\Theta}_{-j,j}\|_0
\leq \tilde{s}$.  
Together with control of the smoothing discrepancy between $\bH^{(h)}$ and $\tilde{\bH}$, this condition provides a sparse population target and facilitates estimation of the decorrelating direction through a Dantzig-type program; see \cite{ning2017general}. Our formulation differs from the inference framework of \citet{yan2023confidence}, which imposes both row sparsity and a uniform matrix $\ell_1$-norm bound directly on the bandwidth-dependent inverse Hessian $\{\bH^{(h)}\}^{-1}$. Finally, when $f_{\varepsilon\,|\,\bx}(0)$ does not depend on $\bx$, we have $\tilde{\bH}^{-1} = \{f_{\varepsilon}(0)\}^{-1}\bSigma^{-1}$. In this case, Condition~1(c) reduces to a row-sparsity condition on the precision matrix $\bSigma^{-1}$, as commonly assumed in high-dimensional debiased inference; see, for example, \cite{vandegeer2014asymptotically}. Finally, Condition~1(d) imposes smoothness and moment conditions on kernel function $K$. These conditions are satisfied by several commonly used smooth kernels, including the Gaussian and logistic kernels, and are comparable to those typically imposed in convolution-smoothed QR; see \cite{fernandes2021smoothing} and \cite{tan2022high}.

According to \eqref{eq:local.screening}, $\pr\{j\in\cS^{(m)}\}= \pr\{|\hat{\xi}_j^{(m)}|\geq\rho\}$ for each $j\in\{2,\ldots,p\}$ and $m\in[M]$. To analyze this probability, we first derive a Bahadur representation of the studentized estimation error. For each $j\in[p]$, recall the unsmoothed population decorrelating direction $\tilde{\bw}_{(j)} = (\tilde{\bH}_{-j,-j})^{-1} \tilde{\bH}_{-j,j}$, and let $\bv_{(j)}\in\RR^p$ satisfy $\{\bv_{(j)}\}_j=1$ and $\{\bv_{(j)}\}_{-j}=-\tilde{\bw}_{(j)}$. For each $i\in\cI_m$ and $j\in[p]$, define
$$
\zeta_{i,j} = \{\tau-\bar{K}_h(-\varepsilon_i)\} \bx_i^{\T}\bv_{(j)} \,, ~~ \sigma_j^2={\rm var}(\zeta_{i,j})\,, ~~\mbox{and}~~ Z_j^{(m)}:= \frac{1}{\sqrt{n}}\sum_{i\in\cI_m}\frac{ \zeta_{i,j}-E(\zeta_{i,j}) }{\sigma_j}\,.
$$
In the following analysis, the fixed constants specified in Conditions~1(a), 1(b), and 1(d) are absorbed into the notation $\lesssim$, $\gtrsim$, and $\asymp$. Write 
\begin{align}\label{def.deltan}
    \delta_n  =  h^2 + \frac{(\tilde{s}\vee s)s^2(\log p)^{11/2}}{n^{3/2}h^3}\,.
\end{align}

\begin{proposition}
\label{prop.Bahadur}
    Suppose Condition~{\rm \ref{cond}} holds and the tuning parameters and bandwidth required for constructing \(\{\hat{\xi}_j^{(m)}\}\) in \eqref{statistic} satisfy
    \begin{align*}
         \lambda\asymp   \sigma_{\bx}\sqrt{\frac{\tau(1-\tau)\log p}{n}}\,, ~ \Delta \asymp \frac{s^2(\log p)^5 }{   nh^3}\,, ~~\mbox{and}~~\bigg(  \frac{s\log p}{n}\bigg)^{1/2} \lesssim h \lesssim \bigg(  \frac{s\log p}{n}\bigg)^{1/4} \,,
    \end{align*}
    where $\sigma^2_{\bx}=\max_{j\in[p]}E(x_{j}^2)$. Then, for every  $m\in[M]$,
    \begin{align*}
        \pr \bigg\{ \max_{j\in[p]} \bigg| \hat{\xi}_j^{(m)} -  \frac{\sqrt{n}  \beta_j^*
            \hat{H}^{(m)}_{j|-j} }{\hat{\sigma}^{(m)}_{j}}  -  Z_j^{(m)} \bigg| \geq  \bar{C}\sqrt{n}\delta_n  \bigg\}  \lesssim p^{-\bar{c}}\,,
    \end{align*}
  provided that $\sqrt{n}\delta_n$ is sufficiently small,
where $\delta_n$ is defined in \eqref{def.deltan}, $\bar c>0$ is a sufficiently large constant,
and $\bar C>0$ may depend on $\bar c$ but is
independent of $(n,p,M)$. Moreover, the multiplicative constants for $\lambda$ and $\Delta$ are sufficiently large, and $h$ satisfies the bandwidth conditions of Lemma {\rm \ref{Theorem.tan2022}} in the supplementary material. These constants may depend on $\bar c$.
\end{proposition}

\begin{remark} 
For each local machine, Proposition~{\rm \ref{prop.Bahadur}} approximates the screening statistics uniformly over coordinates by a signal-dependent term and a leading stochastic term $Z_j^{(m)}$, with a controlled remainder. This representation facilitates the analysis of the positive and negative tail probabilities of $\hat\xi_j^{(m)}$ through those of $Z_j^{(m)}$. In particular, characterizing the difference between these tail probabilities is crucial for establishing the separation of expected signed votes between active and inactive coordinates. 
The exponent $\bar{c}$ in Proposition~{\rm \ref{prop.Bahadur}} is introduced to facilitate the subsequent theoretical analysis. 
\end{remark}

 As shown in \eqref{normalZ} of the supplementary material,
the leading stochastic term $Z_j^{(m)}$ satisfies $\sup_{x\in\RR}
    |
    \pr\{Z_j^{(m)}\leq x\}-\Phi(x)
    |
    \lesssim n^{-1/2}$ 
for every $j\in[p]$ and $m\in[M]$.
Combining this normal approximation with the Bahadur
representation in Proposition~\ref{prop.Bahadur}
yields a distributional approximation for the studentized
estimation error associated with $\tilde{\beta}_j^{(m)}$,
as stated in Theorem~\ref{thm.normal}. Asymptotic normality follows when the approximation error tends to zero.

\begin{theorem} \label{thm.normal}
Let the assumptions of Proposition {\rm \ref{prop.Bahadur}}  hold. Then, for each $j\in[p]$ and $m\in[M]$,
    \begin{align*}
		\sup_{x\in\RR}\big| \pr\big\{ 
              \sqrt{n} ( \tilde{\beta}_j^{(m)}-\beta_j^*)\hat{H}^{(m)}_{j|-j}   \leq  \hat{\sigma}^{(m)}_{j} x\big\} -\Phi(x) \big|\lesssim  \sqrt{n}\delta_n + p^{-\bar c} \,,
	\end{align*}
    where  $\delta_n$ is defined in \eqref{def.deltan} and $\bar{c}>0$ is specified in Proposition {\rm \ref{prop.Bahadur}}. 
\end{theorem}

\begin{remark}
Theorem~{\rm \ref{thm.normal}} is closely related to the inference theory for high-dimensional QR developed by \textup{\citet{yan2023confidence}}. Both approaches combine convolution smoothing with debiasing to establish asymptotic normality. \textup{\cite{yan2023confidence}} study inference for general linear projections of a debiased $\ell_1$-penalized smoothed QR estimator. In comparison, we establish coordinatewise distributional approximations at the local machine level and use them to characterize the local selection probabilities underlying the distributed screening procedure. The assumptions imposed in the two analyses also differ. \textup{\cite{yan2023confidence}} additionally require $s\lesssim\log^{1/2}(p\vee n)$ and impose both row sparsity, quantified by \(c_{n,p}\), and a uniform matrix \(\ell_1\)-norm bound on $\{\bH^{(h)}\}^{-1}$. Our formulation instead imposes row sparsity on the bandwidth-free inverse Hessian $\tilde{\bH}^{-1}$ and explicitly tracks the combined effect of the regression sparsity \(s\) and the decorrelation sparsity \(\tilde s\) through \(\tilde s\vee s\). When \(c_{n,p}\) is treated as fixed, the simplified growth condition in \textup{\cite{yan2023confidence}} takes the form $s^8\operatorname{polylog}(n,p)\ll n$, where \(\operatorname{polylog}(n,p)\) refers to a term that depends polynomially on the logarithms of \(n\) and \(p\). For our result, balancing the two terms in \(\delta_n\) yields the sufficient condition $(\tilde{s}\vee s)^4s^8\operatorname{polylog}(p)\ll n$. Under the additional restriction $s\lesssim\log^{1/2}(p\vee n)$ imposed by \textup{\cite{yan2023confidence}}, and with \(\tilde s\) treated as fixed analogously to their treatment of \(c_{n,p}\), the factor \((\tilde s\vee s)^4\) is at most polylogarithmic in $(n,p)$ and can be absorbed into the generic polylogarithmic term. Under this sparsity regime, the sufficient growth conditions agree up to logarithmic factors.
\end{remark}

\begin{theorem}\label{thm.convergence}
    Let the assumptions of Proposition {\rm \ref{prop.Bahadur}}  hold.
    Take the local screening threshold $\rho$ in
\eqref{eq:local.screening} to satisfy
$0\leq\rho\leq\sqrt{2\log p}$, and choose the
voting threshold $v$ in \eqref{eq:global.screening}
as $v\asymp\sqrt{n}\delta_n\phi(\rho)$, where $\delta_n$ is defined in \eqref{def.deltan}.   Then, with $\gamma \asymp \sqrt{ {(\log p)}/{N}}+\delta_n$ and $\sqrt{n}\delta_n=o(1)$, it holds that 
    \[
    \| \mathcal{T}_{\gamma}(\hat\bbeta) - \bbeta^* \|_2^2 = O_{\rm p}\bigg(s \bigg[  \frac{\log p}{N} + \delta_n^2 + \frac{1}{n}\min\bigg\{ \frac{1}{M (1+\rho)\phi(\rho)}, 1+\rho^2 \bigg\}  \bigg]   \bigg)\,.
    \]
Here, the constants implicit in the lower bounds for
$v$ and $\gamma$ are taken to be sufficiently large.
\end{theorem}

\begin{remark}
   The local screening threshold $\rho$ can, in principle, be larger than $\sqrt{2\log p}$. Based on our Bahadur representation (Proposition {\rm \ref{prop.Bahadur}}),  a moderate-deviation argument allows us to control the selection probabilities of inactive coordinates in terms of the standard normal tail probability $\bar\Phi(\rho)$ (see Proposition \ref{prop.screening.size} and its proof). In particular, at $\rho=\sqrt{2\log p}$, this tail probability is already sufficiently small to help control the expected number of false selections. This motivates our focus on $ \rho \in [0,\sqrt{2\log p}]$.
   The voting threshold $v$ is chosen to distinguish active from inactive coordinates based on their expected signed votes. The choice $v\asymp\sqrt{n}\delta_n\phi(\rho)$ accounts for the systematic bias induced by the Bahadur remainder in the expected signed votes of inactive coordinates. In the idealized symmetric setting in which the Bahadur remainder is absent, the expected signed votes of inactive coordinates vanish, whereas those of active coordinates with weak signals satisfying $|\beta_j^*|=o(n^{-1/2})$ have absolute values bounded below by a quantity proportional to $\sqrt{n}|\beta_j^*|\phi(\rho)$. Consequently, to distinguish signals exceeding the target estimation rate $r_n=o(n^{-1/2})$ from inactive coordinates, one could instead choose $v\asymp\sqrt{n}r_n\phi(\rho)$.
\end{remark}

\begin{remark}
    The convergence bound in Theorem {\rm \ref{thm.convergence}} consists of three components, each reflecting a different source of estimation error. The first term, $s(\log p)/N$, represents the standard full-sample estimation error arising from information aggregation across machines. The second term, $s\delta_n^2$, reflects the effect of the local Bahadur remainders. As defined in \eqref{def.deltan}, $\delta_n$ consists of two components: the first captures the bias introduced by convolution smoothing, while the second reflects the estimation errors associated with the initial
penalized estimator and the Dantzig-type program. Notably, $\delta_n$ depends on the local sample size $n$, rather than the total sample size $N$. This dependence arises because the initial estimators and decorrelating directions are estimated separately on each machine. Finally, the third term, $sn^{-1}\min[\{M(1+\rho)\phi(\rho)\}^{-1},\,1+\rho^2]$ 
captures the additional estimation error arising from signal loss during screening. A smaller value of this term indicates less screening-induced error. Increasing $\rho$ imposes more stringent local screening, potentially excluding more active coordinates and increasing the resulting estimation error. Holding the local sample size and screening threshold fixed, aggregation across more machines can mitigate this effect, as reflected by the decreasing term $ \{M(1+\rho)\phi(\rho)\}^{-1}$ in the error bound.
\end{remark}

\begin{remark}
 When $\tilde{s}\asymp s$, balancing the two terms in $\delta_n$  
gives the optimal bandwidth order
$h\asymp n^{-3/10}s^{3/5}(\log p)^{11/10}$,
which yields $\delta_n^2 \asymp \{(\log p)/N\} \cdot \{n^{-1}M^5 s^{12}(\log p)^{17} \}^{1/5}$. Consequently, $\delta_n^2=o\{(\log p)/n\}$ whenever
$s^{12}(\log p)^{17}=o(n)$. For any fixed small $\rho\geq 0$, local screening retains a large proportion of the covariates, as all covariates are retained when $\rho=0$ and the number of selected covariates is nonincreasing in $\rho$ (see Proposition \ref{prop.screening.size} below). In this case, under the conditions of Theorem \ref{thm.convergence}, our estimator achieves the convergence rate $O_{\rm p}\{\sqrt{s(\log p)/N}\}$ provided additionally that the number of machines is not too large, namely, $M^5s^{12}(\log p)^{17}\lesssim n$. This matches the standard full-sample convergence rate for high-dimensional sparse estimation and is comparable, up to logarithmic factors, to those established by \cite{tan2022communication}, \cite{wang2024distributed}, and \cite{yang2026}.  However, achieving this rate with a small screening threshold offers limited communication savings, as a substantial proportion of the covariates must be transmitted.
For larger screening thresholds, local screening becomes more stringent, reducing the communication cost per machine. Although the full-sample convergence rate may no longer be attainable, Theorem {\rm \ref{thm.convergence}} shows that, with sufficiently many machines, information aggregation can still yield a convergence rate faster than the local rate $O_{\rm p}\{\sqrt{s(\log p)/n}\}$.
\end{remark}

Proposition \ref{prop.screening.size} characterizes the per-machine communication costs of the proposed method in the screening and estimation stages.

\begin{proposition}\label{prop.screening.size}
    Let the assumptions of Proposition {\rm \ref{prop.Bahadur}}  hold, and suppose that the thresholds $(\rho,v)$
satisfy the conditions specified in
Theorem {\rm \ref{thm.convergence}}. If $\delta_n\sqrt{n \log p}= O(1)$, we have
    \begin{align*}
        &~~~~~~~~~~~~~~~~ \max_{m\in[M]}E\{ |\cS^{(m)}| \}\lesssim  s + (p-s)\bar\Phi
( \rho)\,,
\\
 & E( |\hat\cS|) \lesssim  s+(p-s) \min \big [ M \bar\Phi
({\rho})
,  \exp\{ -  C Mn \delta_n^2 \phi(\rho) (1+ \rho)\}   \big ] \,.
    \end{align*} 
\end{proposition}

\begin{remark}
Proposition \ref{prop.screening.size} provides bounds on
the expected screening sizes at both the local and global
stages. 
The local screening size is nonincreasing in $\rho$, and its expectation is $O(s)$ whenever  $\rho\geq \sqrt{2\log \{(p-s)/s \}}$.
For the global screening size, the exponential term in the bound captures the benefit of aggregating signed votes. Holding the remaining quantities fixed, this term decreases exponentially with $M$. Alternatively, sufficiently stringent local screening (with a larger $\rho$) controls the global screening size through the term $M\bar\Phi(\rho)$, even when relatively few machines are available.
Accordingly, the expected global screening size is
$O(s)$ if either $M\lesssim s/\{(p-s)\bar\Phi(\rho)\}$ or $ M\gtrsim \log\{(p-s)/s\}/\{n\delta_n^2
(1+\rho)\phi(\rho)\}$.
\end{remark}

\section{ Numerical studies}
\label{sec.simulation}

We evaluate the finite-sample performance of the proposed method through numerical experiments. For each local machine $m\in[M]$, we independently generate $n$ observations $\{(y_i^{(m)},\bx_i^{(m)})\}_{i\in[n]}$ according to the linear model $y_i^{(m)} =  \bx_i^{(m),\T} \bbeta^* +\varepsilon_i^{(m)}$, where $\bx_i^{(m),\T}=\{1,\bz_i^{(m),\T}\}$. Here,  $\bz_i^{(m)} \overset{\mathrm{i.i.d.}}{\sim}N(\mathbf{0},\bSigma)$ with $\bSigma=(0.5^{|j-k|})_{j,k\in[p-1]}$. For quantile level $\tau\in(0,1)$, we generate the errors as
$\varepsilon_i^{(m)} =\eta_i^{(m)} - F_{\eta}^{-1}(\tau)$,
where $F_{\eta}^{-1}(\tau)$ denotes the $\tau$-quantile of the distribution of $\eta_i^{(m)}$. The variables $\eta_i^{(m)}$ are independent of the covariates. This construction ensures that $\pr\{\varepsilon_i^{(m)} \leq 0\}=\tau$. We consider two error distributions: $\eta_i^{(m)}\overset{\mathrm{i.i.d.}}{\sim}{N}(0,1)$ and $\eta_i^{(m)}\overset{\mathrm{i.i.d.}}{\sim}t_{2.25}$. For sparsity level $s\geq 2$, the true coefficient vector $\bbeta^* =(1,\beta_2^*,\ldots,\beta_{p}^*)^{\T}$ is specified as $\beta_j^*=a\{1-(j-2)/(2s-2)\}$ for {$j=2,\ldots,s+1$}, and $\beta_j^*=0$ for $j>s+1$, where $a>0$ controls the overall signal strength. 
Throughout this section, we set the quantile level at $\tau=0.7$ and use the Gaussian smoothing kernel $K(u)=(2\pi)^{-1/2}e^{-u^2/2}$ in \eqref{l1.est}. Observations are independent within and across machines.

We measure estimation accuracy by the relative $\ell_2$ estimation error
$\|\hat{\bbeta}^{(r)}-\bbeta^*\|_2/\|\bbeta^*\|_2$,
which is summarized using means or boxplots over 500 Monte Carlo replications.
Here, $\hat{\bbeta}^{(r)}$ denotes the estimate under consideration in the $r$th replication.
To assess communication cost and screening performance of proposed method, we also report the average local screening size
$(500M)^{-1}\sum_{r=1}^{500} \sum_{m=1}^M
|\cS^{(m,r)}|$, the average global screening size
$500^{-1}\sum_{r=1}^{500}|\hat{\cS}^{(r)}|$,
and the average true positive rate (TPR)
$500^{-1}\sum_{r=1}^{500}
|\hat{\cS}^{(r)}\cap\cS|/s$,
where $\cS^{(m,r)}$ and $\hat{\cS}^{(r)}$
denote the local candidate set on machine $m$
and the global screening set, respectively,
in the $r$th replication.

\subsection{Selection of tuning parameters}\label{sec.parameter_selection}

We describe the tuning choices for the local statistics $\{(\tilde{\beta}_j^{(m)},\hat\xi_j^{(m)})\}$ in Section \ref{sec.debiase} and the estimation procedure in Section \ref{sec.mainprocedure}. These choices are motivated by our theoretical analysis, with practical adaptations for finite-sample implementation.

\begin{itemize}
    \item {\sl Tuning parameters for constructing local statistics   in Section \ref{sec.debiase}.}  Guided by Proposition~\ref{prop.Bahadur} and adapting the specification of \citet{tan2022high}, we set
$h=\max\{0.05,\\\sqrt{\tau(1-\tau)}(n^{-1}\log p)^{0.3}\}$. The regularization parameters $\lambda=\lambda_m$ in \eqref{l1.est} and $\Delta=\Delta_m$ in \eqref{Ch3dantzig} are selected separately on each machine using local data. We suppress the machine index $m$ where unambiguous. To select $\lambda$, we use simulation-based calibration motivated by \cite{belloni2011l1}. Conditional on the observed covariates, for each simulation draw \(b\in[200]\), we independently generate $U_1^{(b)},\ldots,U_n^{(b)}\overset{\mathrm{i.i.d.}}{\sim}{\rm Unif}(0,1)$ and compute $\lambda^{(b)}
    =  \max_{2\leq j\leq p }
    | {n}^{-1} \sum_{i=1}^{n} {x}_{i,j}^{(m)}
    \{ \tau- \mathbf{1} (U_i^{(b)}\leq\tau ) \}
    |$, where $\bx_i^{(m)}=\{x_{i,1}^{(m)},\ldots,x_{i,p}^{(m)}\}^{\T}$ is the $i$th covariate vector on machine $m$.
We then define $\tilde{\lambda}=\max_{b\in[200]}\lambda^{(b)}$ and construct a grid of 20 equally spaced candidate values over $[0.25\tilde{\lambda},\tilde{\lambda}]$. We select \(\lambda\) by three-fold cross-validation using the quantile check loss. 
For $\Delta$, Proposition~\ref{prop.Bahadur} suggests the theoretical order $\Delta=c_{\Delta}(\log p)^5/(nh^3)$. We select $c_{\Delta}$ by three-fold cross-validation over the grid $10^{-5}\times\{0.5,1,2,5,10,20,50\}$. Specifically, let $\mathcal{U}_1,\mathcal{U}_2,\mathcal{U}_3$ denote the validation folds. For each candidate value of $\Delta$, we compute $$\operatorname{CV}(\Delta)
=
 \frac{1}{n}\sum_{k=1}^{3}\sum_{i\in\mathcal{U}_k}
\rho_{\tau}\{
y_i-\bx_i^{\T}\tilde{\bbeta}_{h}^{(-k)}(\Delta)
\}\,,$$ 
where $\tilde{\bbeta}_{h}^{(-k)}(\Delta)$ is the full debiased estimator, including the intercept, fitted after excluding the $k$th fold. We choose the value of $\Delta$ minimizing $\operatorname{CV}(\Delta)$. We solve the Dantzig programs using Lasso-initialized column generation inspired by \citet{mazumder2019computing}, while preserving the optimization problem and cross-validation criterion. These local tuning procedures require no additional communication.

\item {\sl Selection of the thresholds  in Section 
\ref{sec.mainprocedure}.} We set the local screening threshold $\rho=c_\rho\sqrt{2\log p}$, and vary $c_\rho\in[0,1]$ across experiments. 
Motivated by the false-selection calibration principle of stability
selection \citep{meinshausen2010stability}, we select the voting threshold
$v$ using the empirical signed votes across local machines.
For a given $\rho\geq 0$, we estimate the local selection probability of inactive coordinates by the pooled selection proportion $\hat{\pi}_0
=
 \{M(p-1)\}^{-1}
\sum_{m=1}^{M}\sum_{j=2}^{p}\mathbf{1}\{j\in\mathcal{S}^{(m)}\}$. 
This working estimate relies on most coordinates being inactive under sparsity.
For calibration, we approximate inactive-coordinate votes as independent across machines, with a common selection probability and symmetric signs. Conditional on $\hat{\pi}_0$,
we therefore construct independent reference votes
$D_0^{(1)},\ldots,D_0^{(M)}$ satisfying $\pr \{D_0^{(m)}=1\}=
 {\hat{\pi}_0}/{2}
=
\pr \{D_0^{(m)}=-1\}
$ and $\pr \{D_0^{(m)}=0\}
=
1-\hat{\pi}_0$. 
The distribution of their sum is computed exactly by convolution.
Let $\mathcal{C} \subseteq [5\times 10^{-3}, 10]$ be a prespecified finite grid of positive
multipliers. As suggested by Theorem \ref{thm.convergence}, for each $c\in\mathcal{C}$, define $v(c)
=
\min\{
c  {n^{-1/10}(\log p)^{11/5}}{}\phi(\rho),\,1
\}$. Given a prespecified reference false-selection budget $q>0$, we choose  $v=v(\hat{c})$ with
\begin{align*}
    \hat{c}
=
\min\bigg\{
c\in\mathcal{C}: \pr \bigg\{
\bigg|\frac{1}{M}\sum_{m=1}^{M}D_0^{(m)}\bigg|
\ge v(c)
\bigg\} \le \frac{q}{p-1}
\bigg\} \,.
\end{align*}
The smallest admissible multiplier favors signal retention. If no candidate satisfies the criterion, we use
$v= M^{-1}$ to avoid screening out potentially relevant variables. We use $q=3$ throughout, based on its stable performance across extensive numerical experiments.  Finally, following the rate suggested by Theorem \ref{thm.convergence}
but with a practically chosen multiplicative constant, we set the
hard-thresholding parameter to
$\gamma=0.02\{\sqrt{(\log p)/N}+n^{-3/5}(\log p)^{11/5}\}$
throughout the numerical studies.
\end{itemize}

\subsection{Simulation results}\label{sec.simulationresults}

\begin{figure}[!htbp]
    \centering
    \includegraphics[width= 0.9\textwidth]{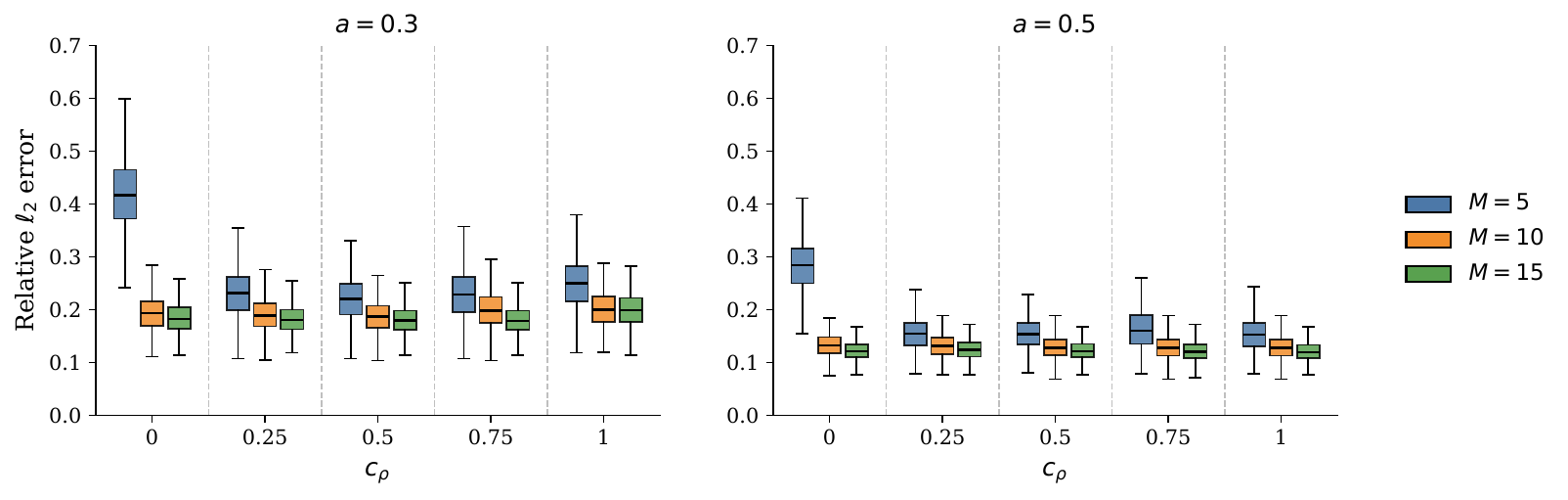}
    \caption{Boxplots of estimation performance over 500 Monte Carlo replications across different signal strengths $a$, numbers of machines $M$, and local screening thresholds $\rho$, with $ (n,p ,s)=(200,500,10)$, $\tau=0.7$, and standard normal errors.  }
    \label{fig:error}
\end{figure}

We first examine how estimation accuracy varies with signal strength $a$, the number of machines $M$,
and the local screening threshold $\rho$.
We fix $(n,p,s)=(200,500,10)$ and consider standard
normal errors.
Fig. \ref{fig:error} shows that when the signal is
relatively weak ($a=0.3$) and the number of machines is small,
the estimation error exhibits a roughly U-shaped pattern
as $\rho$ increases. A small $\rho$ retains many irrelevant
covariates, potentially increasing the estimation error,
whereas an excessively large $\rho$ may exclude relevant
covariates with weak signals. 
Stronger signals or more machines
$M$ reduce this sensitivity to the screening threshold.
Table \ref{tab:screening} reports the corresponding screening results. The local screening size decreases markedly as $\rho$ increases. Except
when $\rho=0$ and $M$ is small, the global screening size remains
close to the true sparsity level $s=10$ even when local screening sets are large. This supports the theoretical finding that signed-vote aggregation or stringent local screening can control the global screening size.
For weak signals and large $\rho$, lower TPRs indicate missed relevant covariates, which may explain the larger estimation errors. Stronger signals improve both TPR and estimation accuracy.
The small global screening sizes and generally high TPRs support the effectiveness of the tuning strategy for $v$.
In particular, $c_\rho$ around $0.75$ keeps both screening sizes small without appreciably reducing estimation accuracy across the settings considered.  Additional results under $t_{2.25}$ errors are presented in Fig.~\ref{fig:error_t} and Table \ref{tab:screening_t} of the supplementary material, which show similar patterns.

\begin{table}[htbp]
\centering
\caption{
Average screening performance of the proposed method over
500 Monte Carlo replications under different combinations
of $(a,M,c_\rho)$, with $(n,p,s)=(200,500,10)$, $\tau=0.7$,
and standard normal errors.
}
\vspace{3pt}
\label{tab:screening}
\small
\setlength{\tabcolsep}{2.5pt}
\renewcommand{\arraystretch}{1 }
\begin{tabular}{cc *{9}{c}}
\toprule
& & \multicolumn{3}{c}{Local screening size}
  & \multicolumn{3}{c}{Global screening size}
  & \multicolumn{3}{c}{TPR} \\
\cmidrule(lr){3-5}
\cmidrule(lr){6-8}
\cmidrule(lr){9-11}
$c_\rho$ & $a$
& $M=5$ & $M=10$ & $M=15$
& $M=5$ & $M=10$ & $M=15$
& $M=5$ & $M=10$ & $M=15$ \\
\midrule

{$0$}
& $0.3$ & 499.00 & 499.00 & 499.00
        & 499.00 & 11.10 & 10.00
        & 1.000 & 0.992 & 0.990 \\
& $0.5$ & 499.00 & 499.00 & 499.00
        & 499.00 & 11.27 & 10.14
        & 1.000 & 1.000 & 0.999 \\
\addlinespace[2pt]

{$0.25$}
& $0.3$ & 194.00 & 194.03 & 193.96
        & 9.84 & 10.50 & 10.67
        & 0.954 & 1.000 & 1.000 \\
& $0.5$ & 193.53 & 193.37 & 193.42
        & 10.25 & 10.55 & 10.92
        & 0.995 & 1.000 & 1.000 \\
\addlinespace[2pt]

{$0.50$}
& $0.3$ & 46.18 & 46.10 & 46.11
        & 10.35 & 9.99 & 10.21
        & 0.981 & 0.996 & 1.000 \\
& $0.5$ & 46.61 & 46.65 & 46.62
        & 10.53 & 10.03 & 10.27
        & 1.000 & 1.000 & 1.000 \\
\addlinespace[2pt]

{$0.75$}
& $0.3$ & 14.49 & 14.55 & 14.49
        & 9.81 & 9.63 & 10.11
        & 0.945 & 0.963 & 0.998 \\
& $0.5$ & 17.43 & 17.38 & 17.27
        & 10.07 & 10.00 & 10.22
        & 0.983 & 1.000 & 1.000 \\
\addlinespace[2pt]

{$1.00$}
& $0.3$ & 8.34 & 8.35 & 8.34
        & 8.89 & 9.66 & 9.46
        & 0.885 & 0.961 & 0.946 \\
& $0.5$ & 11.61 & 11.65 & 11.55
        & 9.97 & 9.99 & 10.00
        & 0.990 & 0.999 & 1.000 \\

\bottomrule
\end{tabular}
\end{table}

\begin{figure}[!htbp]
    \centering
    \includegraphics[width= 0.9\textwidth]{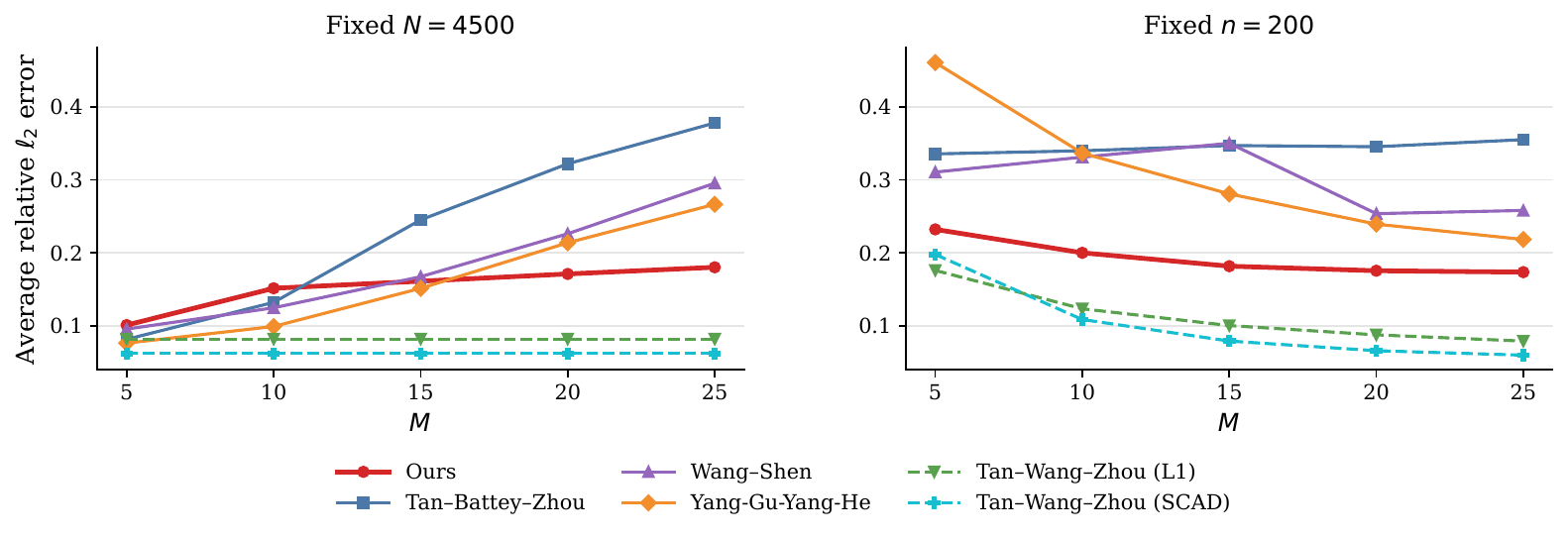}
    \caption{Average relative $\ell_2$ estimation errors over 500 Monte Carlo
replications for the proposed method with $c_\rho=0.75$ and competing centralized
and distributed high-dimensional QR methods
across different numbers of machines $M$. The left and right
panels correspond to fixed total sample size $N=4500$ and
fixed local sample size $n=200$, respectively, with
$(p,s)=(500,10)$, $\tau=0.7$, $a=0.3$, and standard normal errors. The solid lines represent the proposed method  and three
competing distributed methods: Tan-Battey-Zhou
\citep{tan2022communication}, Wang-Shen
\citep{wang2024distributed}, and Yang-Gu-Yang-He
\citep{yang2026}.
The dashed lines represent the full-sample 
$\ell_1$-penalized and SCAD-penalized estimators
of Tan-Wang-Zhou \citep{tan2022high}. }
    \label{fig:comparison}
\end{figure}

We next compare the proposed method with the full-sample $\ell_1$-penalized and SCAD-penalized estimators of \citet{tan2022high} and the distributed QR methods of \citet{tan2022communication}, \citet{wang2024distributed}, and \citet{yang2026}.\footnote{The full-sample benchmarks use the authors' Python package at \url{https://github.com/WenxinZhou/conquer}. Our implementation of \citet{tan2022communication} is based on code provided by the authors. We implement \citet{wang2024distributed} and \citet{yang2026} in Python with AI assistance, based on their public R and R/Rcpp code at \url{https://github.com/WangCaixing-96/DHSQR} and \url{https://github.com/Yif-99/IMSA-for-distributed-HD-QR}, respectively.
For \citet{wang2024distributed}, initialization follows Algorithm~1 of the paper. For the methods of \citet{tan2022communication} and \citet{wang2024distributed}, we select regularization parameters by three-fold cross-validation using the available training observations, without generating additional validation samples, to facilitate comparison under a common data budget and application to real data. In addition, we estimate $s$ using each method's own initial estimator, counting nonzero coefficients including the intercept for \citet{tan2022communication} and nonzero slope coefficients for \citet{wang2024distributed}. This use of initial support sizes follows the approach of \citet{yang2026}. Further implementation details are provided in Section~\ref{sec.addsimulation} of the supplementary material. } We set $c_\rho=0.75$, $(p,s)=(500,10)$, and $a=0.3$,
and consider $M\in\{5,10,15,20,25\}$ under two scenarios:
fixed total sample size $N=4500$ and fixed local sample size $n=200$. 
Fig. \ref{fig:comparison} shows that when the total sample
size $N=4500$ is fixed, all distributed methods achieve
estimation errors close to those of the full-sample
benchmarks for small $M$. As $M$ increases and local samples shrink, estimation errors increase. Our method deteriorates more gradually and achieves the lowest error among the distributed methods when local samples are small. 
With fixed local sample size $n=200$, our method improves as $M$ increases and performs more stably than the competing distributed methods, indicating effective aggregation across machines. The limited gains of \citet{tan2022communication} and \citet{wang2024distributed} may partly reflect their sensitivity to initialization, as the relatively small local sample size may yield insufficiently accurate initial estimators.
Together with Table \ref{tab:screening}, these results show that the proposed method achieves competitive estimation accuracy relative to the competing distributed methods while substantially reducing per-machine communication costs.  Results under $t_{2.25}$ errors show a similar pattern. See Fig. \ref{fig:comparison_t} of the supplementary material for details.

\section{ Real data analysis}
\label{sec.RD}

We evaluate the proposed method using the `NIR Shootout 2002' dataset,\footnote{\url{https://www.eigenvector.com/data/tablets/index.html}} which contains near-infrared (NIR) spectra and active ingredient assay values for pharmaceutical tablets. NIR spectroscopy enables rapid, nondestructive assessment of active ingredient content in intact tablets.  We study conditional quantile prediction of assay values from spectral measurements. Of the two instruments used to collect the spectra, we use only Instrument 1  and combine its original calibration, validation, and test subsets. The resulting dataset comprises $N=655$ tablets, each with 650 spectral measurements from 600 to 1898 \texttt{nm} in 2 \texttt{nm} increments, giving \(p=651\) including the intercept. The assay value is the response, and the spectral measurements are the predictors. In each replication, we randomly allocate 80\% of the observations to training and 20\% to testing, using the same split for all methods. Predictors are standardized using their training means and standard deviations, and the response is divided by its training standard deviation. These transformations are also applied to the test data.  We repeat this procedure 500 times.  
Tuning parameters are selected as described in Section~\ref{sec.parameter_selection}.

\begin{figure}[!htbp]
    \centering
    \includegraphics[width= 0.9\textwidth]{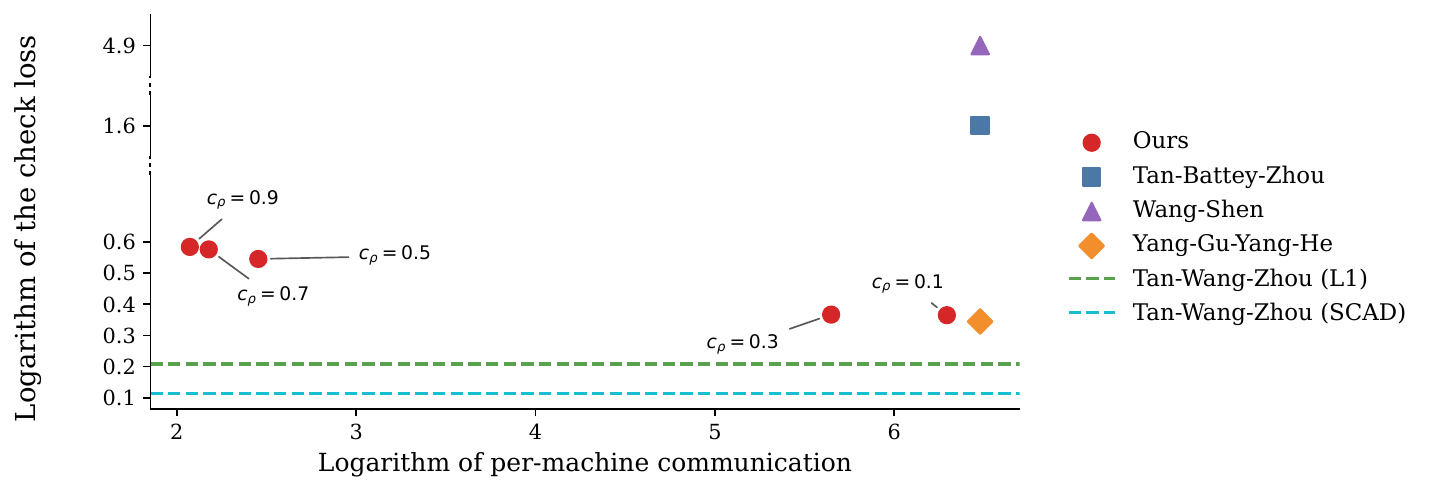}
    \caption{Comparison of logarithmic check loss and per-machine communication cost across different methods on the `NIR Shootout 2002' dataset over 500 replications, with $\tau=0.7$ and $M=3$. For the proposed method, the communication proxy is the replication average of one plus the maximum of the mean local screening size across machines and the global screening size. For the distributed competitors, it is the full covariate dimension, including the intercept. The markers represent the proposed method  and
three competing distributed methods:
Tan-Battey-Zhou \citep{tan2022communication},
Wang-Shen \citep{wang2024distributed}, and
Yang-Gu-Yang-He \citep{yang2026}.
The dashed horizontal lines represent the full-sample
$\ell_1$-penalized and SCAD-penalized estimators of
Tan-Wang-Zhou \citep{tan2022high}. }
    \label{fig:RD_loss}
\end{figure}

 We focus on $\tau=0.7$, consistent with the simulation studies, to examine the trade-off between predictive accuracy and communication costs. To emulate a distributed setting,  
we randomly partition the training observations across $M=3$ machines, with approximately $n=175$ observations per machine. We use the same competitors as in the simulations: two full-sample benchmarks from \cite{tan2022high} and three distributed high-dimensional QR methods \citep{tan2022communication,wang2024distributed,yang2026}. All methods are fitted on the training data, and their predictive performance is evaluated using the out-of-sample check loss on the held-out test data.
Fig. \ref{fig:RD_loss} shows the trade-off between predictive accuracy and a proxy for per-machine communication cost. For the proposed method, the communication proxy is the replication average of one plus the maximum of the mean local screening size across machines and the global screening size. For the distributed competitors, it is the full covariate dimension, including the intercept. At small local screening thresholds,  our method incurs relatively high communication costs, with prediction loss comparable to that of \cite{yang2026} but higher than those of the full-sample benchmarks. Increasing the threshold substantially reduces communication cost but raises prediction loss, which remains well below those of \cite{tan2022communication} and \cite{wang2024distributed}.
Table \ref{tab:selection_size} compares our screening sizes with the competitors' final model sizes, defined as the numbers of nonzero slope coefficients.  Although the competing distributed methods yield smaller final models, this sparsity does not reduce their communication costs during estimation, as they transmit full-dimensional summaries.

\begin{table}[t]
\centering
\caption{Average screening sizes of the proposed method and final model sizes of competing methods on the `NIR Shootout 2002' dataset over 500 replications.}
\label{tab:selection_size}
\small
\setlength{\tabcolsep}{5pt}
\begin{tabular}{@{}ccc@{\hspace{12pt}}lc@{}}
\toprule
\multicolumn{3}{c}{\textit{Proposed method}}
& \multicolumn{2}{c}{\textit{Competing methods}} \\
\cmidrule(lr){1-3}\cmidrule(l){4-5}
$c_\rho$ & Local screening size & Global screening size
& Method & Final size \\
\midrule
0.1 & 496.66 (10.74) & 539.81 (14.64)
& Tan--Wang--Zhou (L1) & 17.52 (5.47) \\
0.3 & 222.05 (22.97) & 278.90 (47.31)
& Tan--Wang--Zhou (SCAD) & 12.88 (1.79) \\
0.5 & 10.64 (3.02) & 8.13 (1.12)
& Tan--Battey--Zhou & 3.63 (2.54) \\
0.7 & 7.78 (0.74) & 6.92 (0.91)
& Wang--Shen & 2.98 (4.97) \\
0.9 & 6.86 (0.64) & 6.33 (0.81)
& Yang--Gu--Yang--He & 3.31 (0.99) \\
\bottomrule
\end{tabular}

\vspace{3pt}
\parbox{0.95\linewidth}{\footnotesize
~~~~~~~~~~~~~~  Values are averages, with standard
deviations in parentheses.}
\end{table}

\begin{figure}[!htbp]
    \centering
    \includegraphics[width= 0.9\textwidth]{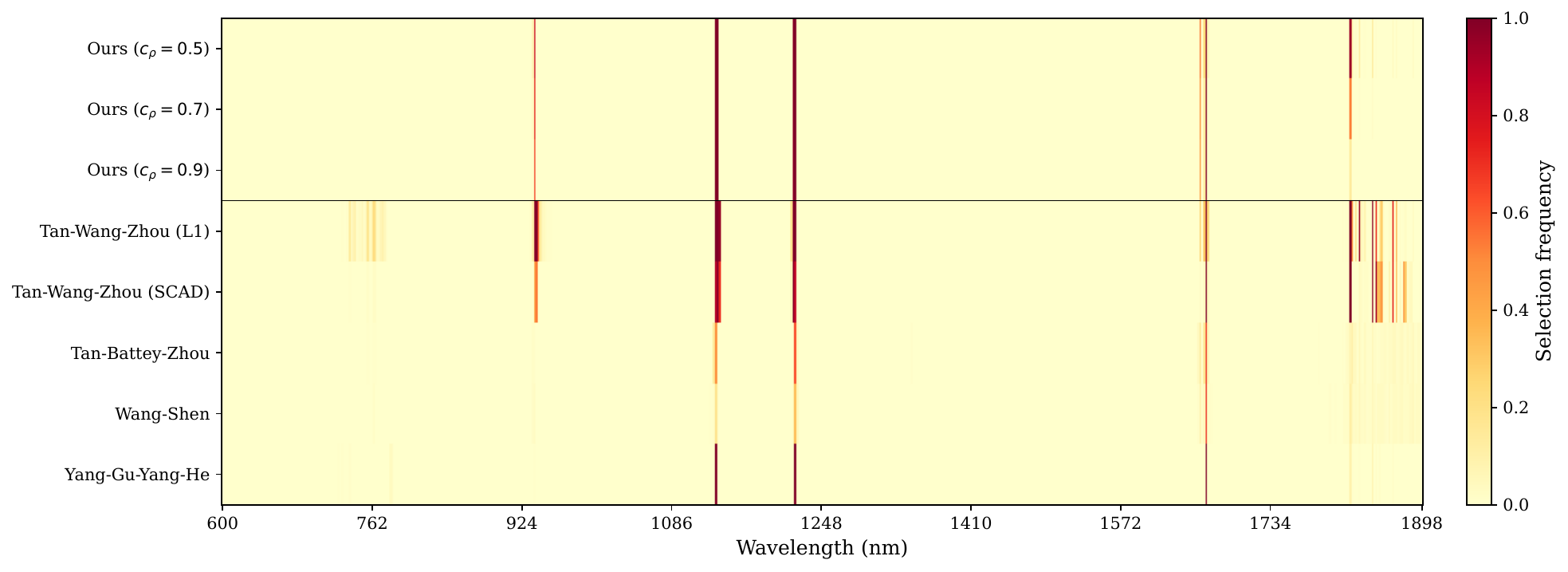}
    \caption{Selection frequencies of the 650 wavelength variables on the `NIR Shootout 2002' dataset over 500 replications for each method, with $\tau=0.7$ and $M=3$. }
    \label{fig:RD_selection}
\end{figure}

Fig. \ref{fig:RD_selection} shows selection frequencies across wavelengths. For our method, selection refers to inclusion in the global screening set under $c_\rho\in\{0.5,0.7,0.9\}$; for the competitors, it denotes a nonzero slope coefficient in the final estimator. Our method frequently selects wavelengths near 1134 and 1220 \texttt{nm} across values of $c_\rho$, as well as wavelengths near 1664 \texttt{nm}. Competing methods identify similar regions, suggesting agreement on potentially important predictors.  The clustered selection patterns are consistent with the broad, overlapping absorption bands of NIR spectra \citep{reich2005near}. In a separate pharmaceutical study, \cite{vanarase2010realtime} reported spectral responses near 1130, 1220, 1490, and 1650 \texttt{nm} associated with composition changes during NIR analysis of active-ingredient concentration. The regions near 1130, 1220, and 1650 \texttt{nm} are close to frequently selected wavelengths in our analysis. This correspondence suggests practical relevance to pharmaceutical NIR analysis but does not establish the selected predictors' chemical identity.

\section{Discussion} 

This paper develops a screening-then-estimation procedure for high-dimensional distributed QR that combines local debiased screening with support aggregation. Although the local screening procedure is tailored to QR, the aggregation theory relies on probabilistic guarantees for local screening rather than the particular form of the screening statistics. The framework may therefore extend to other high-dimensional distributed models, including robust and logistic regression, provided that suitable local screening procedures satisfy these guarantees. Relative to logistic regression and robust regression with differentiable losses, high-dimensional QR presents additional analytical challenges: the check loss is nonsmooth, and the population curvature depends on the unknown conditional error density at the target quantile. In our smoothed QR construction, the bandwidth must balance smoothing bias against the stochastic error in curvature estimation. These errors must then be controlled in the uniform Bahadur representation and coordinatewise distributional approximations used to analyze signed voting.

Our procedure aggregates both selection indicators and sign information. A natural extension is to incorporate quantitative information, such as local test statistics, debiased estimates, or screening scores, to capture the strength of evidence for individual coordinates. This information may help distinguish consistently relevant variables from those supported by weaker or less consistent evidence across machines. Another direction is to accommodate heterogeneity across machines through machine-specific aggregation weights \citep{gu2023distributed}. Such weights could account for differences in the reliability of local screening results arising from unequal sample sizes, covariate distributions, or noise levels. These extensions would allow the aggregation rule to reflect the type and quality of local information available to the master machine.

\section*{Supplementary material}

The Supplementary Material contains a review of convolution smoothing,
additional numerical results, and proofs of all theoretical results.

	{\small{
			\setlength{\bibsep}{3pt}
			\bibliographystyle{apalike}
			\bibliography{refs}

\begin{thebibliography}{}

\bibitem[Chen et~al., 2013]{chen2013steinS}
Chen, L. H.~Y., Fang, X., and Shao, Q.-M. (2013).
\newblock From {Stein} identities to moderate deviations.
\newblock {\em The Annals of Probability}, 41(1):262--293.

\bibitem[Fernandes et~al., 2021]{fernandes2021smoothingS}
Fernandes, M., Guerre, E., and Horta, E. (2021).
\newblock Smoothing quantile regressions.
\newblock {\em Journal of Business \& Economic Statistics}, 39(1):338--357.

\bibitem[Kuchibhotla and Chakrabortty, 2022]{Kuchibhotla2022S}
Kuchibhotla, A.~K. and Chakrabortty, A. (2022).
\newblock Moving beyond sub-gaussianity in high-dimensional statistics:
  applications in covariance estimation and linear regression.
\newblock {\em Information and Inference: A Journal of the IMA},
  11(4):1389--1456.

\bibitem[Shevtsova, 2014]{Shevtsova2014S}
Shevtsova, I. (2014).
\newblock On the absolute constants in the {Berry-Esseen-type} inequalities.
\newblock {\em Doklady Mathematics}, 89(3):378--381.

\bibitem[Tan et~al., 2022]{tan2022highS}
Tan, K.~M., Wang, L., and Zhou, W.-X. (2022).
\newblock High-dimensional quantile regression: Convolution smoothing and
  concave regularization.
\newblock {\em Journal of the Royal Statistical Society, Series B},
  84(1):205--233.

\bibitem[Vershynin, 2018]{vershynin2018highS}
Vershynin, R. (2018).
\newblock {\em High-dimensional probability: An introduction with applications
  in data science}, volume~47.
\newblock {Cambridge University Press}.

\end{thebibliography}


\begin{thebibliography}{}

\bibitem[Adrian and Brunnermeier, 2016]{adrian2016covar}
Adrian, T. and Brunnermeier, M.~K. (2016).
\newblock {CoVaR}.
\newblock {\em American Economic Review}, 106(7):1705--1741.

\bibitem[Battey et~al., 2018]{battey2018distributed}
Battey, H., Fan, J., Liu, H., Lu, J., and Zhu, Z. (2018).
\newblock Distributed testing and estimation under sparse high dimensional
  models.
\newblock {\em The Annals of Statistics}, 46(3):1352--1382.

\bibitem[Belloni and Chernozhukov, 2011]{belloni2011l1}
Belloni, A. and Chernozhukov, V. (2011).
\newblock {$\ell_1$}-penalized quantile regression in high-dimensional sparse
  models.
\newblock {\em The Annals of Statistics}, 39(1):82--130.

\bibitem[Cand{\`e}s and Tao, 2007]{candes2007dantzig}
Cand{\`e}s, E.~J. and Tao, T. (2007).
\newblock The {Dantzig} selector: Statistical estimation when $p$ is much
  larger than $n$.
\newblock {\em The Annals of Statistics}, 35(6):2313--2351.

\bibitem[Chang et~al., 2021]{chang2021high}
Chang, J., Chen, S.~X., Tang, C.~Y., and Wu, T.~T. (2021).
\newblock High-dimensional empirical likelihood inference.
\newblock {\em Biometrika}, 108(1):127--147.

\bibitem[Chen et~al., 2020]{chen2020}
Chen, X., Liu, W., Mao, X., and Yang, Z. (2020).
\newblock Distributed high-dimensional regression under a quantile loss
  function.
\newblock {\em Journal of Machine Learning Research}, 21(182):1--43.

\bibitem[Chen et~al., 2019]{chen2019quantile}
Chen, X., Liu, W., and Zhang, Y. (2019).
\newblock Quantile regression under memory constraint.
\newblock {\em The Annals of Statistics}, 47(6):3244--3273.

\bibitem[Engle and Manganelli, 2004]{engle2004caviar}
Engle, R.~F. and Manganelli, S. (2004).
\newblock {CAViaR}: Conditional autoregressive value at risk by regression
  quantiles.
\newblock {\em Journal of Business \& Economic Statistics}, 22(4):367--381.

\bibitem[Fan et~al., 2014]{fan2014adaptive}
Fan, J., Fan, Y., and Barut, E. (2014).
\newblock Adaptive robust variable selection.
\newblock {\em The Annals of Statistics}, 42(1):324--351.

\bibitem[Fan et~al., 2023]{fan2023communication}
Fan, J., Guo, Y., and Wang, K. (2023).
\newblock Communication-efficient accurate statistical estimation.
\newblock {\em Journal of the American Statistical Association},
  118(542):1000--1010.

\bibitem[Fan et~al., 2020]{fan2020statistical}
Fan, J., Li, R., Zhang, C.-H., and Zou, H. (2020).
\newblock {\em Statistical Foundations of Data Science}.
\newblock Chapman and Hall/CRC.

\bibitem[Fernandes et~al., 2021]{fernandes2021smoothing}
Fernandes, M., Guerre, E., and Horta, E. (2021).
\newblock Smoothing quantile regressions.
\newblock {\em Journal of Business \& Economic Statistics}, 39(1):338--357.

\bibitem[Fonseca and Nadler, 2023]{fonseca2023distributed}
Fonseca, R. and Nadler, B. (2023).
\newblock Distributed sparse linear regression under communication constraints.
\newblock {\em arXiv preprint arXiv:2301.04022}.

\bibitem[Gao et~al., 2022]{gao2022review}
Gao, Y., Liu, W., Wang, H., Wang, X., Yan, Y., and Zhang, R. (2022).
\newblock A review of distributed statistical inference.
\newblock {\em Statistical Theory and Related Fields}, 6(2):89--99.

\bibitem[Gu and Chen, 2023]{gu2023distributed}
Gu, J. and Chen, S.~X. (2023).
\newblock Distributed statistical inference under heterogeneity.
\newblock {\em Journal of Machine Learning Research}, 24(387):1--57.

\bibitem[Jordan et~al., 2019]{jordan2019communication}
Jordan, M.~I., Lee, J.~D., and Yang, Y. (2019).
\newblock Communication-efficient distributed statistical inference.
\newblock {\em Journal of the American Statistical Association},
  114(526):668--681.

\bibitem[Koenker and Bassett, 1978]{koenker1978}
Koenker, R. and Bassett, Jr., G. (1978).
\newblock Regression quantiles.
\newblock {\em Econometrica}, 46(1):33--50.

\bibitem[Koenker and Hallock, 2001]{koenker2001quantile}
Koenker, R. and Hallock, K.~F. (2001).
\newblock Quantile regression.
\newblock {\em Journal of Economic Perspectives}, 15(4):143--156.

\bibitem[Lee et~al., 2017]{lee2017communication}
Lee, J.~D., Liu, Q., Sun, Y., and Taylor, J.~E. (2017).
\newblock Communication-efficient sparse regression.
\newblock {\em Journal of Machine Learning Research}, 18(5):1--30.

\bibitem[Li et~al., 2020]{li2020distributed}
Li, X., Li, R., Xia, Z., and Xu, C. (2020).
\newblock Distributed feature screening via componentwise debiasing.
\newblock {\em Journal of Machine Learning Research}, 21(24):1--32.

\bibitem[Liu et~al., 2025]{liu2025communication}
Liu, W., Mao, X., and Tu, J. (2025).
\newblock Communication-efficient distributed sparse learning with oracle
  property and geometric convergence.
\newblock {\em Journal of the American Statistical Association},
  120(552):2606--2618.

\bibitem[Liu et~al., 2024]{liu2024majority}
Liu, W., Tu, J., Mao, X., and Chen, X. (2024).
\newblock Majority vote for distributed differentially private sign selection.
\newblock {\em The Annals of Statistics}, 52(4):1671--1690.

\bibitem[Mazumder et~al., 2019]{mazumder2019computing}
Mazumder, R., Wright, S., and Zheng, A. (2019).
\newblock Computing estimators of {Dantzig} selector type via column and
  constraint generation.
\newblock {\em arXiv preprint arXiv:1908.06515}.

\bibitem[Meinshausen and B{\"u}hlmann, 2010]{meinshausen2010stability}
Meinshausen, N. and B{\"u}hlmann, P. (2010).
\newblock Stability selection.
\newblock {\em Journal of the Royal Statistical Society, Series B},
  72(4):417--473.

\bibitem[Ning and Liu, 2017]{ning2017general}
Ning, Y. and Liu, H. (2017).
\newblock A general theory of hypothesis tests and confidence regions for
  sparse high dimensional models.
\newblock {\em The Annals of Statistics}, 45(1):158--195.

\bibitem[Pan et~al., 2022]{pan2022distributed}
Pan, R., Ren, T., Guo, B., Li, F., Li, G., and Wang, H. (2022).
\newblock A note on distributed quantile regression by pilot sampling and
  one-step updating.
\newblock {\em Journal of Business \& Economic Statistics}, 40(4):1691--1700.

\bibitem[Reich, 2005]{reich2005near}
Reich, G. (2005).
\newblock Near-infrared spectroscopy and imaging: Basic principles and
  pharmaceutical applications.
\newblock {\em Advanced Drug Delivery Reviews}, 57(8):1109--1143.

\bibitem[Tan et~al., 2022a]{tan2022communication}
Tan, K.~M., Battey, H., and Zhou, W.-X. (2022a).
\newblock Communication-constrained distributed quantile regression with
  optimal statistical guarantees.
\newblock {\em Journal of Machine Learning Research}, 23(272):1--61.

\bibitem[Tan et~al., 2022b]{tan2022high}
Tan, K.~M., Wang, L., and Zhou, W.-X. (2022b).
\newblock High-dimensional quantile regression: Convolution smoothing and
  concave regularization.
\newblock {\em Journal of the Royal Statistical Society, Series B},
  84(1):205--233.

\bibitem[van~de Geer et~al., 2014]{vandegeer2014asymptotically}
van~de Geer, S., B{\"u}hlmann, P., Ritov, Y., and Dezeure, R. (2014).
\newblock On asymptotically optimal confidence regions and tests for
  high-dimensional models.
\newblock {\em The Annals of Statistics}, 42(3):1166--1202.

\bibitem[Vanarase et~al., 2010]{vanarase2010realtime}
Vanarase, A.~U., Alcal{\`a}, M., Jerez~Rozo, J.~I., Muzzio, F.~J., and
  Roma{\~n}ach, R.~J. (2010).
\newblock Real-time monitoring of drug concentration in a continuous powder
  mixing process using {NIR} spectroscopy.
\newblock {\em Chemical Engineering Science}, 65(21):5728--5733.

\bibitem[Volgushev et~al., 2019]{volgushev2019distributed}
Volgushev, S., Chao, S.-K., and Cheng, G. (2019).
\newblock Distributed inference for quantile regression processes.
\newblock {\em The Annals of Statistics}, 47(3):1634--1662.

\bibitem[Wang and Shen, 2024]{wang2024distributed}
Wang, C. and Shen, Z. (2024).
\newblock Distributed high-dimensional quantile regression: Estimation
  efficiency and support recovery.
\newblock In {\em Proceedings of the 41st International Conference on Machine
  Learning (ICML)}.

\bibitem[Wang et~al., 2012]{wang2012quantile}
Wang, L., Wu, Y., and Li, R. (2012).
\newblock Quantile regression for analyzing heterogeneity in ultra-high
  dimension.
\newblock {\em Journal of the American Statistical Association},
  107(497):214--222.

\bibitem[Yan et~al., 2023]{yan2023confidence}
Yan, Y., Wang, X., and Zhang, R. (2023).
\newblock Confidence intervals and hypothesis testing for high-dimensional
  quantile regression: Convolution smoothing and debiasing.
\newblock {\em Journal of Machine Learning Research}, 24(245):1--49.

\bibitem[Yang et~al., 2026]{yang2026}
Yang, S., Gu, Y., Yang, H., and He, X. (2026).
\newblock Communication-efficient and distributed-oracle estimation for
  high-dimensional quantile regression.
\newblock {\em The Annals of Statistics}, 54(1):442--465.

\end{thebibliography}
		}
	}
	
	\newpage
	\appendix

\bigskip
\baselineskip 24pt

\setlength{\abovedisplayskip}{0.2\baselineskip}
\setlength{\belowdisplayskip}{0.2\baselineskip}
\setlength{\abovedisplayshortskip}{0.2\baselineskip}
\setlength{\belowdisplayshortskip}{0.2\baselineskip}
{\begin{center}
		{\noindent \bf \Large Supplementary Material for ``Communication-efficient distributed quantile regression via signed voting''}\\
\end{center}}

\renewcommand{\thepage}{S\arabic{page}}  
\setcounter{page}{1}

\setcounter{section}{0}
\renewcommand\thesection{\Alph{section}}  

\setcounter{lemma}{0}
\renewcommand{\thelemma}{\thesection.\arabic{lemma}} 

\setcounter{equation}{0}  
\renewcommand{\theequation}{S.\arabic{equation}}

\setcounter{definition}{0}
\renewcommand{\thedefinition}{\thesection.\arabic{definition}}

\setcounter{proposition}{0}
\renewcommand{\theproposition}{\thesection.\arabic{proposition}}

\setcounter{theorem}{0}
\renewcommand{\thetheorem}{\thesection.\arabic{theorem}} 

\setcounter{corollary}{0}
\renewcommand{\thecorollary}{\thesection.\arabic{corollary}}  

\setcounter{table}{0} 
\setcounter{figure}{0}  
\renewcommand{\thefigure}{S\arabic{figure}}
\renewcommand{\thetable}{S\arabic{table}}


\renewcommand{\theHequation}{supp.equation.\arabic{equation}}

\renewcommand{\theHlemma}{supp.lemma.\theHsection.\arabic{lemma}}
\renewcommand{\theHdefinition}{supp.definition.\theHsection.\arabic{definition}}
\renewcommand{\theHproposition}{supp.proposition.\theHsection.\arabic{proposition}}
\renewcommand{\theHtheorem}{supp.theorem.\theHsection.\arabic{theorem}}
\renewcommand{\theHcorollary}{supp.corollary.\theHsection.\arabic{corollary}}

\renewcommand{\theHfigure}{supp.figure.\arabic{figure}}
\renewcommand{\theHtable}{supp.table.\arabic{table}}

\counterwithin{lemma}{section}

 Throughout this supplement, $C \in (0,\infty)$ denotes a generic universal constant, independent of $(n,p,M)$, whose value may vary from line to line. For an integer $q\geq2$, we write $[q]=\{1,\ldots,q\}$. Let $\mathbf{1}(\cdot)$ denote the indicator function, and write $(1-E)(X)=X-E(X)$ for any random variable $X\in\RR$. For $\iota > 0$, define $\psi_{\iota}(x) := \exp(x^{\iota}) - 1$ for $x > 0$. For a real-valued random variable $\xi$, define
$\|\xi\|_{\psi_{\iota}} := \inf[\lambda > 0 : E\{\psi_{\iota}(|\xi|/\lambda)\} \leq 1]$ and  $\|\xi\|_{L_q}=\{ E(|\xi|^q)\}^{1/q}$ for $\iota,q>0$.  For a $p\times p$ symmetric matrix $\bA=(A_{j,k})_{p\times p}$ and a $p$-dimensional vector $\bv=(v_1,\ldots,v_p)^{\T}$, let $\|\bA\|_{2} = \max\{|\lambda_{\min}(\bA)|, |\lambda_{\max}(\bA)|\}$, $\|\bA\|_\infty = \max_{j,k\in[p]}|A_{j,k}| $,  $\|\bv\|_1 =  \sum_{j=1}^p |v_j|$,  $\|\bv\|_2 = \sqrt{\sum_{j=1}^p v_j^2}$, and  ${\rm supp}(\bv)=\{j\in[p]:v_j \neq 0\}$. For $j\in[p]$, write $\bv_{-j}=(v_k)_{k\in[p]\setminus\{j\}}\in\RR^{p-1}$ for a vector $\bv=(v_1,\ldots,v_p)^{\T}\in\RR^p$, and define $\bA_{j,j}=A_{j,j}$, $\bA_{j,-j}=(A_{j,k})_{k\in[p]\setminus\{j\}}\in\RR^{1\times(p-1)}$, $\bA_{-j,j}=(A_{k,j})_{k\in[p]\setminus\{j\}}\in\RR^{p-1}$ and $\bA_{-j,-j}=(A_{k,l})_{k,l\in[p]\setminus\{j\}}\in\RR^{(p-1)\times(p-1)}$ for a matrix $\bA=(A_{k,l})_{1\leq k,l\leq p}$.  For a $p\times p$ positive semidefinite matrix $\bSigma$, let
 $\|\cdot\|_{\bSigma}$ denote the seminorm induced by $\bSigma$, defined by $\|{\bu}\|_{\bSigma}= \|\bSigma^{1/2}\bu\|_2,\bu \in \RR^p$. For positive sequences $\{a_n\}$ and $\{b_n\}$, we write $a_n\lesssim b_n$ or $b_n\gtrsim a_n$ if $\limsup_{n\rightarrow\infty}a_n/b_n\leqslant c $ for some positive constant $c $. Write $a_n\asymp b_n$ if $a_n\lesssim b_n$ and $b_n\lesssim a_n$ both hold.
We write $a_n\ll b_n$ or $b_n\gg a_n $ if $\limsup_{n\rightarrow\infty}a_n/b_n=0$. Finally, $|\cI|$ denotes the cardinality of a set $\cI$, and $x_+=\max\{x,0\}$ denotes the positive part of $x\in\mathbb{R}$.

\section{Review of convolution smoothing}\label{sec.smoothing}

For residuals $\{r_i(\bbeta)= y_i-\bx_i^{\T}\bbeta\}_{i\in[n]}$ and quantile level $\tau$, the empirical quantile loss is
\begin{align} \label{Ch3eloss}
    \hat{Q}(\bbeta)= \frac{1}{n} \sum_{i=1}^n\rho_\tau(r_i(\bbeta))  = \int_{-\infty}^{\infty}\rho_\tau(u)\, \mbox{d} \hat{F}(u;\bbeta)\,,
\end{align} 
where $\rho_\tau(u)=u\{\tau - \mathbf{1}(u<0)\}$ is the check loss, and $\hat{F}(\cdot ;\bbeta)$ is the empirical cumulative distribution function (CDF) of the residuals, given by 
\begin{align*}
    \hat{F}(u;\bm{\beta}) = \frac{1}{n}\sum_{i=1}^n \mathbf{1}\{r_i(\bm{\beta})\leq u\}
\end{align*}
for any $u\in\mathbb{R}$. The empirical quantile loss $\hat{Q}(\bbeta)$ is nonsmooth, which complicates computation and statistical analysis.
To address this difficulty, \citeS{fernandes2021smoothingS} proposed smoothing the empirical CDF. For bandwidth $h$, the resulting CDF is 
\begin{align*}
    \hat{F}_h(u;\bbeta)
=
 \frac{1}{n}\sum_{i=1}^n
\int_{-\infty}^u
K_h(v-r_i(\bbeta))\,\mbox{d}v \,,
\end{align*} 
where $K_h(\cdot)=h^{-1}K(\cdot/h)$ and $K(\cdot)$ is a symmetric, nonnegative kernel integrating to one. Replacing $\hat{F}(u;\bbeta)$ in \eqref{Ch3eloss} with $\hat{F}_h(u;\bbeta)$ yields the smoothed empirical quantile loss
\begin{align*}
\hat{Q}_{h}(\bbeta)
&:=
\int_{-\infty}^{\infty}
\rho_\tau(u)\,\mbox{d}\hat{F}_h(u;\bbeta)   =
\frac{1}{n }\sum_{i=1}^n
\int_{-\infty}^{\infty}
\rho_\tau(u)
K_h\big(u-r_i(\bbeta)\big)\,\mbox{d}u\,.
\end{align*}   
Let $\bar{K}_h(u)=\int_{-\infty}^{u/h}K(v)\,\mbox{d}v$. Under the kernel smoothness assumptions, $\hat{Q}_{h}(\bbeta)$ is twice continuously differentiable, with gradient and Hessian given by
\begin{align}\label{grad.hess}
    \nabla\hat{Q}_{h}(\bbeta) = \frac{1}{n}\sum_{i=1}^n\big\{\bar{K}_h(-r_i(\bbeta))-\tau\big\}\bx_i~~ \mathrm{and}~~ \nabla^2 \hat{Q}_{h}(\bbeta) = \frac{1}{n}\sum_{i=1}^n K_h (-r_i(\bbeta))\bx_i{\bx}_i^{\T}\,.
\end{align}

\section{Additional simulation results}\label{sec.addsimulation}

We provide additional implementation details for the competing methods considered in Section~\ref{sec.simulationresults}. For the full-sample benchmarks, both estimators use five-fold cross-validation. The L1 and SCAD estimators use 20-point and 50-point regularization grids, respectively. Internal predictor standardization is disabled for L1 and enabled for SCAD.
For \citet{tan2022communication}, we initialize with the local L1-penalized smoothed QR estimator on the first machine and use a practical multiscale CV grid. We perform 10 distributed updates for both \citet{tan2022communication} and \citet{wang2024distributed}. For these two methods, each CV fold computes its own initial estimator using only that fold's training observations, and the resulting sparsity estimate remains fixed during subsequent updates.
For \citet{yang2026}, we perform eight contraction updates followed by nine tightening updates. The bandwidth specification and initial support-size estimate follow the published paper, without truncating the estimated support size.
As numerical safeguards for the distributed update subproblems, candidates failing the boundedness check are excluded during tuning. Numerical failures are retried at the same regularization parameter and with the same objective. The grid is expanded upward only when no candidate passes the boundedness check. No additional ridge penalties or coefficient constraints are introduced. 

\begin{figure}[!htbp]
    \centering
    \includegraphics[width= 0.9\textwidth]{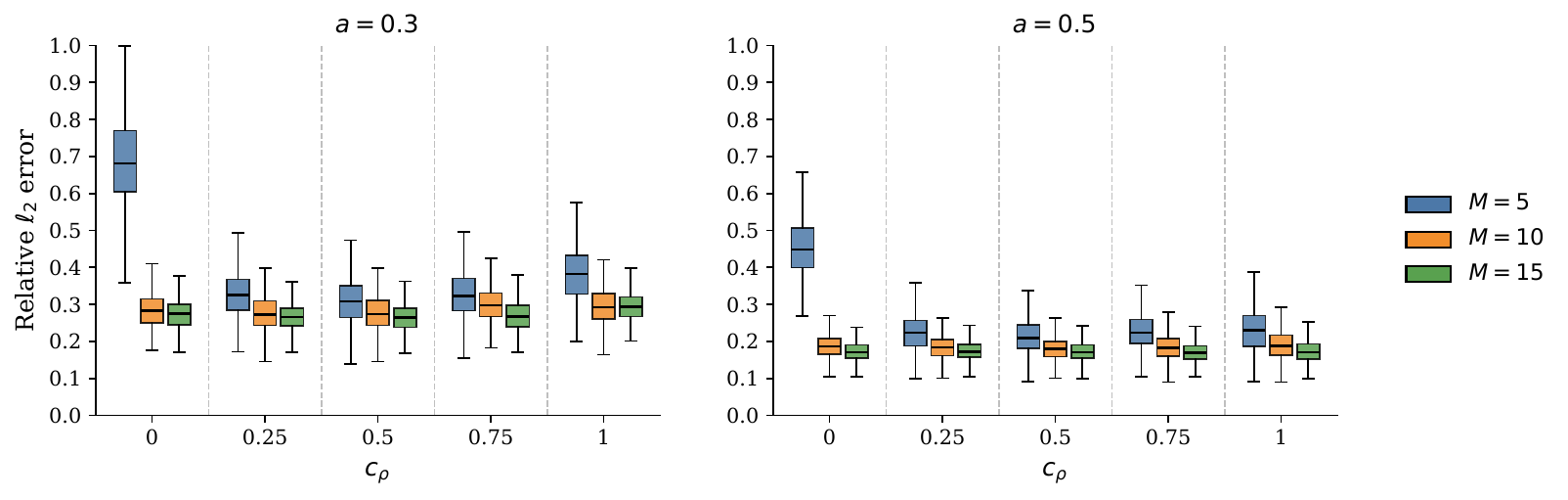}
    \caption{Boxplots of estimation performance over 500 Monte Carlo replications for varying signal strengths $a$, numbers of machines $M$, and local screening thresholds $\rho$, with $ (n,p ,s)=(200,500,10)$, $\tau=0.7$, and $t_{2.25}$ errors.}
    \label{fig:error_t}
\end{figure}

\begin{figure}[!htbp]
    \centering
    \includegraphics[width= 0.9\textwidth]{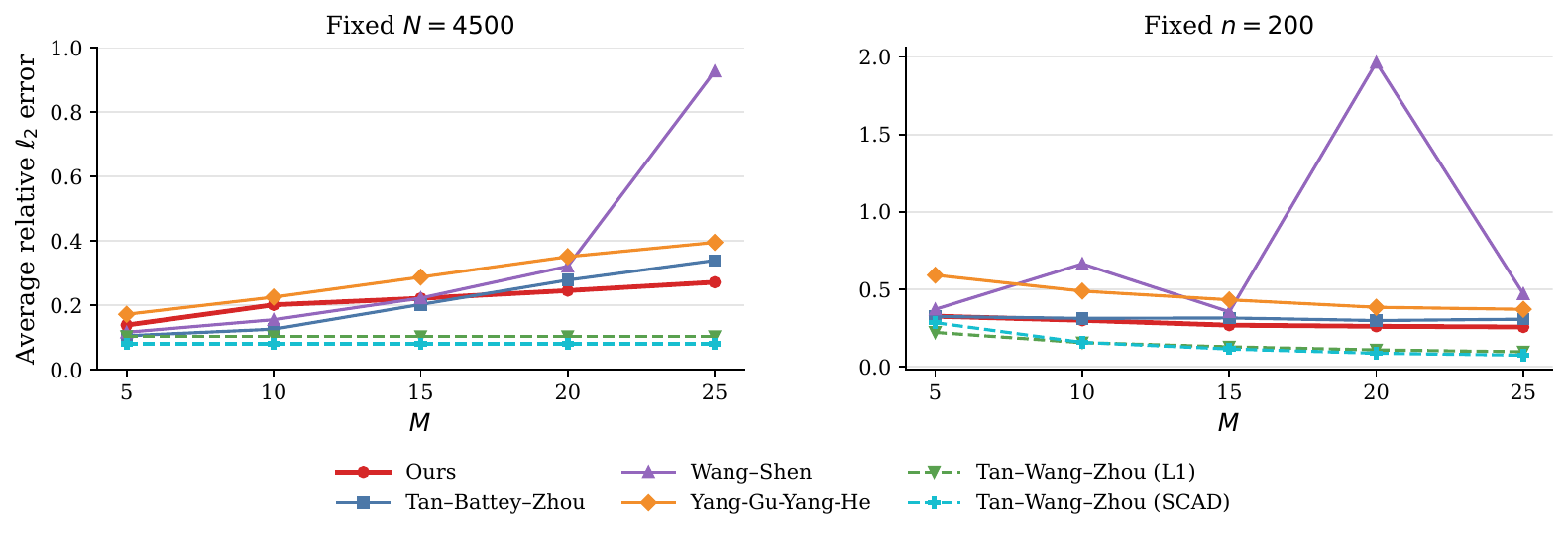}
    \caption{Average relative $\ell_2$ estimation errors over 500 Monte Carlo
replications for the proposed method with $c_\rho=0.75$ and competing centralized
and distributed high-dimensional QR methods
for varying numbers of machines $M$. The left and right
panels correspond to fixed total sample size $N=4500$ and
fixed local sample size $n=200$, respectively, with
$(p,s)=(500,10)$, $\tau=0.7$, $a=0.3$, and $t_{2.25}$ errors. The solid lines represent the proposed method  and three
competing distributed methods: Tan-Battey-Zhou
\citep{tan2022communication}, Wang-Shen
\citep{wang2024distributed}, and Yang-Gu-Yang-He
\citep{yang2026}.
The dashed lines represent the full-sample 
$\ell_1$-penalized and SCAD-penalized estimators of Tan-Wang-Zhou \citep{tan2022high}.}
    \label{fig:comparison_t}
\end{figure}

The following results under $t_{2.25}$ errors serve as supplementary experiments to complement those in Section \ref{sec.simulationresults} of the main text.  Fig. \ref{fig:error_t} and Table \ref{tab:screening_t} summarize the estimation and screening performance of the proposed method, respectively, with $(n,p,s)=(200,500,10)$ and $\tau=0.7$. Fig. \ref{fig:comparison_t} compares its estimation performance with that of centralized and distributed competitors under fixed total and fixed local sample sizes.

\begin{table}[htbp]
\centering
\caption{
Average screening performance of the proposed method over
500 Monte Carlo replications for combinations
of $(a,M,c_\rho)$, with $(n,p,s)=(200,500,10)$, $\tau=0.7$,
and $t_{2.25}$ errors.
}
\label{tab:screening_t}
\small
\setlength{\tabcolsep}{4.5pt}
\renewcommand{\arraystretch}{1.15}
\begin{tabular}{cc *{9}{c}}
\toprule
& & \multicolumn{3}{c}{Local screening size}
  & \multicolumn{3}{c}{Global screening size}
  & \multicolumn{3}{c}{TPR} \\
\cmidrule(lr){3-5}
\cmidrule(lr){6-8}
\cmidrule(lr){9-11}
$c_\rho$ & $a$
& $M=5$ & $M=10$ & $M=15$
& $M=5$ & $M=10$ & $M=15$
& $M=5$ & $M=10$ & $M=15$ \\
\midrule

{$0$}
& $0.3$ & 499.00 & 499.00 & 499.00
        & 499.00 & 10.83 & 9.73
        & 1.000 & 0.976 & 0.964 \\
& $0.5$ & 499.00 & 499.00 & 499.00
        & 499.00 & 11.18 & 10.10
        & 1.000 & 0.998 & 0.996 \\
\addlinespace[2pt]

{$0.25$}
& $0.3$ & 194.31 & 194.38 & 194.35
        & 9.27 & 10.43 & 10.57
        & 0.902 & 0.998 & 1.000 \\
& $0.5$ & 193.82 & 193.99 & 194.04
        & 10.07 & 10.52 & 10.78
        & 0.976 & 1.000 & 1.000 \\
\addlinespace[2pt]

{$0.50$}
& $0.3$ & 45.51 & 45.51 & 45.64
        & 10.02 & 9.78 & 10.20
        & 0.955 & 0.975 & 0.998 \\
& $0.5$ & 46.13 & 46.12 & 46.26
        & 10.40 & 10.05 & 10.29
        & 0.995 & 1.000 & 1.000 \\
\addlinespace[2pt]

{$0.75$}
& $0.3$ & 12.19 & 12.04 & 12.11
        & 9.14 & 8.89 & 9.88
        & 0.884 & 0.889 & 0.980 \\
& $0.5$ & 15.67 & 15.55 & 15.71
        & 9.98 & 9.89 & 10.19
        & 0.967 & 0.989 & 1.000 \\
\addlinespace[2pt]

{$1.00$}
& $0.3$ & 6.20 & 6.11 & 6.13
        & 7.21 & 9.14 & 8.89
        & 0.719 & 0.905 & 0.887 \\
& $0.5$ & 10.12 & 10.01 & 10.14
        & 9.59 & 9.81 & 9.87
        & 0.955 & 0.980 & 0.987 \\

\bottomrule
\end{tabular}
\end{table}

 \section{Proof of Proposition \ref{prop.Bahadur}}

\subsection{Preliminaries}\label{sec.pre}

We introduce the notation and decomposition used to prove Proposition~\ref{prop.Bahadur}. Since the argument applies to each machine, we consider a generic local sample $\{(y_i,\bx_i)\}_{i\in[n]}$ and suppress the machine index $m$ throughout. Our goal is to establish a Bahadur representation for the studentized statistic 
 \begin{align*}
     \frac{\sqrt{n}(\tilde{\beta}_j-{\beta}_j^*)\hat{H}_{j|-j}}{{\hat\sigma}_{j}}\,,
 \end{align*}
 where $\tilde{\beta}_j 
    =
    \hat{\beta}_{h,j} 
    -
    \hat{H}_{j|-j}^{-1}
    \hat{S}_j(\hat\bbeta_h)$,  $\hat{\sigma}_{j}^2= {n}^{-1}\sn  [ \{  \tau-\bar{K}_h( -r_i(\hat{\bbeta}_h)) \}  \bx_i^{\T}\hat{\bv}_{(j)}  ]^2$, and 
\begin{align}\label{def.H}
    \hat{H}_{j|-j}
    = 
    \{\nabla^2\hat{Q}_h(\hat{\bbeta}_h)\}_{j,j}
    -
    \hat{\bw}_{(j)}^{\T}
    \{\nabla^2\hat{Q}_h(\hat{\bbeta}_h)\}_{-j,j}  
    = 
    \frac{1}{n}\sn
    K_h(-r_i(\hat{\bbeta}_h))
    x_{i,j}\bx_i^{\T}\hat{\bv}_{(j)}\, .
\end{align}
Here, $r_i(\bbeta)=y_i-\bx_i^{\T}\bbeta$, $\hat{\bbeta}_h=(\hat{\beta}_{h,1},\ldots, \hat{\beta}_{h,p} )^{\T}$ is the estimator in \eqref{l1.est} computed from $\{(y_i,\bx_i)\}_{i\in[n]}$,
\begin{align}\label{def.hatS}
    \hat{S}_j(\hat{\bbeta}_h)
    =&~
    \{\nabla\hat{Q}_h(\hat{\bbeta}_h)\}_j
    -
    \hat{\bw}_{(j)}^{\T}
    \{\nabla\hat{Q}_h(\hat{\bbeta}_h)\}_{-j}   
    = 
    \frac{1}{n}\sum_{i=1}^n
    \{ \bar{K}_h( -r_i(\hat{\bbeta}_h)) -\tau\}
    \bx_i^{\T}\hat{\bv}_{(j)}\,,
\end{align}
and the estimated decorrelation vector $\hat{\bv}_{(j)}\in\RR^p$ is defined by  
        $\{\hat{\bv}_{(j)}\}_j=1$ and
        $\{\hat{\bv}_{(j)}\}_{-j}=-\hat{\bw}_{(j)}$ for each $j\in [p]$, where 
 \begin{align}\label{def.hatw}
             \hat{\bw}_{(j)} 
            \in  
            \arg\min_{\bw\in\RR^{p-1}}
            \|\bw\|_1
            ~ \mathrm{s.t.}~
            \big \|\{\nabla^2\hat{Q}_h(\hat{\bbeta}_h)\}_{j,-j} -  {\bw}^{\T}\{\nabla^2\hat{Q}_h(\hat{\bbeta}_h)\}_{-j,-j}\big\|_\infty\leq \Delta \, .
        \end{align}
 For $i\in[n]$ and $j\in [p]$, define $ \zeta_{i,j}  = \{   \tau-\bar{K}_h(  -\varepsilon_i)  \}\bx_i^{\T} {\bv}_{(j)} $, where ${\bv}_{(j)}\in\RR^p$ is defined by 
        $\{{\bv}_{(j)}\}_j=1$ and
        $\{{\bv}_{(j)}\}_{-j}=-\tilde{\bw}_{(j)}$. For $j\in [p]$, define the standardized sum
  \begin{align}\label{def.Z}
      Z_j = \frac{\mathring{S}_{n,(j)}}{{\sigma}_{j}}= \frac{(1-E)\{S_{n,(j)}\}}{{\sigma}_{j}}\,,
  \end{align}      
 where  $S_{n,(j)}= n^{-1/2}\sum_{i\in[n]} \zeta_{i,j}$ and   ${\sigma}_{j}^2  =  {\rm var}(\zeta_{i,j} )$.  The following decomposition holds for each $j\in [p]$:  
\begin{align}\label{div.bahadur}
   &~\bigg| \frac{\sqrt{n}(\tilde{\beta}_j-{\beta}_j^*)\hat{H}_{j|-j}}{{\hat\sigma}_{j}} -Z_j\bigg| 
   = \bigg| \frac{- \sqrt{n} \hat{S}_j(\hat{\bbeta}_h)+\sqrt{n}(\hat{\beta}_{h,j}-\beta_j^*)\hat{H}_{j|-j} }{{\hat\sigma}_{j}}  -Z_j\bigg|\nonumber\\
    &~~~~= \bigg|\frac{\mathring{S}_{n,(j)}}{{\hat\sigma}_{j}}-Z_j +  \frac{E\{{S}_{n,(j)}\}  -S_{n,(j)} -  \sqrt{n} \hat{S}_j(\hat{\bbeta}_h)     +\sqrt{n}(\hat{\beta}_{h,j}-\beta_j^*)\hat{H}_{j|-j}   }{{\hat\sigma}_{j}} \bigg|\nonumber\\
    &~~~~=  \bigg|  Z_j\bigg( \frac{\sigma_j}{\hat\sigma_j}-1 \bigg) +  \frac{E\{{S}_{n,(j)}\}  -S_{n,(j)} -  \sqrt{n} \hat{S}_j(\hat{\bbeta}_h)     +\sqrt{n}(\hat{\beta}_{h,j}-\beta_j^*)\hat{H}_{j|-j}   }{{  \hat\sigma}_{j}}   \bigg|  \,.
\end{align} 
This decomposition separates the error due to estimating the standard deviation from the remaining approximation error. The auxiliary results in the next subsection provide the bounds needed to prove Proposition \ref{prop.Bahadur}.

 \subsection{Auxiliary results}\label{sec.lemmas}

The proof of Proposition \ref{prop.Bahadur} requires error bounds for the preliminary estimator $\hat{\bbeta}_h$ and the estimated decorrelation vector $\hat{\bw}_{(j)}$, and the estimated standard deviation $\hat{\sigma}_j$. The following result, adapted from \citeS{tan2022highS}, controls $\hat{\bbeta}_h$.
Although the original result in \citeS{tan2022highS}
holds with probability at least $1-p^{-1}$,
adjusting the concentration parameters and associated constants in its proof yields a probability of at least $1-O(p^{-c_0})$ for any sufficiently
large fixed constant $c_0 = \bar{c}+1$, without changing the convergence rate. Bounds for $\hat{\bw}_{(j)}$ and $\hat{\sigma}_j$ are given below. Proofs of the technical lemmas appear in Section~\ref{sec.lemma.proof}.

\begin{lemma}[Adapted from Theorem 1 of \citeS{tan2022highS}]\label{Theorem.tan2022}
    Under model \eqref{linear.model} with an $(s+1)$-sparse coefficient vector $\bbeta^*\in \RR^p$, assume Conditions {\rm 1(a), 1(b)} and {\rm1(d)} hold. Let $\sigma_{\bx}^2=\max_{j\in[p]}E(x_{j}^2)$. If the smoothing bandwidth satisfies
    \begin{align}\label{h.req}
        \max\bigg(\frac{\sigma_{\bx}}{f_l}\sqrt{\frac{s \log p}{n}},\frac{\sigma_{\bx}^2f_u}{f_l^2}\frac{s \log p}{n}\bigg)\lesssim h \leq \min\{f_l/(2l_0),(\sqrt{s }\lambda)^{1/2}\}\,,
    \end{align}  
    then the estimator $\hat{\bbeta}_h$ with $\lambda \asymp \sigma_{\bx}\sqrt{\tau(1-\tau)(\log p)/n}$ satisfies
    \begin{align*}
         \|{\hat{\bbeta}_h  - \bbeta^*}\|_1\leq C_{\beta} \frac{s}{f_l}  \sqrt{\frac{ \log p}{n}}
    \end{align*}
    with probability at least $1-O(p^{-c_0})$,  
    where $c_0>0$ is any fixed constant
    and $C_{\beta}>0$ may depend on $(c_0,l_0,\upsilon_0,\gamma_1,\gamma_p,\underline{\kappa},\kappa_2)$. The constants implicit in $\asymp$ and $\lesssim$
may depend on $c_0$. Moreover, the multiplicative constant in the choice of $\lambda$ and the implicit constant in the lower bound for $h$ are taken sufficiently large.
\end{lemma}

Lemma \ref{lemma.triangle} establishes subadditivity of the $\psi_\zeta$-Orlicz quasi-norm up to a multiplicative constant. Its proof is given in Section \ref{proof.lemma.triangle}.

\begin{lemma}\label{lemma.triangle}
    Let $X_1,\ldots,X_n$ be random variables satisfying
$\|X_i\|_{\psi_\zeta}<\infty$ for all $i\in[n]$ and some
$\zeta>0$. Then,
    \begin{align*} 
     \bigg\|\sum_{i=1}^n  X_i\bigg\|_{\psi_{\zeta}}   \leq C_\zeta   \sum_{i=1}^n\|  X_i\|_{\psi_{\zeta}}\,,
 \end{align*}
 where $C_\zeta>0$ depends only on $\zeta$.
\end{lemma}

Lemma \ref{lemma.w.cons} establishes an $\ell_1$-error bound for the estimated decorrelation vector $\hat{\bw}_{(j)}$. Its proof is given in Section \ref{proof.lemma.w.cons}.

\begin{lemma}\label{lemma.w.cons}
    Assume Condition  {\rm 1} and \eqref{h.req} hold with $\lambda \asymp \sigma_{\bx}\sqrt{\tau(1-\tau)(\log p)/n}$, and let
    \begin{align}\label{def.Delta}
        \Delta =   C_{\Delta} \sqrt{ \frac{f_u}{f_l}} \bigg\{      \frac{f_u}{f_l}   \sqrt{\frac{ s^2\log p}{nh^2}}+  \frac{1}{f_l^2}  \frac{s^2(\log p)^5 }{   nh^3}  \bigg\}\,,
    \end{align} 
   where $C_{\Delta}>0$ is a large constant depending on $(c_0,l_0,\upsilon_0,\gamma_1,\gamma_p,\underline{\kappa},\kappa_1,\kappa_2,C_K,\kappa_u,\kappa_u',\kappa_u'')$. 
 If $8 \tilde{s} \Delta  \leq f_l \gamma_p$, then, for each $j\in [p]$,
\begin{align*}
    \|\hat{\bw}_{(j)}-\tilde{\bw}_{(j)}\|_1 \lesssim \frac{\tilde{s}\Delta}{f_l}  
\end{align*} 
with probability at least $1- O(p^{-c_0})$.
\end{lemma}

Lemma \ref{lemma.sigma.cons} bounds $|\hat{\sigma}_{j}-{\sigma}_{j}|$. Its proof is given in Section \ref{proof.lemma.sigma.cons}.

 \begin{lemma}\label{lemma.sigma.cons}
    Suppose the assumptions of Lemma {\rm \ref{lemma.w.cons}} hold. For each $j\in [p]$, if, in addition, $\max\{ l_0^2  \kappa_2^2h^4, 4 f_u\kappa_1 h + 2\tau l_0\kappa_2 h^2\} \leq \tau(1-\tau)$, we have $\sigma_j^2 \gtrsim   \|\bv_{(j)}\|_{\bSigma}^2 \geq \gamma_p$. Moreover, with probability at least $1- O(p^{-c_0})$,
    \begin{align}\label{def.tildedelta}
    |\hat{\sigma}_{j}-{\sigma}_{j}| \lesssim  \tilde{\Delta}   :=  \frac{f_u^{1/2}}{f_l^{3/2}}\tilde{s}\Delta + C_{\tilde{\Delta}} \frac{f_u}{f_l^2}   \sqrt{\frac{ s^2(\log p)^7}{nh^2}}\,,
\end{align}
provided that $\tilde{\Delta} \leq C$ for some fixed constant $C>0$, where $C_{\tilde{\Delta}}>0$ is a large constant depending on $(c_0,l_0,\upsilon_0,\gamma_1,\gamma_p,\underline{\kappa}, \kappa_u,\kappa_2)$.
\end{lemma}

\subsection{Proof of Proposition \ref{prop.Bahadur}}\label{proof.prop.Bahadur}

Throughout the proof, constants depending on $(l_0,\upsilon_0,\gamma_1,\gamma_p,\underline{\kappa},\kappa_1,\kappa_2,C_K,\kappa_u,\kappa_u',\kappa_u'',f_l,f_u)$ are absorbed into the
notation $\lesssim$, $\gtrsim$ and $\asymp$. 
For each $j\in [p]$, define the event $\cE_{j}=\{ |\hat{\sigma}_{j}-{\sigma}_{j}| \leq C_0\tilde{\Delta} \}$ for some sufficiently large constant $C_0>0$, where $\tilde{\Delta}$ is specified in \eqref{def.tildedelta}. By Lemma \ref{lemma.sigma.cons}, provided that $\tilde{\Delta} \ll 1$ and 
\begin{align}\label{h.req.1}
    \bigg(  \frac{s\log p}{n}\bigg)^{1/2} \lesssim h \lesssim \bigg(  \frac{s\log p}{n}\bigg)^{1/4}\,,
\end{align}
we have $\sigma_j \gtrsim 1$, and $\cE_{j}$ holds with probability at least $1-O(p^{-c_0})$, which implies $\hat{\sigma}_{j} \gtrsim 1$ on $\cE_{j}$. On $\cE_{j}$, \eqref{div.bahadur} therefore gives 
\begin{align}\label{div.bahadur.2}
     &~\bigg| \frac{\sqrt{n}(\tilde{\beta}_j-{\beta}_j^*)\hat{H}_{j|-j}}{{\hat\sigma}_{j}} -Z_j\bigg| \nonumber\\
   \lesssim &~    \tilde{\Delta} |  Z_j |  +  |  E\{{S}_{n,(j)}\} | + |  S_{n,(j)} + \sqrt{n} \hat{S}_j(\hat{\bbeta}_h)     -\sqrt{n}(\hat{\beta}_{h,j}-\beta_j^*)\hat{H}_{j|-j}  | \,,
\end{align}
where $S_{n,(j)}= n^{-1/2}\sum_{i\in[n]} \zeta_{i,j}$ with $ \zeta_{i,j}  = \{   \tau-\bar{K}_h(  -\varepsilon_i)  \}\bx_i^{\T} {\bv}_{(j)} $. We first bound the standardized sum. By \eqref{psinorm.xv} in Section \ref{proof.lemma.w.cons}, we have $\|\bx_i^{\T} {\bv}_{(j)}\|_{\psi_1} \leq C$ for any $i\in[n]$ and $j\in [p]$. Since $|  \tau-\bar{K}_h(  -\varepsilon_i)|\leq 1$, we obtain
\begin{align}\label{psi1.zeta}
    \|\zeta_{i,j}\|_{\psi_1}\leq C
\end{align}
   for any $i\in[n]$ and $j\in [p]$. Since $|Z_j| \lesssim | n^{-1/2}\sn (1-E)(\zeta_{i,j})|$, for any $j\in [p]$, Bernstein's inequality \citepS[see, e.g., Theorem~2.8.1 of][]{vershynin2018highS} yields 
\begin{align}\label{upper.Z}
    |Z_j|  \lesssim (\log p)^{1/2} 
\end{align}
with probability at least $1-O(p^{-c_0})$, provided that $\log p \ll n$.

Next, we bound the smoothing bias. By \eqref{psi.beta} in Section \ref{proof.lemma.sigma.cons}, we have $|E [ \{   \tau-\bar{K}_h(  -\varepsilon_i)  \}\,|\,\bx_i]| \lesssim h^2$. Since $\|\bx_i^{\T} {\bv}_{(j)}\|_{\psi_1} \leq C$, we obtain $|E(\zeta_{i,j})| \lesssim h^2$,  which implies 
\begin{align}\label{upper.ES}
    |  E\{{S}_{n,(j)}\} |  \lesssim n^{1/2}h^2\,.
\end{align}

To control the remaining approximation error, write $\psi_i(\bbeta)=  \tau-\bar{K}_h( -r_i({\bbeta})) $ for each $i\in[n]$ and $\bbeta\in \RR^p$. By \eqref{def.H} and \eqref{def.hatS}, we have
\begin{align}\label{upper.3.pre}
    &~|  S_{n,(j)} + \sqrt{n} \hat{S}_j(\hat{\bbeta}_h)     -\sqrt{n}(\hat{\beta}_{h,j}-\beta_j^*)\hat{H}_{j|-j}  |\nonumber \\
  &~~~~ =  \sqrt{n}\bigg| \frac{1}{n}\sn\big\{\psi_i(\hat\bbeta_h) \bx_i^{\T}\hat{\bv}_{(j)}  -  \psi_i(\bbeta^*) \bx_i^{\T}{\bv}_{(j)} + (\hat{\beta}_{h,j}-\beta_j^*)  
    K_h(-r_i(\hat{\bbeta}_h))
    x_{i,j}\bx_i^{\T}\hat{\bv}_{(j)}\big\}\bigg|\nonumber\\
    &~~~~ \leq   \underbrace{\sqrt{n}\bigg| \frac{1}{n}\sn\psi_i(\bbeta^*)\bx_i^{\T}\{\hat{\bv}_{(j)}- {\bv}_{(j)}\}   \bigg|}_{ \Pi_1}\nonumber\\
    &~~~~~~~+\underbrace{\sqrt{n}\bigg| \frac{1}{n} \sn \big\{\psi_i(\hat\bbeta_h) -\psi_i(\bbeta^*) + (\hat{\beta}_{h,j}-\beta_j^*) 
    K_h(-r_i(\hat{\bbeta}_h))
    x_{i,j}\big\}\bx_i^{\T}\hat{\bv}_{(j)}\bigg|}_{ \Pi_2}\,.
\end{align}
Since $\|\hat{\bv}_{(j)}- {\bv}_{(j)}\|_1 = \|\hat{\bw}_{(j)}- {\tilde\bw}_{(j)}\|_1$, Lemma \ref{lemma.w.cons} gives
\begin{align*}
   \Pi_1 \leq \sqrt{n} \|\hat{\bv}_{(j)}- {\bv}_{(j)}\|_1 \bigg\|   \frac{1}{n}\sn\psi_i(\bbeta^*)\bx_i\bigg\|_{\infty} \lesssim \sqrt{n}\tilde{s}\Delta \max_{j\in[p]}\bigg |   \frac{1}{n}\sn\psi_i(\bbeta^*) x_{i,j}\bigg | 
\end{align*}
with probability at least $1-O(p^{-c_0})$, where $\Delta$ is specified in \eqref{def.Delta}. By \eqref{psinorm.x} in Section \ref{proof.lemma.w.cons}, we have $\|x_{i,j}\|_{\psi_1}\leq C$ for any $i\in[n]$ and $j\in[p]$. Since $|\psi_i(\bbeta^*)|\leq 1$ and $|E \{ \psi_i({\bbeta}^*)\,|\,\bx_i\}| \lesssim h^2$, it follows from Bernstein's inequality that 
\begin{align*}
    \max_{j\in[p]}\bigg |   \frac{1}{n}\sn\psi_i(\bbeta^*) x_{i,j}\bigg |  \lesssim  \max_{j\in[p]}\bigg |   \frac{1}{n}\sn(1-E)\{\psi_i(\bbeta^*) x_{i,j}\}\bigg | + h^2 \lesssim \sqrt{\frac{  \log p}{n}}  +h^2
\end{align*}
with probability at least $1-O(p^{-c_0})$. Combining these bounds gives
\begin{align}\label{upper.Pi1}
    \Pi_1   \lesssim \sqrt{n}\tilde{s}\Delta\bigg(  \sqrt{\frac{ \log p}{n}} +h^2 \bigg)= \tilde{s}(\log p)^{1/2}\Delta + \tilde{s}\sqrt{nh^4}\Delta
\end{align}
with probability at least $1-O(p^{-c_0})$.

 Write $\bvarpi= \hat\bbeta_h-\bbeta^*$ and $d_i = \bx_i^{\T}\bvarpi$. Then
\begin{align*}
    \psi_i(\hat\bbeta_h) -\psi_i(\bbeta^*)  = &~ \bar{K}_h( -\varepsilon_i)- \bar{K}_h( -\varepsilon_i + d_i)\\
     =&~ - K_h( -\varepsilon_i +d_i)d_i +\underbrace{ \int_{0}^{d_i}\{K_h( -\varepsilon_i +d_i)-K_h( -\varepsilon_i +t) \}\,\mbox{d}t}_{\varrho_i}\,.
\end{align*}
 Since $-\varepsilon_i +d_i = - r_i(\hat\bbeta_h)$,
\begin{align}\label{upper.Pi2.pre}
    \Pi_2  = &~ \sqrt{n}\bigg|\frac{1}{n} \sn \big\{-K_h(-r_i(\hat{\bbeta}_h))\bx_i^{\T}\bvarpi  +  
    K_h(-r_i(\hat{\bbeta}_h))
    x_{i,j}(\hat{\beta}_{h,j}-\beta_j^*) + \varrho_i\big\}\bx_i^{\T}\hat{\bv}_{(j)} \bigg|\nonumber\\
    =&~  \sqrt{n}\bigg| \frac{1}{n} \sn \big\{ 
    -K_h(-r_i(\hat{\bbeta}_h))\bx_{i,-j}^{\T}\bvarpi_{-j} + \varrho_i
     \big\}\bx_i^{\T}\hat{\bv}_{(j)}\bigg|\nonumber\\
      \leq &~  \underbrace{\sqrt{n}\bigg|\frac{1}{n} \sn   
     K_h(-r_i(\hat{\bbeta}_h))\bx_{i,-j}^{\T}\bvarpi_{-j} \bx_i^{\T}\hat{\bv}_{(j)}\bigg|}_{\Pi_{2,1}}+\underbrace{\sqrt{n}\bigg|\frac{1}{n} \sn  \varrho_i
      \bx_i^{\T}\hat{\bv}_{(j)}\bigg|}_{\Pi_{2,2}}\,.
\end{align}
Recall that $\nabla^2\hat{Q}_h(\bbeta) =  {n}^{-1}\sn K_h(-r_i(\bbeta))\bx_i\bx_i^{\T}$ and $\hat{\bv}_{(j)}\in\RR^p$ is defined by 
        $\{\hat{\bv}_{(j)}\}_j=1$ and
        $\{\hat{\bv}_{(j)}\}_{-j}=-\hat{\bw}_{(j)}$. The constraint in \eqref{def.hatw} and Lemma \ref{Theorem.tan2022} yield 
\begin{align*}
   &~ \bigg|\frac{1}{n} \sn   
     K_h(-r_i(\hat{\bbeta}_h))\bx_{i,-j}^{\T}\bvarpi_{-j} \bx_i^{\T}\hat{\bv}_{(j)}\bigg|\\ 
    &~~~~ =  \bigg|-\frac{1}{n} \sn   
     K_h(-r_i(\hat{\bbeta}_h))\bx_{i,-j}^{\T}\bvarpi_{-j} \bx_{i,-j}^{\T}\hat{\bw}_{(j)} + \frac{1}{n} \sn   
     K_h(-r_i(\hat{\bbeta}_h))\bx_{i,-j}^{\T}\bvarpi_{-j} x_{i,j}\bigg|\\
     &~~~~=   \big|\big[\{\nabla^2\hat{Q}_h(\hat{\bbeta}_h)\}_{j,-j}- \hat{\bw}_{(j)}^{\T}\{\nabla^2\hat{Q}_h({\hat\bbeta}_h)\}_{-j,-j}\big]  \bvarpi_{-j}\big|\\
     &~~~~ \leq   \big\|\{\nabla^2\hat{Q}_h(\hat{\bbeta}_h)\}_{j,-j}- \hat{\bw}_{(j)}^{\T}\{\nabla^2\hat{Q}_h({\hat\bbeta}_h)\}_{-j,-j}\big\|_{\infty}\|\bvarpi\|_1 \lesssim s \Delta \sqrt{\frac{\log p}{n}} 
\end{align*}
with probability at least $1-O(p^{-c_0})$, which implies 
\begin{align}\label{upper.Pi21}
    \Pi_{2,1} \lesssim s(\log p)^{1/2} \Delta
\end{align}
with probability at least $1-O(p^{-c_0})$.

 To bound $\Pi_{2,2} $, first write
\begin{align*}
    \varrho_i = &~\int_{0}^{d_i}\{K_h( -\varepsilon_i +d_i)-K_h( -\varepsilon_i +t) \}\,\mbox{d}t\\
    = &~ d_i\int_{0}^{1}\{K_h( -\varepsilon_i +d_i)-K_h( -\varepsilon_i +sd_i) \}\,\mbox{d}s \\
    = &~ d_i^2\int_{0}^{1} \bigg\{  \int_{s}^1  K_h'(-\varepsilon_i + z d_i) \,\mbox{d}z  \bigg\}\,\mbox{d}s  =  d_i^2\int_{0}^{1} z K_h'(-\varepsilon_i + z d_i)  \,\mbox{d}z \,.
\end{align*}
Since $\int_{0}^1 z \,\mbox{d}z = 1/2$,
\begin{align}\label{Pi22.pre}
    \bigg|\frac{1}{n} \sn  \varrho_i
      \bx_i^{\T}\hat{\bv}_{(j)}\bigg| =&~  \bigg|\int_{0}^{1} z \bigg\{ \frac{1}{n} \sn  d_i^2 K_h'(-\varepsilon_i + z d_i) 
      \bx_i^{\T}\hat{\bv}_{(j)}  \bigg\}\,\mbox{d}z  \bigg|\nonumber\\
      \leq  &~  \bigg|\int_{0}^{1} z \bigg[ \frac{1}{n} \sn  d_i^2 \{ K_h'(-\varepsilon_i + z d_i) - K_h'(-\varepsilon_i  )\}
      \bx_i^{\T}\hat{\bv}_{(j)}  \bigg]\,\mbox{d}z  \bigg|\nonumber\\
      &~ +\frac{1}{2}\bigg|   \frac{1}{n} \sn  d_i^2 K_h'(-\varepsilon_i) 
      \bx_i^{\T}{\bv}_{(j)}    \bigg|+\frac{1}{2}\bigg|   \frac{1}{n} \sn  d_i^2 K_h'(-\varepsilon_i) 
      \bx_i^{\T}\{\hat{\bv}_{(j)} - {\bv}_{(j)}\}  \bigg|\,.
\end{align}
Since $K_h'(u)=h^{-2}K'(u/h)$, we obtain
\begin{align*}
      K_h'(-\varepsilon_i + z d_i) - K_h'(-\varepsilon_i  ) = \frac{z  d_i }{h^3} \int_{0}^1 K''\bigg( \frac{-\varepsilon_i+ szd_i}{h}\bigg) \,\mbox{d}s    
\end{align*}
for any $z\in[0,1]$.  Thus,
\begin{align*}
    &~\bigg|\int_{0}^{1} z \bigg[ \frac{1}{n} \sn  d_i^2 \{ K_h'(-\varepsilon_i + z d_i) - K_h'(-\varepsilon_i  )\}
      \bx_i^{\T}\hat{\bv}_{(j)}  \bigg]\,\mbox{d}z  \bigg|  \\
    &~~~~  =   \frac{1}{h^3}\bigg|\int_{0}^{1} z^2 \int_{0}^1  \bigg[ \frac{1}{n} \sn  d_i^3  K''\bigg( \frac{-\varepsilon_i+ szd_i}{h}\bigg) 
      \bx_i^{\T}\hat{\bv}_{(j)}  \bigg]\,\mbox{d}s \,\mbox{d}z  \bigg|  
\end{align*}
By Condition 1(d), we have
\begin{align*}
    &~\bigg| \frac{1}{n} \sn  d_i^3  K''\bigg( \frac{-\varepsilon_i+ szd_i}{h}\bigg) 
      \bx_i^{\T}\hat{\bv}_{(j)}  \bigg|\\
     &~~~~ \leq  \bigg| \frac{1}{n} \sn  d_i^3  K''\bigg( \frac{-\varepsilon_i+ szd_i}{h}\bigg) 
      \bx_i^{\T}{\bv}_{(j)}  \bigg|+ \bigg| \frac{1}{n} \sn  d_i^3  K''\bigg( \frac{-\varepsilon_i+ szd_i}{h}\bigg) 
      \bx_i^{\T}\{\hat{\bv}_{(j)} -{\bv}_{(j)}\} \bigg|  \\
      &~~~~\leq \kappa_u'' \|\bvarpi\|_1^3  \cdot  \max_{k,l,r\in [p]}\frac{1}{n}\sn|x_{i,k}x_{i,l}x_{i,r}\bx_i^{\T}{\bv}_{(j)}| + \kappa_u'' \|\bvarpi\|_1^3 \|\hat{\bv}_{(j)} -{\bv}_{(j)}\|_1 \cdot \max_{k\in[p]}\frac{1}{n}\sn|x_{i,k}|^4\,.
\end{align*}
Recall that $\tilde{\Delta} \ll 1$. Combining Lemmas \ref{Theorem.tan2022} and \ref{lemma.w.cons} with \eqref{abs.moment.x} in Section \ref{proof.lemma.w.cons} gives
\begin{align*}
   &~ \bigg|\int_{0}^{1} z \bigg[ \frac{1}{n} \sn  d_i^2 \{ K_h'(-\varepsilon_i + z d_i) - K_h'(-\varepsilon_i  )\}
      \bx_i^{\T}\hat{\bv}_{(j)}  \bigg]\,\mbox{d}z  \bigg|  \\
      &~~~~\lesssim  \frac{\kappa_u''}{h^3} \|\bvarpi\|_1^3  \cdot  \max_{k,l,r\in [p]}\frac{1}{n}\sn|x_{i,k}x_{i,l}x_{i,r}\bx_i^{\T}{\bv}_{(j)}| \\
      &~~~~~~~+ \frac{\kappa_u''}{h^3} \|\bvarpi\|_1^3 \|\hat{\bv}_{(j)} -{\bv}_{(j)}\|_1 \cdot \max_{k\in[p]}\frac{1}{n}\sn|x_{i,k}|^4 \\
      &~~~~\lesssim   \bigg(\frac{s }{h} \sqrt{\frac{\log p}{n}}\bigg)^3  (\log p)^4 = \frac{s^3 (\log p)^{11/2}}{n^{3/2}h^3}
\end{align*}
with probability at least $1-O(p^{-c_0})$. For the second term on the right-hand side of \eqref{Pi22.pre},
\begin{align*}
     \bigg|   \frac{1}{n} \sn  d_i^2 K_h'(-\varepsilon_i) 
      \bx_i^{\T}{\bv}_{(j)}    \bigg| \lesssim &~ \|\bvarpi\|_1^2 \cdot \max_{ k,l\in[p]} \bigg| \frac{1}{n}\sn  x_{i,k}x_{i,l}\bx_i^{\T}{\bv}_{(j)} K_h'(-\varepsilon_i) \bigg|\\
      \leq &~ \|\bvarpi\|_1^2\bigg\{  \max_{k,l\in[p]} \bigg| \frac{1}{n}\sn (1-E)\{x_{i,k}x_{i,l}\bx_i^{\T}{\bv}_{(j)}  K_h'(-\varepsilon_i)\} \bigg|\\
      &~~~~~~~~~~~~~~~~~~~+\max_{k,l\in[p]} \bigg| \frac{1}{n}\sn E\{x_{i,k}x_{i,l}\bx_i^{\T}{\bv}_{(j)}  K_h'(-\varepsilon_i)\} \bigg| \bigg\}\,.
\end{align*} 
Since $|K_h'(-\varepsilon_i)|\lesssim h^{-2}$, we have $\|x_{i,k}x_{i,l}\bx_i^{\T}{\bv}_{(j)} K_h'(-\varepsilon_i)\|_{\psi_{1/3}}\lesssim  h^{-2}$ for any $i\in[n]$. Then, by Theorem 3.1 of \citeS{Kuchibhotla2022S} with $\alpha=1/3$ and $\ba= (n^{-1},\ldots,n^{-1})^{\T}\in \RR^n$, we obtain
\begin{align*}
     \pr\bigg\{\max_{k,l\in[p]} \bigg| \frac{1}{n}\sn (1-E)\{x_{i,k}x_{i,l}\bx_i^{\T}{\bv}_{(j)}  K_h'(-\varepsilon_i)\} \bigg| \geq \sqrt{\frac{C   z}{nh^4}} + \frac{C  z^3}{nh^{2}}\bigg\} \lesssim  p^2\exp(-z) 
\end{align*}
for any $z\geq 0$, which implies that, with probability at least $1-O(p^{-c_0})$, 
\begin{align*} 
   \max_{k,l\in[p]} \bigg| \frac{1}{n}\sn (1-E)\{x_{i,k}x_{i,l}\bx_i^{\T}{\bv}_{(j)}  K_h'(-\varepsilon_i)\} \bigg|  \lesssim  \sqrt{\frac{\log p}{nh^4}} + \frac{(\log p)^3}{nh^2} \,.
\end{align*}
Integration by parts under Conditions 1(a) and 1(d) gives 
\begin{align*}
    |E\{ K_h'(-\varepsilon_i) \,|\, \bx_i \}| = &~ \bigg|\frac{1}{h^2}\int_{-\infty}^{\infty} K'\bigg(-\frac{u}{h}\bigg)f_{\varepsilon_i|\bx_i}(u)\,\mbox{d}u\bigg|=\bigg| \frac{1}{h }\int_{-\infty}^{\infty} K'( v)f_{\varepsilon_i|\bx_i}(-hv)\,\mbox{d}v\bigg|\\
    = &~  \bigg|\frac{1}{h }\int_{-\infty}^{\infty} f_{\varepsilon_i|\bx_i}(-hv)\,\mbox{d}K( v)\bigg| 
     =   \bigg|\int_{-\infty}^{\infty} K( v)f_{\varepsilon_i|\bx_i}'(-hv)\,\mbox{d}v\bigg| \leq l_0\,,
\end{align*}
which implies that 
\begin{align*}
    \max_{k,l\in[p]} \bigg| \frac{1}{n}\sn E\{x_{i,k}x_{i,l}\bx_i^{\T}{\bv}_{(j)}  K_h'(-\varepsilon_i)\} \bigg|   \lesssim 1\,.
\end{align*}
Combining these bounds with Lemma \ref{Theorem.tan2022} yields
\begin{align}\label{second.Pi22}
    \bigg|   \frac{1}{n} \sn  d_i^2 K_h'(-\varepsilon_i) 
      \bx_i^{\T}{\bv}_{(j)}    \bigg| \lesssim \frac{s^2 \log p}{n}\bigg\{ 1+  \sqrt{\frac{\log p}{nh^4}} + \frac{(\log p)^3}{nh^2}\bigg\}
\end{align}
with probability at least $1-O(p^{-c_0})$. For the last term on the right-hand side of \eqref{Pi22.pre}, note that 
\begin{align*}
   &~ \bigg|   \frac{1}{n} \sn  d_i^2 K_h'(-\varepsilon_i) 
      \bx_i^{\T}\{\hat{\bv}_{(j)} - {\bv}_{(j)}\}   \bigg|\\
      &~~~~\leq   \|\bvarpi\|_1^2 \|\hat{\bv}_{(j)} - {\bv}_{(j)} \|_1 \cdot \max_{k,l,r\in[p]}\bigg|   \frac{1}{n} \sn x_{i,k}x_{i,l}x_{i,r}  K_h'(-\varepsilon_i)    \bigg|\\
     &~~~~ \leq   \|\bvarpi\|_1^2 \|\hat{\bv}_{(j)} - {\bv}_{(j)} \|_1 \cdot  \max_{k,l,r\in[p]}\bigg|   \frac{1}{n} \sn (1-E)\{ x_{i,k}x_{i,l}x_{i,r}  K_h'(-\varepsilon_i) \}   \bigg|\\
      &~~~~~~~ +\|\bvarpi\|_1^2 \|\hat{\bv}_{(j)} - {\bv}_{(j)} \|_1 \cdot \max_{k,l,r\in[p]}\bigg|   \frac{1}{n} \sn E\{x_{i,k}x_{i,l}x_{i,r}  K_h'(-\varepsilon_i) \}   \bigg| \,.
\end{align*}
Applying the argument used for \eqref{second.Pi22}, together with Lemmas \ref{Theorem.tan2022} and \ref{lemma.w.cons}, yields 
\begin{align*}
    \bigg|   \frac{1}{n} \sn  d_i^2 K_h'(-\varepsilon_i) 
      \bx_i^{\T}\{\hat{\bv}_{(j)} - {\bv}_{(j)}\}   \bigg| \lesssim \frac{\tilde{s}s^2 \log p}{n}\Delta\bigg\{ 1+  \sqrt{\frac{\log p}{nh^4}} + \frac{(\log p)^3}{nh^2}\bigg\} 
\end{align*}
with probability at least $1-O(p^{-c_0})$. Combining the three bounds in \eqref{Pi22.pre} gives
\begin{align*}
    \Pi_{2,2}\lesssim \frac{s^3 (\log p)^{11/2}}{n h^3} 
\end{align*}
with probability at least $1-O(p^{-c_0})$. Together with \eqref{h.req.1} and \eqref{upper.3.pre}--\eqref{upper.Pi21}, this yields 
\begin{align*}
   &~ |  S_{n,(j)} + \sqrt{n} \hat{S}_j(\hat{\bbeta}_h)     -\sqrt{n}(\hat{\beta}_{h,j}-\beta_j^*)\hat{H}_{j|-j}  | \\
  &~~~~ \lesssim   \{ \tilde{s}(\log p)^{1/2}+ \tilde{s}\sqrt{nh^4} + {s}(\log p)^{1/2}  \}\Delta + \frac{s^3 (\log p)^{11/2}}{n h^3}  
\end{align*} 
with probability at least $1-O(p^{-c_0})$. It follows from \eqref{div.bahadur.2}--\eqref{upper.ES} that 
\begin{align*}
    &~\bigg| \frac{\sqrt{n}(\tilde{\beta}_j-{\beta}_j^*)\hat{H}_{j|-j}}{{\hat\sigma}_{j}} -Z_j\bigg| \\
   &~~~~ \lesssim    (\log p)^{1/2} \tilde{\Delta} +n^{1/2}h^2 +\{ \tilde{s}(\log p)^{1/2}+ \tilde{s}\sqrt{nh^4} + {s}(\log p)^{1/2}  \}\Delta + \frac{s^3 (\log p)^{11/2}}{n h^3}
\end{align*}
with probability at least $1-O(p^{-c_0})$. Recall that 
\begin{align*}
     \Delta \asymp            \frac{s^2(\log p)^5 }{   nh^3}  \,,~~\mbox{and }~~\tilde{\Delta}  \asymp  \tilde{s}\Delta +  \sqrt{\frac{ s^2(\log p)^7}{nh^2}}\,.
\end{align*}
Substituting these rates gives, for each $j\in[p]$,
\begin{align*}
    \pr\bigg\{
    \bigg|
    \frac{\sqrt{n}(\tilde{\beta}_j-\beta_j^*)
    \hat{H}_{j|-j}}{\hat{\sigma}_j}
    -Z_j
    \bigg|
    \geq \bar{C}\sqrt{n}\delta_n
    \bigg\}
    \lesssim p^{-c_0},
\end{align*}
provided that $\sqrt{n}\delta_n$ is sufficiently small.
Consequently, by the union bound,
\begin{align*}
    \pr\bigg\{
    \max_{j\in[p]}
    \bigg|
    \frac{\sqrt{n}(\tilde{\beta}_j-\beta_j^*)
    \hat{H}_{j|-j}}{\hat{\sigma}_j}
    -Z_j
    \bigg|
    \geq \bar{C}\sqrt{n}\delta_n
    \bigg\}
    \lesssim p^{-\bar{c}},
\end{align*}
where $\bar{c}=c_0-1>0$ and $\bar{C}>0$ may depend
on $\bar{c}$ but is independent of $n$ and $p$.
This completes the proof of Proposition~\ref{prop.Bahadur}.\hfill$\Box$

\section{Proof of Theorem \ref{thm.normal}}

By Proposition \ref{prop.Bahadur} and \eqref{def.Z} in Section \ref{sec.pre}, it suffices to bound the normal approximation error for 
\begin{align*}
    Z_j^{(m)} = \frac{1}{\sqrt{n}}\sum_{i\in\cI_m}\frac{(1-E)( \zeta_{i,j})}{\sigma_j}\,,
\end{align*}
where $ \zeta_{i,j}  = \{   \tau-\bar{K}_h(  -\varepsilon_i)  \}\bx_i^{\T} {\bv}_{(j)} $ and ${\sigma}_{j}^2  =  {\rm var}(\zeta_{i,j} )$. Applying the Berry--Esseen inequality (see, e.g., \citealpS{Shevtsova2014S}) to the independent random variables $\{  {n}^{-1/2}\sigma_j^{-1}
(1-E)(\zeta_{i,j})  \}_{i\in\cI_m}$ yields 
 \begin{align}\label{berryesseen}
     \sup_{x\in\RR} \big|  \pr\{  Z_j^{(m)}  \leq x  \} - \Phi(x)  \big| \lesssim  \frac{ \sum_{i\in\cI_m} E\{| (1-E) (\zeta_{i,j}) |^3\} } {n^{3/2}\sigma_j^3} 
 \end{align}
 for any $j\in [p]$. Condition 1(b) implies $ E\{|\bx_i^{\T} {\bv}_{(j)}|^3\}\lesssim \|{\bv}_{(j)}\|_{\bSigma}^3$ for any $i\in \cI_m$. Moreover, $| \tau-\bar{K}_h(  -\varepsilon_i) | \leq \max\{\tau,1-\tau\}\leq  1$ and  
  \begin{align*} 
    E\{ |\tau-\bar{K}_h(  -\varepsilon_i)|^2 \,|\,\bx_i\}  = \tau^2+E\{ | \bar{K}_h(  -\varepsilon_i)|^2 \,|\,\bx_i\}-2\tau [E\{  \bar{K}_h(  -\varepsilon_i)  \,|\,\bx_i\}-\tau]-2\tau^2\,. 
\end{align*}
Under Conditions~1(a) and~1(d), equations~(G.4) and (G.5) in the supplementary material of \citeS{tan2022highS} yield $E\{ | \bar{K}_h(  -\varepsilon_i)|^2 \,|\,\bx_i\} \leq \tau + l_0\kappa_2 h^2$ and $|E\{  \bar{K}_h(  -\varepsilon_i)  \,|\,\bx_i\} -\tau | \leq l_0\kappa_2 h^2/2$, which imply
$ E \{ |  \tau-\bar{K}_h(  -\varepsilon_i) |^3 | \bx_i\} \leq   E \{ |  \tau-\bar{K}_h(  -\varepsilon_i) |^2 | \bx_i\} \leq \tau(1-\tau)+(\tau+1) l_0 \kappa_2 h^2$ for any $i\in\cI_m$. Combining these bounds gives 
\begin{align}\label{thrid.upper}
      E\{| (1-E) (\zeta_{i,j}) |^3\} \lesssim   E(|  \zeta_{i,j} |^3) \lesssim   \|{\bv}_{(j)}\|_{\bSigma}^3\{\tau(1-\tau)+(\tau+1) l_0 \kappa_2 h^2  \} \lesssim  \|{\bv}_{(j)}\|_{\bSigma}^3 
\end{align}
 for any $i\in\cI_m$, where the last inequality follows from $h\lesssim  1$. Lemma \ref{lemma.sigma.cons} gives $\sigma_j^2 \gtrsim   \|\bv_{(j)}\|_{\bSigma}^2$. Substituting this lower bound and \eqref{thrid.upper} into \eqref{berryesseen} yields
\begin{align} \label{normalZ}
    \sup_{x\in\RR} \big|  \pr\{  Z_j^{(m)}  \leq x  \} - \Phi(x)  \big|  \lesssim  n^{-1/2}\,. 
\end{align} 
Finally, combining Proposition \ref{prop.Bahadur} with the standard normal anti-concentration inequality and using $n^{-1/2}\lesssim \sqrt{n}\delta_n$, we obtain
\begin{align*} 
    \sup_{x\in\RR}\bigg| \pr\bigg\{ 
             \frac{\sqrt{n} ( \tilde{\beta}_j^{(m)}-\beta_j^*)\hat{H}^{(m)}_{j|-j}}{\hat{\sigma}^{(m)}_{j}}  \leq x\bigg\}-\Phi(x) \bigg|  \lesssim       \sqrt{n}\delta_n + p^{-\bar{c}} 
\end{align*} 
for any $j\in [p]$ and $m\in[M]$. This completes the proof of Theorem \ref{thm.normal}.  \hfill$\Box$

\section{Proof of Theorem \ref{thm.convergence}}

Write $\check{\cS}= \{1\}\cup \hat\cS$ and $\check{\cS}^{\rm c}=[p]\setminus \check{\cS}$.  Observe that 
 \begin{align}\label{target}
     \|\cT_{\gamma}(\hat{\bbeta})-\bbeta^*\|_2^2=  \|\{\cT_{\gamma}(\hat{\bbeta})\}_{\check{\cS}}-\bbeta^*_{\check{\cS}}\|_2^2  +  \|\bbeta^*_{\check{\cS}^{\rm c}}\|_2^2  \,.
 \end{align}
The first term on the right-hand side of \eqref{target}
represents the estimation error over $\check{\cS}$,
while the second captures the screening error
arising from missed signals. For each $j\in[p]$, write $\bar{\beta}_j = M^{-1}\sum_{m=1}^M \tilde{\beta}_j^{(m)}$, where $\tilde{\beta}_j^{(m)}$ is defined in \eqref{eq:debiased-beta-j}. The following two lemmas establish the bounds needed for the convergence rate. Their proofs appear
in Sections~\ref{proof.term1} and \ref{proof.term2},
respectively. Throughout the proof, we choose the constant
$\bar c$ in Proposition~\ref{prop.Bahadur}
to be sufficiently large, so that the
probability remainder $O(p^{-\bar c})$
is negligible relative to the relevant
normal tail probabilities. This is possible
since the proposition holds for any
prespecified $\bar c>0$.

\begin{lemma}\label{term1}
    Suppose the assumptions of Proposition {\rm \ref{prop.Bahadur}} hold. Assume further that $M\leq p^{\iota}$ for some fixed $\iota>0$. Provided that $\sqrt{n}\delta_n = o(1)$, we have 
    \begin{align*}
    \pr\bigg\{  \max_{j\in [p]} | \bar{\beta}_j-\beta_j^* | \leq C\bigg(\sqrt{\frac{\log p}{N}}+ \delta_n\bigg) \bigg\} \geq 1-O( p^{- \check{c}})\,,
\end{align*}
where $\check{c}=\bar{c}-\iota-1>0$.  
\end{lemma}

\begin{lemma}\label{term2}
    Suppose the assumptions of Proposition {\rm \ref{prop.Bahadur}} hold. Take the local screening threshold $\rho$ in
\eqref{eq:local.screening} to satisfy
$0\leq\rho\leq\sqrt{2\log p}$, and choose the
voting threshold $v$ in \eqref{eq:global.screening}
as $v\asymp\sqrt{n}\delta_n\phi(\rho)$. Then, provided that $\sqrt{n}\delta_n=o(1)$, the aggregate set $\hat\cS$ defined in \eqref{eq:global.screening} satisfies 
\begin{align*}
    E\big(\|\bbeta^*_{\check{\cS}^{\rm c}}\|_2^2\big) \lesssim s \bigg[ \delta_n^2 + \frac{1}{n}\min\bigg\{ \frac{1  }{M (1+\rho)\phi (\rho)} ,1+\rho^2 \bigg\} \bigg]\,.
\end{align*}
\end{lemma}

By construction, $\hat{\beta}_j=\bar{\beta}_j$ for
$j\in \check{\cS}$, and $\hat{\beta}_j=0$ otherwise.  Recall that $\cS = {\rm supp}(\bbeta^*_{-1})$ with $s=|\cS|$. For each $j\in \check{\cS}\cap (\{1\}\cup\cS) ^{\rm c}$, we have
\begin{align*}
    |\{\cT_{\gamma}(\hat{\bbeta})\}_{j}-\beta^*_{j}|^2 = |\{\cT_{\gamma}(\hat{\bbeta})\}_{j} |^2 = \hat{\beta}_j^2\mathbf 1\{|\hat\beta_j|\ge\gamma\}\,.
\end{align*}
Moreover, for any $j\in[p]$,
 \begin{align*} 
     |\{\cT_{\gamma}(\hat{\bbeta})\}_{j}-\beta^*_{j}| \leq |\hat{\beta}_j-\beta^*_{j}|+|\beta^*_{j}|\mathbf{1}(|\hat{\beta}_j|<\gamma) \leq 2|\hat{\beta}_j-\beta^*_{j}| + \gamma\,,
 \end{align*} 
which yields 
\begin{align*} 
  \|\{\cT_{\gamma}(\hat{\bbeta})\}_{\check{\cS}}-\bbeta^*_{\check{\cS}}\|_2^2 = &~\sum_{j \in \check{\cS}\cap (\{1\}\cup\cS)} |\{\cT_{\gamma}(\hat{\bbeta})\}_{j}-\beta^*_{j}|^2 + \sum_{j \in \check{\cS}\cap (\{1\}\cup\cS)^{\rm c}}\hat{\beta}_j^2\mathbf 1\{|\hat\beta_j|\ge\gamma\}\nonumber\\
   \lesssim &~ s\gamma^2 +  \sum_{j \in    \{1\} \cup \cS } | \bar{\beta}_{j}-\beta^*_{j}|^2  +   \sum_{j \in (\{1\}\cup\cS)^{\rm c}}\bar{\beta}_j^2\mathbf 1\{|\bar\beta_j|\ge\gamma\}   \,.
\end{align*}
Define the event $\cE_n=
[
\max_{j\in[p]}
|\bar\beta_j-\beta_j^*|\leq  C\{\sqrt{(\log p)/{N}} +\delta_n\}
]$. Since $\bar{c}>0$ is a sufficiently large fixed constant,
it follows from Lemma \ref{term1} that $\pr(\cE_n^{\rm c})=o(1)$. On $\cE_n$,
\begin{align*}
 \sum_{j \in    \{1\} \cup \cS } | \bar{\beta}_{j}-\beta^*_{j}|^2
\lesssim  s \bigg(  \frac{ \log p}{N} +\delta_n^2 \bigg)\,.
\end{align*}
Moreover, choosing
$\gamma=\tilde{C}\{\sqrt{(\log p)/N}+\delta_n\}$
with a sufficiently large constant $\tilde{C}>0$ yields
\[
\sum_{j \in (\{1\}\cup\cS)^{\rm c}}\bar{\beta}_j^2\mathbf 1\{|\bar\beta_j|\ge\gamma\} =0
\]
on $\cE_n$.  Consequently,
\begin{align}\label{rate.I}
    \|\{\cT_{\gamma}(\hat{\bbeta})\}_{\check{\cS}}-\bbeta^*_{\check{\cS}}\|_2^2 =O_{\rm p}\bigg \{s \bigg(  \frac{ \log p}{N} +\delta_n^2 \bigg) \bigg\}\,.
\end{align} 
This, together with \eqref{target} and Lemma \ref{term2}, yields 
\[
\|\cT_{\gamma}(\hat{\bbeta})-\bbeta^*\|_2^2=  O_{\rm p}\bigg(s \bigg[\frac{ \log p}{N} + \delta_n^2 + \frac{1}{n}\min\bigg\{ \frac{1  }{M (1+\rho)\phi (\rho)} ,1+\rho^2 \bigg\} \bigg]\bigg)\,.
\]
This completes the proof of Theorem \ref{thm.convergence}.  \hfill$\Box$

\subsection{Proof of Lemma \ref{term1}}\label{proof.term1}

By Proposition \ref{prop.Bahadur}, we have
\[
\tilde{\beta}_j^{(m)}-\beta_j^* =  \frac{\hat{\sigma}_j^{(m)}}{\sqrt{n}\hat{H}^{(m)}_{j|-j}} \{ Z_j^{(m)} + R_j^{(m)}  \}\,,
\]
where 
\[
Z_j^{(m)} = \frac{1}{\sqrt{n}}\sum_{i\in \cI_m} \frac{ (1-E)[ \{\tau-\bar{K}_h(-\varepsilon_i)\}\bx_i^{\T}\bv_{(j)} ]   }{\sigma_j} \,, 
\]
and the remainder term $R_j^{(m)}$ satisfies
\[
\pr\bigg\{\max_{j\in[p]}|R_j^{(m)}|>\bar{C}\sqrt{n}\delta_n\bigg\}\lesssim p^{-\bar{c}}\,.
\]
Therefore, 
\begin{align*}
    \bar{\beta}_j-\beta_j^* = \frac{1}{M}\sum_{m=1}^M \{\tilde{\beta}_j^{(m)}-\beta_j^*\}= &~ \frac{1}{N}\sum_{i=1}^N \tilde{H}_{j|-j}^{-1} (1-E)[ \{\tau-\bar{K}_h(-\varepsilon_i)\}\bx_i^{\T}\bv_{(j)} ] \\
    &~+  \frac{1}{M\sqrt{n}}\sum_{m=1}^M \bigg(  \frac{\hat{\sigma}_j^{(m)}}{ \hat{H}^{(m)}_{j|-j}} - \frac{ {\sigma}_j }{ \tilde{H}_{j|-j}} \bigg) Z_j^{(m)} + \frac{1}{M\sqrt{n}}\sum_{m=1}^M \frac{\hat{\sigma}_j^{(m)}}{\hat{H}^{(m)}_{j|-j}}  R_j^{(m)}\,,
\end{align*}
where $ \tilde{H}_{j|-j} = E\{f_{\varepsilon_i|\bx_i}(0)
    x_{i,j}\bx_i^{\T}{\bv}_{(j)}   \}$. By the definition of ${\bv}_{(j)}$ and
$\tilde{\bH}$, we have  $\tilde{H}_{j|-j} = {\bv}_{(j)}^{\T}\tilde{\bH}{\bv}_{(j)}   $. Since $\|{\bv}_{(j)}  \|_2 \geq 1$, $\tilde{H}_{j|-j} \geq \lambda_{\min}(\tilde{\bH})$. By Conditions 1(a) and 1(b), we further have $\lambda_{\min}(\tilde{\bH})\geq f_l\gamma_p$, which implies $\tilde{H}_{j|-j} \geq f_l\gamma_p$. For each $j\in[p]$, since $\|\{\tau-\bar{K}_h(-\varepsilon_i)\}\bx_i^{\T} {\bv}_{(j)}\|_{\psi_1} \leq C$, provided that $\log p\ll N$,  Bernstein's inequality yields 
\begin{align}\label{term.lead}
    \bigg| \frac{1}{N}\sum_{i=1}^N \tilde{H}_{j|-j}^{-1} (1-E)[ \{\tau-\bar{K}_h(-\varepsilon_i)\}\bx_i^{\T}\bv_{(j)} ] \bigg| \lesssim \sqrt{\frac{\log p}{N}} 
\end{align}
with probability at least $1-O(p^{-\bar{c}})$, where $\bar{c}$ is a large constant as in Proposition \ref{prop.Bahadur}. Note that  
\begin{align*} 
   \big|\hat{H}_{j|-j}^{(m)}-\tilde{H}_{j|-j}\big| = &~  \bigg|\frac{1}{n}\sum_{i\in\cI_m}
    K_h(-r_i(\hat{\bbeta}_h^{(m)}))
    x_{i,j}\bx_i^{\T}\hat{\bv}_{(j)}^{(m)} -  E\{f_{\varepsilon_i|\bx_i}(0)
    x_{i,j}\bx_i^{\T}{\bv}_{(j)}   \}\bigg|\nonumber\\
    \leq &~ \underbrace{\bigg|\frac{1}{n}\sum_{i\in\cI_m}
    \{ K_h(-r_i(\hat{\bbeta}_h^{(m)}))
    x_{i,j}\bx_i^{\T}\hat{\bv}_{(j)}^{(m)} -   K_h(-r_i(\hat{\bbeta}_h^{(m)}))
    x_{i,j}\bx_i^{\T}{\bv}_{(j)}  \}\bigg|}_{{\rm V}_1}\nonumber\\
    &~ +\underbrace{ \bigg|\frac{1}{n}\sum_{i\in\cI_m}
    K_h(-r_i(\hat{\bbeta}_h^{(m)}))
    x_{i,j}\bx_i^{\T}{\bv}_{(j)}  -  E\{f_{\varepsilon_i|\bx_i}(0)
    x_{i,j}\bx_i^{\T}{\bv}_{(j)}   \}\bigg|}_{{\rm V}_2}\,.
\end{align*}
Using \eqref{tilde.w} and \eqref{tilde.w.1} and the argument for \eqref{w.solution} in Section \ref{proof.lemma.w.cons}, we obtain ${\rm V}_2 \leq \Delta$ with probability at least $1-O(p^{-\bar{c}})$. Moreover,  by \eqref{grad.hess}, we have
\begin{align*}
    {\rm V}_1 \leq &~ \|\hat{\bv}_{(j)}^{(m)} -{\bv}_{(j)}\|_1 \cdot \max_{k\in[p]}\bigg| \frac{1}{n}\sum_{i\in\cI_m}
    K_h(-r_i(\hat{\bbeta}_h^{(m)}))
    x_{i,j}x_{i,k} \bigg| \\
    \leq &~ \|\hat{\bv}_{(j)}^{(m)} -{\bv}_{(j)}\|_1 \big\{\|\nabla^2\hat{Q}_h^{(m)}(\hat{\bbeta}_h^{(m)}) - \tilde{\bH}\|_{\infty}+  \|  \tilde{\bH}\|_{\infty}\big\}\,.
\end{align*}
The argument for \eqref{cons.hatQ} in Section \ref{proof.lemma.w.cons} also gives $\|\nabla^2\hat{Q}_h^{(m)}(\hat{\bbeta}_h^{(m)}) - \tilde{\bH}\|_{\infty} \leq \Delta$ with probability at least $1-O(p^{-\bar{c}})$. By Conditions 1(a) and 1(b), we have $ \|  \tilde{\bH}\|_{\infty} \lesssim f_u\upsilon_0^2\gamma_1$. Since $\tilde{s}\Delta \ll 1$, by Lemma \ref{lemma.w.cons}, we obtain ${\rm V}_1 \lesssim \tilde{s}\Delta$ with probability at least $1-O(p^{-\bar{c}})$. Then, for each $(j,m)$,
\begin{align*}
    \big|\hat{H}_{j|-j}^{(m)}-\tilde{H}_{j|-j}\big| \lesssim \tilde{s}\Delta
\end{align*}
with probability at least $1-O(p^{-\bar{c}})$. By Conditions 1(a) and 1(b), we have $\max_{j\in[p]}\tilde{H}_{j|-j}\leq C$. Provided that $\tilde{s}\Delta = o(1)$, the estimator $\hat{H}_{j|-j}^{(m)}$ is bounded away from zero and infinity with probability at least $1-O(p^{-\bar{c}})$. Moreover, since $\sigma_j \leq C$ for any $j\in[p]$, by Lemma \ref{lemma.sigma.cons},  we have
\begin{align*}
    \bigg|\frac{\hat{\sigma}_j^{(m)}}{\hat{H}^{(m)}_{j|-j}}
    -\frac{\sigma_j}{\tilde{H}_{j|-j}} \bigg|
    &=
     \bigg|\frac{\hat{\sigma}_j^{(m)}-\sigma_j}
    {\hat{H}^{(m)}_{j|-j}}
    +
    \frac{\sigma_j(\tilde{H}_{j|-j}-\hat{H}^{(m)}_{j|-j})}
    {\hat{H}^{(m)}_{j|-j}\tilde{H}_{j|-j}} \bigg| \lesssim \tilde{s}\Delta +  \sqrt{\frac{ s^2(\log p)^7}{nh^2}}  \,,
\end{align*}
with probability at least $1-O(p^{-\bar{c}})$. Provided that $\sqrt{n}\delta_n=o(1)$, there exist
constants $0<c_1<C_1<\infty$ such that, for every
$j\in[p]$ and $m\in[M]$,
\begin{align}\label{control1}
    \pr\bigg\{
    c_1\leq
     \frac{\hat{H}_{j|-j}^{(m)}}{\hat{\sigma}_j^{(m)}}
    \leq C_1
    \bigg\}
    \geq 1-O(p^{-\bar{c}})\,.
\end{align}
By \eqref{upper.Z} and Proposition \ref{prop.Bahadur}, we have $|Z_j^{(m)}| \lesssim \sqrt{\log p}$ and $|R_j^{(m)}| \lesssim \sqrt{n}\delta_n$ with probability at least $1-O(p^{-\bar{c}})$, respectively. Therefore, for any $j\in[p]$ and $m\in[M]$,
\begin{align*}
    \bigg|  \frac{1}{ \sqrt{n}}  \bigg(  \frac{\hat{\sigma}_j^{(m)}}{ \hat{H}^{(m)}_{j|-j}} - \frac{ {\sigma}_j }{ \tilde{H}_{j|-j}} \bigg) Z_j^{(m)} + \frac{1}{ \sqrt{n}}  \frac{\hat{\sigma}_j^{(m)}}{\hat{H}^{(m)}_{j|-j}}  R_j^{(m)} \bigg| \lesssim \delta_n 
\end{align*}
with probability at least $1-O(p^{-\bar{c}})$. Since $M\leq p^{\iota}$, a union bound over machines gives, for each $j\in[p]$,
\begin{align*}
     \bigg|  \frac{1}{M\sqrt{n}}\sum_{m=1}^M \bigg(  \frac{\hat{\sigma}_j^{(m)}}{ \hat{H}^{(m)}_{j|-j}} - \frac{ {\sigma}_j }{ \tilde{H}_{j|-j}} \bigg) Z_j^{(m)} + \frac{1}{M\sqrt{n}}\sum_{m=1}^M \frac{\hat{\sigma}_j^{(m)}}{\hat{H}^{(m)}_{j|-j}}  R_j^{(m)}\bigg| \lesssim \delta_n
\end{align*}
with probability at least $1-O\{ p^{-(\bar{c}-\iota)}\}$. Combining \eqref{term.lead} with this bound and applying a union bound over coordinates, we obtain
\begin{align*}
    \pr\bigg\{  \max_{j\in [p]} | \bar{\beta}_j-\beta_j^* | \leq C\bigg(\sqrt{\frac{\log p}{N}}+ \delta_n\bigg) \bigg\} \geq 1-O\{ p^{-(\bar{c}-\iota-1)}\}\,.
\end{align*}
 This completes the proof of Lemma \ref{term1}. \hfill$\Box$

\subsection{Proof of Lemma \ref{term2}}\label{proof.term2}
 
Since $\beta_j^*=0$ for all $j\in (\{1\} \cup \cS)^{\rm c}$,
we have
\begin{align*}
    \|\bbeta^*_{\check{\cS}^{\rm c}}\|_2^2
    &=
    \sum_{j\in(\{1\} \cup \cS)\cap\check{\cS}^{\rm c}}
    (\beta_j^*)^2 =
    \sum_{j\in\cS}
    (\beta_j^*)^2
    \mathbf 1\{j\notin\hat{\cS}\}\,.
\end{align*}
Consequently,
\begin{align}
    E\big(\|\bbeta^*_{\check{\cS}^{\rm c}}\|_2^2\big) 
     =
    \sum_{j\in\cS}
    (\beta_j^*)^2
    \pr\{|V_j|<v \}\,,
    \label{eq:II.expectation}
\end{align}
where $V_j=M^{-1}\sum_{m=1}^M D_j^{(m)}$, with
$D_j^{(m)}=\mathbf{1}\{\hat{\xi}_j^{(m)}\geq\rho\}
-\mathbf{1}\{\hat{\xi}_j^{(m)}\leq-\rho\}$
for $j\in\cS^{(m)}$, and $D_j^{(m)}=0$ otherwise. 
For each $j\in \cS$ and $m\in[M]$, recall the local
Bahadur representation
\begin{align*}
    \hat\xi_j^{(m)}
    =
    Z_j^{(m)}
    +\frac{\sqrt{n}\beta_j^*\hat{H}_{j|-j}^{(m)}}
    {\hat{\sigma}_j^{(m)}}
    +R_j^{(m)}\,.
\end{align*}
By \eqref{control1}, there exist constants
$0<c_1<C_1<\infty$ such that
$c_1\leq\hat{H}_{j|-j}^{(m)}/\hat{\sigma}_j^{(m)}
\leq C_1$
with probability at least $1-O(p^{-\bar{c}})$.
Moreover, the remainder satisfies
$|R_j^{(m)}|\leq\bar{C}\sqrt{n}\delta_n$
with probability at least $1-O(p^{-\bar{c}})$. For $j\in\cS$, let
$s_j=\operatorname{sign}(\beta_j^*)$ and define
\begin{align*}
    a_j
     =
    \sqrt n(c_1|\beta_j^*|-\bar{C}\delta_n) ~~\mbox{and}~~
    b_j
     =
    \sqrt n(C_1|\beta_j^*|+\bar{C}\delta_n)\,.
\end{align*}
Then, with probability at least $1-O(p^{-\bar{c}})$, we have 
\begin{align}
    s_jZ_j^{(m)}+a_j
    \leq
    s_j\hat\xi_j^{(m)}
    \leq
    s_jZ_j^{(m)}+b_j\,.
    \label{eq:II.score.bounds}
\end{align}
We establish two complementary bounds on $E(\|\bbeta^*_{\check{\cS}^{\rm c}}\|_2^2)$. The first applies generally, whereas the second exploits aggregation across
$M$ machines to yield a sharper bound for small $\rho$.
Taking their minimum yields the desired result.

\vspace{5pt}
\noindent\underline{\sl General bound.} Let
\begin{align*}
    r_{n,\rho}
    =
    A_0\bigg(
        \delta_n+\frac{1+\rho}{\sqrt n}
    \bigg)\,,
\end{align*}
where $A_0>0$ is a sufficiently large fixed constant.
Then, whenever $|\beta_j^*|>r_{n,\rho}$, we have
\begin{align*}
    a_j-\rho
    &=
    c_1\sqrt n|\beta_j^*|
    -\bar{C}\sqrt n\delta_n-\rho \geq
    \frac{c_1}{2}\sqrt n|\beta_j^*|\,.
\end{align*}
Define $\cS_{r_{n,\rho}}= \{j\in\cS: |\beta_j^*| > r_{n,\rho}\}$.  Therefore,
\begin{align}
    &\sum_{ j\in\cS }
    (\beta_j^*)^2
    \pr\{|V_j|<v \}  \leq
    sr_{n,\rho}^2 + \sum_{ j\in \cS_{r_{n,\rho}}}
    (\beta_j^*)^2
    \pr\{|V_j|<v \} \,.
    \label{eq:II.robust.small}
\end{align}
Since $ s_j =  \operatorname{sign}(\beta_j^*) \in \{-1,1\}$, we have
\begin{align}\label{eqiv}
    \pr\{|V_j|<v \} = \pr\bigg\{ \bigg| \frac{1}{M}\sum_{m=1}^M D_j^{(m)} \bigg| <v  \bigg\} = \pr\bigg\{ \bigg| \frac{1}{M}\sum_{m=1}^M s_j D_j^{(m)} \bigg| <v  \bigg\}\,.
\end{align}
Since $s_j D_j^{(m)} \in \{-1,0,1\}$, $s_j D_j^{(m)} \geq 1-2 \mathbf{1}\{s_j D_j^{(m)} \neq 1\}$. Consequently, 
\begin{align*}
    \frac{1}{M}\sum_{m=1}^M s_j D_j^{(m)} \geq 1 - \frac{2}{M}\sum_{m=1}^M \mathbf{1}\{s_j D_j^{(m)} \neq 1\}\,.
\end{align*}
Therefore, if $|V_j|< v$, we have $1- 2M^{-1}\sum_{m=1}^M \mathbf{1}\{s_j D_j^{(m)} \neq 1\} < v$. Recall that $v\asymp \sqrt{n}\delta_n \phi(\rho)=o(1)\leq 1/2$. This implies $ \sum_{m=1}^M \mathbf{1}\{s_j D_j^{(m)} \neq 1\} > M/4$. Consequently, 
\begin{align*}
    \pr\{|V_j|<v \} \leq \pr\bigg[ \sum_{m=1}^M \mathbf{1}\{s_j D_j^{(m)} \neq 1\} >  \frac{M}{4}  \bigg]\,.
\end{align*}
Let $q_j = \pr\{s_j D_j^{(m)} \neq 1\}$ be the local voting failure probability.  
Independence across machines implies that
$\{\mathbf{1}\{s_jD_j^{(m)}\neq 1\}\}_{m=1}^M$
are independent Bernoulli variables with parameter $q_j$.
Consequently,
$\sum_{m=1}^M\mathbf{1}\{s_jD_j^{(m)}\neq 1\}$
follows a binomial distribution with parameters
$M$ and $q_j$. It follows that
\begin{align*}
 \pr\bigg[
\sum_{m=1}^M
\mathbf{1}\{s_jD_j^{(m)}\neq 1\}
>\frac{M}{4}
\bigg] = &~
\sum_{k=\lfloor M/4\rfloor+1}^{M}
\binom{M}{k}q_j^k(1-q_j)^{M-k}\\
 \leq &~ 
\sum_{k=\lfloor M/4\rfloor+1}^{M}
\binom{M}{k}q_j^k \leq
q_j^{M/4}
\sum_{k=\lfloor M/4\rfloor+1}^{M}
\binom{M}{k} \leq
2^Mq_j^{M/4}\,,
\end{align*}
where the second inequality follows from
$0\leq q_j\leq 1$ and $k>M/4$,
while the last inequality uses
$\sum_{k=0}^M\binom{M}{k}=2^M$. By \eqref{eq:II.score.bounds}, we have $s_j\hat\xi_j^{(m)} \geq s_jZ_j^{(m)}+a_j 
    $ with probability at least $1-O(p^{-\bar{c}})$. If $s_jD_j^{(m)}\neq 1$, then $s_j\hat\xi_j^{(m)}  \leq \rho$. On the event where \eqref{eq:II.score.bounds} holds, $s_jZ_j^{(m)} \leq -(a_j-\rho) \leq - c_1\sqrt{n}|\beta_j^*|/2$. Since $\max_{j\in\cS}|\beta_j^*|=O(1)$, \eqref{psi1.zeta} and Bernstein's inequality give
\begin{align*}
    q_j \leq \pr\big\{ s_jZ_j^{(m)}   \leq - c_1\sqrt{n}|\beta_j^*|/2 \big\} + O(p^{-\bar{c}}) \leq 2\exp\{-cn(\beta_j^*)^2\}+ O(p^{-\bar{c}})  
\end{align*}
for some constant $c>0$. Consequently, for $j\in \cS_{r_{n,\rho}}$, 
\[
\pr\{|V_j|<v \} \leq C^M \exp\{-cN(\beta_j^*)^2\} + C^M p^{-\bar{c}M/4} \leq \exp\{-c_2N(\beta_j^*)^2\} + C^M p^{-\bar{c}M/4}\,,
\]
where $0<c_2<c$. Since $\bar{c}>0$ is a sufficiently
large fixed constant, we have
$C^Mp^{-\bar{c}M/4}=o(N^{-1})$. Since $xe^{-x}\leq 1$ for all $x\geq0$, we have
\begin{align}\label{rate.high}
    \sum_{ j\in \cS_{r_{n,\rho}}}
    (\beta_j^*)^2
    \pr\{|V_j|<v \} \lesssim \sum_{ j\in \cS_{r_{n,\rho}}}
    (\beta_j^*)^2\exp\{-c_2N(\beta_j^*)^2\} + s C^M p^{-\bar{c}M/4}\lesssim  \frac{s}{N} \,.
\end{align}
Note that $N\geq n$. This, together with \eqref{eq:II.robust.small}, yields
\begin{align}\label{general}
   E\big(\|\bbeta^*_{\check{\cS}^{\rm c}}\|_2^2\big) 
     = \sum_{ j\in\cS }
    (\beta_j^*)^2
    \pr\{|V_j|<v \}  \lesssim s \bigg( \delta_n^2 + \frac{1+\rho^2}{n}  \bigg)\,.
\end{align}

 \vspace{5pt}
\noindent\underline{\sl Refined bound.} For sufficiently large fixed constants
$\tilde{C}_\beta > 2\bar{C}/c_1$ and $K_0$, partition $\cS$ into
\begin{align*}
    \cS_1
    &=
    \{j\in\cS:|\beta_j^*|<\tilde{C}_\beta\delta_n\}\,,\\
    \cS_2
    &=
    \{j\in\cS:|\beta_j^*|\geq \tilde{C}_\beta\delta_n,\,
    a_j\leq K_0(1+\rho)\}\,,\\
    \cS_3
    &=
    \{j\in\cS:|\beta_j^*|\geq \tilde{C}_\beta\delta_n,\,
    a_j>K_0(1+\rho)\}\,.
\end{align*}
Empty subsets contribute zero to the sums below. We have 
\begin{align}\label{div.three}
    E\big(\|\bbeta^*_{\check{\cS}^{\rm c}}\|_2^2\big) 
      = & \sum_{ j\in\cS_1 }
    (\beta_j^*)^2
    \pr\{|V_j|<v \} + \sum_{ j\in\cS_2 }
    (\beta_j^*)^2
    \pr\{|V_j|<v \} + \sum_{ j\in\cS_3 }
    (\beta_j^*)^2
    \pr\{|V_j|<v \}  \nonumber\\
    \lesssim &~ s\delta_n^2 + \sum_{ j\in\cS_2 }
    (\beta_j^*)^2
    \pr\{|V_j|<v \} + \sum_{ j\in\cS_3 }
    (\beta_j^*)^2
    \pr\{|V_j|<v \}\,.
\end{align}
Since $\tilde{C}_\beta > 2\bar{C}/c_1$ and $K_0$ is large, we have, for any $j\in\cS_3$, $ a_j  -\rho \gtrsim \sqrt{n}|\beta_j^*|$. Applying the argument used to establish \eqref{rate.high}, we obtain
\begin{align}\label{rate.S3}
    \sum_{ j\in\cS_3 }
    (\beta_j^*)^2
    \pr\{|V_j|<v \} \lesssim  \frac{s}{N}\,.
\end{align}
It remains to bound $\sum_{ j\in\cS_2 }
    (\beta_j^*)^2
    \pr\{|V_j|<v \}$. Recall that $ s_j =  \operatorname{sign}(\beta_j^*)  \in \{-1,1\}$. By \eqref{eqiv}, we have 
\begin{align*} 
    \pr\{|V_j|<v \}=\pr\{|s_jV_j|<v \}   = \pr\bigg\{ \bigg| \frac{1}{M}\sum_{m=1}^M s_j D_j^{(m)} \bigg| <v  \bigg\} \leq \pr\bigg\{   \frac{1}{M}\sum_{m=1}^M s_j D_j^{(m)}   <v  \bigg\}\,.
\end{align*}
Write $v_j = E\{s_j D_j^{(m)}\}$ and $\mu_j = \pr\{j\in \cS^{(m)}\}$. The following argument is presented for $\rho>0$. When $\rho=0$, $\mu_j=1$ and, for $j\in S_2$, \eqref{normalZ} and \eqref{eq:II.score.bounds} yield $v_j \gtrsim a_j$, and hence $v_j^2/\mu_j\gtrsim a_j^2$.
We have $|s_j D_j^{(m)}-v_j|\leq 2$. Then, if
\begin{align}\label{req.v}
    v_j \geq 2v\,,
\end{align}
Bernstein's inequality \citepS[see, e.g., Theorem~2.8.4 of][]{vershynin2018highS} yields 
\begin{align}\label{tailtt}
   \pr\{|V_j|<v \}\leq  \pr\bigg\{   \frac{1}{M}\sum_{m=1}^M s_j D_j^{(m)}   <v  \bigg\} \leq \exp\bigg\{  -\frac{M(v_j-v)^2}{2\mu_j + 2(v_j-v) }  \bigg\} \leq \exp\bigg( -\frac{Mv_j^2}{12\mu_j}  \bigg)\,,
\end{align}
where the second inequality follows from the fact that ${\rm var}\{s_j D_j^{(m)}\} \leq E[\{D_j^{(m)}\}^2] = \mu_j$, and the third inequality follows from $v\leq v_j/2$ and $v_j \leq \mu_j$.  Write 
\begin{align*}
    q_{j,+} 
     = &~
    \pr\{s_jD_j^{(m)}=1\}=\pr\{s_j \hat{\xi}_j^{(m)}\geq \rho\}\,,\\
    q_{j,-} = &~
    \pr\{s_jD_j^{(m)}=-1\}=\pr\{s_j \hat{\xi}_j^{(m)}\leq  -\rho\}\,.
\end{align*}
Then 
\begin{align*}
    \mu_j =  q_{j,+} + q_{j,-} ~~\mbox{and}~~
    v_j =  q_{j,+} - q_{j,-}\,.
\end{align*}
Since $\tilde{C}_\beta > 2\bar{C}/c_1$, we have, for any $j\in\cS_2$, $ c_1\sqrt{n}|\beta_j^*|/2 \leq a_j \lesssim b_j$. By \eqref{eq:II.score.bounds}, we have
\begin{align*}
    \pr\{ s_j Z_j^{(m)}\geq \rho-a_j \} - O(p^{-\bar{c}})  \leq q_{j,+}  \leq &~ \pr\{s_j Z_j^{(m)}\geq \rho-b_j \} + O(p^{-\bar{c}}) \,,\\
    q_{j,-}  \leq &~ \pr\{s_j Z_j^{(m)}\leq -\rho-a_j \} + O(p^{-\bar{c}})
\end{align*} 
Write $\bar{\Phi}(x)=1-\Phi(x)$. By \eqref{normalZ}, we have 
\begin{align*}
    \pr\{ s_j Z_j^{(m)}\geq \rho-a_j \} \geq \bar{\Phi}(\rho-a_j) - O(n^{-1/2})\,.
\end{align*}
Moreover, when $\rho-a_j>0$, Proposition~4.6
of \citeS{chen2013steinS}, together with
\eqref{psi1.zeta}, yields
\begin{align*}
    \pr\{ s_j Z_j^{(m)}\geq \rho-a_j \} \geq \bar{\Phi}(\rho-a_j) - O\bigg\{ \frac{1+(\rho-a_j)^3}{\sqrt{n}} \bar{\Phi}(\rho-a_j) \bigg\}\,,
\end{align*}
provided that $\rho-a_j =o(n^{1/6})$. 
Combining the two bounds, we obtain
\begin{align*}
    \pr\{ s_j Z_j^{(m)}\geq \rho-a_j \} \geq \bar{\Phi}(\rho-a_j) - O\{\mathcal R_n(\rho-a_j)\}  \,,
\end{align*}
provided that $(\rho-a_j)_{+} =o(n^{1/6})$, where
\[
\mathcal R_n(x)
=
\frac{1+(x_+)^3}{\sqrt n}\bar\Phi(x_+)\,,
\qquad x\in\RR\,.
\]
Similarly, provided that
$(\rho-b_j)_+=o(n^{1/6})$ and
$\rho+a_j=o(n^{1/6})$, we have
\begin{align*}
    \pr\{s_jZ_j^{(m)}\geq\rho-b_j\}
    &\leq
    \bar{\Phi}(\rho-b_j)
    +O\{\mathcal R_n(\rho-b_j)\}\,,\\
    \pr\{s_jZ_j^{(m)}\leq-\rho-a_j\}
    &\leq
    \bar{\Phi}(\rho+a_j)
    +O\{\mathcal R_n(\rho+a_j)\}\,.
\end{align*}
Since $a_j\lesssim 1+\rho$, $b_j>0$, and
$\rho\leq\sqrt{2\log p}$, we have
\begin{align*}
    \max\{(\rho-a_j)_{+},(\rho-b_j)_+,\rho+a_j\}
    \lesssim 1+\rho
    \lesssim \sqrt{\log p}
    =o(n^{1/6})\,,
\end{align*}
where the last relation follows from
$\sqrt n\delta_n=o(1)$, which implies
$\log p=o(n^{1/3})$. Therefore, 
\begin{align}\label{qjbound}
    q_{j,+} \geq &~  \bar{\Phi}(\rho-a_j) - O\{\mathcal R_n(\rho-a_j)+p^{-\bar{c}}\}\,,\nonumber\\
    q_{j,+} \leq &~ \bar{\Phi}(\rho-b_j)
    +O\{\mathcal R_n(\rho-b_j) + p^{-\bar{c}} \}\,,\nonumber\\
    q_{j,-} \leq &~  \bar{\Phi}(\rho+a_j)
    +O\{\mathcal R_n(\rho+a_j)+p^{-\bar{c}}\}\,.
\end{align}

To establish a lower bound for $v_j^2/\mu_j$,
we consider two cases according to whether
$a_j\leq 1/(1+\rho)$ or $1/(1+\rho)<a_j\leq K_0(1+\rho)$. We first consider $a_j\leq 1/(1+\rho)$. By \eqref{qjbound},
\begin{align*} 
    v_j \geq \bar{\Phi}(\rho-a_j)- \bar{\Phi}(\rho+a_j)
    -O\{\mathcal R_n(\rho+a_j) +\mathcal R_n(\rho-a_j)+p^{-\bar{c}}\}\,,\nonumber\\
    \mu_j \leq \bar{\Phi}(\rho-b_j)+ 
    \bar{\Phi}(\rho+a_j)
    +O\{\mathcal R_n(\rho+a_j)+\mathcal R_n(\rho-b_j)+p^{-\bar{c}}\}\,.
\end{align*}
For $u\in [-a_j,a_j]$, since $a_j\leq 1/(1+\rho)$, we have $|\rho u|\leq 1$ and $u^2/2 \leq 1/2$, which implies 
\begin{align*}
    \bar{\Phi}(\rho-a_j)- \bar{\Phi}(\rho+a_j) = \int_{-a_j}^{a_j}\phi(\rho-u)\,\mbox{d}u= \phi(\rho) \int_{-a_j}^{a_j}\exp\bigg( \rho u -\frac{u^2}{2} \bigg)\,\mbox{d}u\asymp a_j \phi(\rho)  \,.
\end{align*} 
Since $\rho\leq \sqrt{2\log p}$, $\phi(\rho) \gtrsim p^{-1}$. Since $n\leq p^{\iota}$ and $\bar{c}>0$ is sufficiently large, we have $p^{-\bar{c}}=o(\sqrt{n}\delta_n \phi(\rho))$.    Moreover, we have 
\begin{align*}
    \mathcal R_n(\rho+a_j) + \mathcal R_n(\rho-a_j) \asymp \frac{(1+\rho)^3}{\sqrt{n}} \bar\Phi(\rho)\lesssim  \frac{(1+\rho)^2}{\sqrt{n}} \phi(\rho)\,.
\end{align*}
Since $a_j \geq c_1\sqrt{n}|\beta_j^*|/2 $ for any $j\in\cS_2$, $\rho\leq \sqrt{2\log p}$, and $\sqrt{n}\delta_n=o(1)$, we have $(1+\rho)^2 = o(n\delta_n)$ and
\begin{align}\label{vjorder}
    v_j \gtrsim a_j \phi(\rho)\,.
\end{align} 
A similar argument gives 
\begin{align*}
    \mu_j \lesssim \bar\Phi(\rho)\,.
\end{align*} 
Then, if $a_j\leq 1/(1+\rho)$, we have 
\begin{align}\label{ratio1}
     \frac{v_j^2}{\mu_j} \gtrsim \frac{a_j^2 \phi^2(\rho)}{\bar\Phi(\rho)}\,.
\end{align}

Next, consider the case $1/(1+\rho)<a_j\leq K_0(1+\rho)$. Since $a_j\leq K_0(1+\rho)$,
$\rho\leq\sqrt{2\log p}$, and
$\bar c$ is sufficiently large
depending on $K_0$, the probability bounds
in \eqref{qjbound} yield
\begin{align*}
    q_{j,+} \geq \{1-o(1)\}\bar{\Phi}(\rho-a_j)~~\mbox{and}~~q_{j,-} \leq \{1+o(1)\}\bar{\Phi}(\rho+a_j)\,.
\end{align*}
Write $g(x)=\phi(x)/\bar\Phi(x)$ for the
standard normal hazard function.
By its monotonicity and the Mills ratio
inequality, $g(\rho)\gtrsim 1+\rho$ for
$\rho\geq0$.  Consequently,
\begin{align*}
    \frac{\bar\Phi(\rho+a_j)}
    {\bar\Phi(\rho-a_j)}
    &=
    \exp\bigg\{
    -\int_{\rho-a_j}^{\rho+a_j}
    g(t)\,\mbox{d}t
    \bigg\} \leq
    \exp\bigg\{
    -\int_\rho^{\rho+a_j}
    g(t)\,\mbox{d}t
    \bigg\}\\
    &\leq
    \exp\{-a_jg(\rho)\} \leq
    \exp\{-c a_j(1+\rho)\} \leq e^{-c}<1\,,
\end{align*}
where $c>0$ is an absolute constant. Therefore, 
\begin{align*}
    \frac{q_{j,-}}{q_{j,+}}
    &\leq
    \{1+o(1)\}
    \frac{\bar\Phi(\rho+a_j)}
    {\bar\Phi(\rho-a_j)} \leq
    \{1+o(1)\}e^{-c}\,.
\end{align*}
Thus, for all sufficiently large $n$,
there exists a fixed constant
$\eta\in(e^{-c},1)$ such that
\begin{align*}
    q_{j,-}\leq\eta q_{j,+}\,.
\end{align*} 
Recalling that
$v_j=q_{j,+}-q_{j,-}$ and
$\mu_j=q_{j,+}+q_{j,-}$, we obtain
\begin{align}
    \frac{v_j^2}{\mu_j}
     =
    \frac{(q_{j,+}-q_{j,-})^2}
    {q_{j,+}+q_{j,-}} \geq
    \frac{(1-\eta)^2}{1+\eta}
    q_{j,+} \gtrsim
    \bar\Phi(\rho-a_j)\,.
    \label{eq:II.ratio.case2.first}
\end{align}
If $a_j\leq\rho$, the Mills ratio inequality yields
\begin{align*}
     \frac{a_j^2\phi^2(\rho)}
    {\bar\Phi(\rho)\bar\Phi(\rho-a_j)} \lesssim &~
    a_j^2(1+\rho)(1+\rho-a_j)
    \frac{\phi(\rho)}{\phi(\rho-a_j)}\\
     = &~ 
    a_j^2(1+\rho)(1+\rho-a_j)
    \exp\bigg(-\rho a_j+\frac{a_j^2}{2}\bigg)\\
    \leq &~ 
    a_j^2(1+\rho)^2
    \exp\left(-\frac{\rho a_j}{2}\right)
    \lesssim 1\,.
\end{align*}
If instead $a_j>\rho$, then
$\bar\Phi(\rho-a_j)\geq1/2$. Since
$a_j\leq K_0(1+\rho)$, the Mills ratio
inequality yields
\begin{align*}
    \frac{a_j^2\phi^2(\rho)}
    {\bar\Phi(\rho)\bar\Phi(\rho-a_j)}
    &\lesssim
    a_j^2(1+\rho)\phi(\rho) \lesssim
    K_0^2(1+\rho)^3\phi(\rho)
    \lesssim 1\,.
\end{align*} 
Therefore, by \eqref{eq:II.ratio.case2.first}, we obtain 
\begin{align}\label{ratio2}
    \frac{v_j^2}{\mu_j} \gtrsim \frac{a_j^2\phi^2(\rho)}
    {\bar\Phi(\rho) } \,. 
\end{align}
Moreover, the preceding probability
bounds imply
\begin{align*}
    v_j
     =q_{j,+}-q_{j,-} \geq(1-\eta)q_{j,+} \gtrsim\bar\Phi(\rho-a_j)\,.
\end{align*}
If $a_j\leq\rho$, the Mills ratio
inequality gives
\begin{align*}
    \frac{a_j\phi(\rho)}
    {\bar\Phi(\rho-a_j)}
    \lesssim
    a_j(1+\rho-a_j)
    \exp\bigg(-\rho a_j+\frac{a_j^2}{2}\bigg) \leq
    a_j(1+\rho)
    e^{-\rho a_j/2}
    \lesssim 1\,.
\end{align*}
If $a_j>\rho$, then
\begin{align*}
    \frac{a_j\phi(\rho)}
    {\bar\Phi(\rho-a_j)}
     \leq2a_j\phi(\rho) \leq2K_0(1+\rho)\phi(\rho)
    \lesssim 1\,.
\end{align*}
Therefore, we obtain
\begin{align}
    v_j
    \gtrsim
    \bar\Phi(\rho-a_j)
    \gtrsim
    a_j\phi(\rho) \,.\label{eq:II.mean.case2}
\end{align}

Recall that, for any $j\in\cS_2$, $a_j
\geq  {c_1} \sqrt n|\beta_j^*|/2
\geq  c_1\tilde{C}_\beta \sqrt n\delta_n/2$. Under $\sqrt n\delta_n=o(1)$,
combining \eqref{vjorder}, \eqref{ratio1},
\eqref{ratio2}, and \eqref{eq:II.mean.case2}
yields
\begin{align*}
    v_j \gtrsim a_j\phi(\rho) \geq \frac{c_1\tilde{C}_\beta}{2}\sqrt n\delta_n\phi(\rho)~~\mbox{and}~~\frac{v_j^2}{\mu_j} \gtrsim \frac{a_j^2\phi^2(\rho)}
    {\bar\Phi(\rho) } \gtrsim  \frac{n(\beta_j^*)^2 \phi^2(\rho)}
    {\bar\Phi(\rho) }\,.
\end{align*}
Then \eqref{req.v} holds for some sufficiently large constant $\tilde{C}_\beta>0$, as $v\asymp \sqrt n\delta_n\phi(\rho)$. By \eqref{tailtt},
\begin{align*} 
   \sum_{ j\in\cS_2 }
    (\beta_j^*)^2 \pr\{|V_j|<v \}\leq  \sum_{ j\in\cS_2 }
    (\beta_j^*)^2\exp\bigg( -C \frac{N(\beta_j^*)^2 \phi^2(\rho)}
    {\bar\Phi(\rho) }  \bigg) \lesssim \frac{s\bar\Phi(\rho) }{N \phi^2(\rho)} \lesssim \frac{s  }{N (1+\rho)\phi (\rho)}\,. 
\end{align*}
This, together with \eqref{div.three} and \eqref{rate.S3}, yields
\begin{align}\label{refined}
    E\big(\|\bbeta^*_{\check{\cS}^{\rm c}}\|_2^2\big) \lesssim s \bigg\{ \delta_n^2 +  \frac{1  }{N (1+\rho)\phi (\rho)}\bigg\}\,.
\end{align}

\vspace{5pt}
\noindent\underline{\sl Combining the two bounds.} Combining \eqref{general} and \eqref{refined},
and recalling that $N=nM$, we obtain 
\begin{align*}
    E\big(\|\bbeta^*_{\check{\cS}^{\rm c}}\|_2^2\big) \lesssim s \bigg[\delta_n^2 + \frac{1}{n}\min\bigg\{ \frac{1  }{M (1+\rho)\phi (\rho)} ,1+\rho^2 \bigg\}\bigg] \,.
\end{align*}
This completes the proof of Lemma \ref{term2}. \hfill$\Box$

\section{Proof of Proposition \ref{prop.screening.size}}

Write $\pi_{\cS^{\rm c}} = \max_{m\in[M]}\max_{j\in\cS^{\rm c}}\pr\{j\in \cS^{(m)} \}$, where $\cS^{\rm c}= \{2,\ldots,p\}\!\setminus\! \cS$. 
Then, for each $m\in [M]$,
\begin{align} \label{E_Sm_1}
E\{ |\cS^{(m)}| \} = \sum_{j\in \cS} \pr\{j\in \cS^{(m)} \} + \sum_{j\in \cS^{\rm c}} \pr\{j\in \cS^{(m)} \} \leq s + (p-s) \pi_{\cS^{\rm c}}\,. 
\end{align}
For notational simplicity, write $\tilde{\rho} = \rho - \bar{C}\sqrt{n}\delta_n$, where $\bar{C}$ is specified in Proposition \ref{prop.Bahadur}. Then, Proposition \ref{prop.Bahadur} gives, for any $j\in\cS^{\rm c}$, 
\begin{align*}
    \pr\{j\in \cS^{(m)} \} = \pr\{ |\hat\xi_j^{(m)}|\geq \rho \} \leq \pr\big\{
     |Z_j^{(m)}  |
    \geq \tilde{\rho}
    \big\}+O(p^{-\bar{c}})\,.
\end{align*}
Recall that $Z_j^{(m)} =n^{-1/2}\sum_{i\in\cI_m} \{ \zeta_{i,j}-E(\zeta_{i,j}) \}/{\sigma_j}$ is a sum of independent, centered random variables, and the sum has unit variance. By Proposition 4.6 of
\citeS{chen2013steinS} and \eqref{psi1.zeta}, for any $\tilde{\rho}>0$
satisfying $\tilde{\rho}=o(n^{1/6})$, we have
\begin{align*}
    \pr\big\{|Z_j^{(m)}|\geq\tilde{\rho}\big\}
    \leq 2\{1+o(1)\}\bar\Phi(\tilde{\rho})\,.
\end{align*}
Note that $\pr\big\{|Z_j^{(m)}|
\geq \tilde{\rho}\big\} =1$ if $\tilde{\rho} \leq 0$. Therefore, provided that
$(\tilde{\rho})_+=o(n^{1/6})$,  we obtain
\begin{align*}
\pi_{\cS^{\rm c}}
\leq 
2\{1+o(1)\}
\bar\Phi\{
(\tilde{\rho})_+
\}
+O(p^{-\bar{c}}) 
\,.
\end{align*}
Since $\rho\leq \sqrt{2\log p}$, we have $\bar\Phi\{
(\tilde{\rho})_+
\}\gtrsim (p\log p)^{-1}$ and $(\tilde{\rho})_+=o(n^{1/6})$ holds whenever $\log p = o(n^{1/3})$. Consequently, 
\begin{align}\label{upper.pisc}
    \pi_{\cS^{\rm c}} \lesssim \bar\Phi\{
(\tilde{\rho})_+
\}\,.
\end{align} 
Since $\sqrt{n\log p}\delta_n=O(1)$, we have $\bar\Phi\{
(\tilde{\rho})_+
\}\asymp \bar\Phi 
({\rho})$. By \eqref{E_Sm_1}, we obtain
\begin{align}\label{E_Sm_2}
\max_{m\in[M]}E\{ |\cS^{(m)}| \}  \lesssim s + (p-s)\bar\Phi
({\rho})
\,.
\end{align}
This completes the first part of Proposition \ref{prop.screening.size}.

Since $\hat\cS \subseteq \cup_{m=1}^M \cS^{(m)}$, it follows from \eqref{upper.pisc} that
\begin{align}\label{E_cS_1}
    E(|\hat\cS|)   \leq s + \sum_{j\in\cS^{\rm c}}\pr(j \in \hat{\cS}) \leq s + \sum_{j\in\cS^{\rm c}}\pr\{j \in \cup_{m=1}^M \cS^{(m)}\} \lesssim s+ (p-s)M\bar\Phi\{
(\tilde{\rho})_+
\}\,.
\end{align} 
Alternatively, the construction of $\hat\cS$ implies 
\begin{align*}
    E(|\hat\cS|) = \sum_{j=2}^p \pr(|V_j|\geq v) \leq s + \sum_{j\in\cS^{\rm c}}\pr(|V_j|\geq v)\,,
\end{align*}
where $V_j = M^{-1}\sum_{m=1}^M D_{j}^{(m)}$. Note that $\max_{m\in[M]} |D_{j}^{(m)} - E\{D_{j}^{(m)}\}| \leq 2$ and  ${\rm var}\{D_{j}^{(m)}\} \leq E\{|D_{j}^{(m)}|^2\} \leq  \pr\{j\in\cS^{(m)}\} \leq \pi_{\cS^{\rm c}}$ for any $j\in\cS^{\rm c}$. Write $v_{\cS^{\rm c}}(\rho) = \max_{j\in\cS^{\rm c}}\max_{m\in[M]}|E\{D_{j}^{(m)}\}|$. Then, provided that $v > v_{\cS^{\rm c}}(\rho)$, Bernstein's inequality for bounded random variables yields
\begin{align*} 
    \pr(|V_j|\geq v) \leq &~ \pr\bigg( \bigg|\sum_{m=1}^{M}[ D_{j}^{(m)} - E\{D_{j}^{(m)}\} ]\bigg| \geq  M\{v- v_{\cS^{\rm c}}(\rho) \}   \bigg)\nonumber\\
     \leq &~ 2\exp\bigg[ - \frac{ M \{v- v_{\cS^{\rm c}}(\rho) \} ^2 } {2  \pi_{\cS^c} + 2 \{v- v_{\cS^{\rm c}}(\rho) \}  }  \bigg]\,.
\end{align*}
Consequently, 
\begin{align}\label{E_cS_2}
    E(|\hat\cS|) \lesssim  s + (p-s)\exp\bigg[ - \frac{ M \{v- v_{\cS^{\rm c}}(\rho) \} ^2 } {2  \pi_{\cS^c} + 2 \{v- v_{\cS^{\rm c}}(\rho) \}  }  \bigg]\,.
\end{align} 
Recall that
$D_j^{(m)}=\mathbf 1\{\hat\xi_j^{(m)}\geq\rho\}
-\mathbf 1\{\hat\xi_j^{(m)}\leq-\rho\}$.
For any $j\in\cS^{\rm c}$ and $m\in[M]$,
Proposition \ref{prop.Bahadur} implies
\begin{align*}
E\{D_j^{(m)}\}
\leq &~
\pr\{Z_j^{(m)}\geq\tilde\rho\}
-\pr\{-Z_j^{(m)}\geq
\rho+\bar C\sqrt n\delta_n\}
+O(p^{-\bar c}),
\end{align*}
and
\begin{align*}
-E\{D_j^{(m)}\}
\leq &~
\pr\{-Z_j^{(m)}\geq\tilde\rho\}
-\pr\{Z_j^{(m)}\geq
\rho+\bar C\sqrt n\delta_n\}
+O(p^{-\bar c}).
\end{align*}
Since both $Z_j^{(m)}$ and $-Z_j^{(m)}$
satisfy the same normal approximation bounds,
Proposition 4.6 of \citeS{chen2013steinS}
and \eqref{normalZ} yield
\begin{align*}
|E\{D_j^{(m)}\}|
\leq &~
\bar\Phi(\tilde\rho)
-\bar\Phi(\rho+\bar C\sqrt n\delta_n) +
O\bigg[
\frac{1+(\tilde\rho)_+^3}{\sqrt n}
\bar\Phi\{(\tilde\rho)_+\}
\bigg]
+O(p^{-\bar c})\,.
\end{align*}
Moreover,
\begin{align*}
&\bar\Phi(\tilde\rho)
-\bar\Phi(\rho+\bar C\sqrt n\delta_n) \leq
2\bar C\sqrt n\delta_n
\phi\{(\tilde\rho)_+\}\,.
\end{align*}
By the Mills ratio inequality and the conditions
$\log p=o(n^{1/3})$ and
$\delta_n\gtrsim n^{-3/5}$, we have
\begin{align*}
\frac{1+(\tilde\rho)_+^3}{\sqrt n}
\bar\Phi\{(\tilde\rho)_+\}
=o\big[
\sqrt n\delta_n\phi\{(\tilde\rho)_+\}
\big].
\end{align*}
Consequently, for sufficiently large $\bar c$,
\begin{align*}
v_{\cS^{\rm c}}(\rho)
\leq 3\bar C\sqrt n\delta_n
\phi\{(\tilde\rho)_+\}.
\end{align*}
Then, choosing $v=v(\rho)=C\sqrt n\delta_n 
    \phi\{(\tilde\rho)_+\} \asymp \sqrt n\delta_n 
    \phi(\rho)$ for some constant $C>3\bar{C}$, and combining \eqref{upper.pisc} with \eqref{E_cS_2}, we obtain 
 \begin{align*}
     E(|\hat\cS|) \lesssim  s + (p-s)\exp\big[ -  CN \delta_n^2 \phi\{(\tilde\rho)_+\} \{1+ (\tilde\rho)_+\} \big]\,.
 \end{align*} 
Since $\sqrt{n\log p}\delta_n=O(1)$, by \eqref{E_cS_1}, we have
 \begin{align*}
    E(|\hat\cS|) \lesssim s+ (p-s) \min \big \{ M \bar\Phi
({\rho})
,  \exp\{ -  CN \delta_n^2 \phi(\rho) (1+ \rho) \}  \big \} \,.
 \end{align*}
This proves the second part and completes the proof of Proposition \ref{prop.screening.size}. \hfill$\Box$

\section{Proofs of technical lemmas in Section \ref{sec.lemmas} }\label{sec.lemma.proof}

\subsection{Proof of Lemma \ref{lemma.triangle}}\label{proof.lemma.triangle}

It suffices to prove Lemma \ref{lemma.triangle} under $\zeta\in(0,1)$, since the desired result follows directly from the
triangle inequality for the $\psi_\zeta$-Orlicz norm when $\zeta\geq1$.  Throughout the proof, let $C_\zeta>0$ denote a generic constant depending only on $\zeta$, whose value may vary from line to line.
Write $K_i  = \|  X_i\|_{\psi_{\zeta}}$ and $\tilde{K} = \sum_{i=1}^n K_i$. Note that $K_i = \||X_i|^{\zeta}\|_{\psi_1}^{1/\zeta}$. The equivalence in Proposition 2.7.1 of \citeS{vershynin2018highS} gives, for any $q\geq 1$,
\begin{align*}
    \|X_i\|_{L_q} = \||X_i|^{\zeta}\|_{L_{q/\zeta}}^{1/\zeta} \leq C_{\zeta} \{ ( q/\zeta) K_i^{\zeta}\}^{1/\zeta}  =C_{\zeta} ( q/\zeta)^{1/\zeta} K_i\,.
\end{align*}
Therefore, by Minkowski's inequality, we have $\|  \sum_{i=1}^n  X_i \|_{L_q} \leq \sum_{i=1}^n   \|    X_i \|_{L_q} \leq C_{\zeta} ( q/\zeta)^{1/\zeta} \tilde{K}$ for any $q\geq 1$, which implies  
\begin{align*}
    \bigg\| \bigg| \sum_{i=1}^n  X_i \bigg|^{\zeta}   \bigg\|_{L_q} = \bigg\|  \sum_{i=1}^n  X_i \bigg\|_{L_{q\zeta}}^{\zeta} \leq C_{\zeta} q \tilde{K}^{\zeta}  
\end{align*}
 for any $q \geq 1/\zeta$. In addition, for $1\leq  q < 1/\zeta$, we also have $\|  | \sum_{i=1}^n  X_i |^{\zeta}    \|_{L_q} \leq \|  | \sum_{i=1}^n  X_i |^{\zeta}    \|_{L_{1/\zeta}} \leq C_{\zeta} q \tilde{K}^{\zeta}$, where the last inequality follows from $1/\zeta \leq C_{\zeta} q$. Applying the same equivalence in Proposition 2.7.1 of \citeS{vershynin2018highS} yields $\| | \sum_{i=1}^n  X_i |^{\zeta} \|_{\psi_{1}} \leq C_{\zeta} \tilde{K}^{\zeta}$. Hence, 
 \begin{align*} 
     \bigg\|\sum_{i=1}^n  X_i\bigg\|_{\psi_{\zeta}} = \bigg\|\bigg| \sum_{i=1}^n  X_i \bigg|^{\zeta} \bigg\|_{\psi_{1}}^{1/\zeta} \leq C_{\zeta} \tilde{K} = C_{\zeta} \sum_{i=1}^n\|  X_i\|_{\psi_{\zeta}}\,.
 \end{align*}
 This completes the proof of Lemma \ref{lemma.triangle}. \hfill$\Box$

 \subsection{Proof of Lemma \ref{lemma.w.cons}}\label{proof.lemma.w.cons}

We first show that $\tilde{\bw}_{(j)}=(\tilde{\bH}_{-j,-j})^{-1}\tilde{\bH}_{-j,j}$ is a feasible solution to the optimization problem \eqref{Ch3dantzig}. Recall that ${\bv}_{(j)}\in\RR^p$ satisfies 
        $\{{\bv}_{(j)}\}_j=1$ and
        $\{{\bv}_{(j)}\}_{-j}=-\tilde{\bw}_{(j)}$. Since $\tilde{\bw}_{(j)}^{\T}\tilde{\bH}_{-j,-j} =  \tilde{\bH}_{ j,-j}$, we have
\begin{align}\label{tilde.w}
   &~ \big\|\tilde{\bw}_{(j)}^{\T}\{\nabla^2\hat{Q}_h({\hat\bbeta}_h)\}_{-j,-j}-  \{\nabla^2\hat{Q}_h(\hat{\bbeta}_h)\}_{j,-j}  \big\|_{\infty}\nonumber\\
  &~~~~ =  \big\|\tilde{\bw}_{(j)}^{\T}[\{\nabla^2\hat{Q}_h({\hat\bbeta}_h)\}_{-j,-j} -\tilde{\bH}_{-j,-j} ] -  [\{\nabla^2\hat{Q}_h(\hat{\bbeta}_h)\}_{j,-j}-\tilde{\bH}_{j,-j}] \big\|_{\infty}\nonumber\\
  &~~~~ =   \|[\{\nabla^2\hat{Q}_h({\hat\bbeta}_h) -\tilde{\bH}\}{\bv}_{(j)}]_{-j}  \|_{\infty} \leq \| \{\nabla^2\hat{Q}_h({\hat\bbeta}_h) -\tilde{\bH}\}{\bv}_{(j)}   \|_{\infty}\,.
\end{align}
Recall that $\tilde{\bH} = E\{ f_{\varepsilon|\bx}(0)\bx\bx^{\T}  \}$. Let $r_i(\bbeta) = y_i-\bx_i^{\T}{\bbeta}$. By \eqref{grad.hess}, we have 
\begin{align}\label{tilde.w.1}
     &~\|\{\nabla^2\hat{Q}_h({\hat\bbeta}_h) -\tilde{\bH}\}{\bv}_{(j)} \|_{\infty}\nonumber \\
     &~~~~ =   \bigg\| \frac{1}{n}\sum_{i=1}^n K_h (-r_i(\hat\bbeta_h))\bx_{i}\bx_{i}^{\T}{\bv}_{(j)}- E\{ f_{\varepsilon_i|\bx_i}(0)\bx_{i}\bx_{i}^{\T}{\bv}_{(j)}  \} \bigg\|_{\infty} \nonumber\\
      &~~~~ =  \max_{k\in[p]}\bigg | \frac{1}{n}\sum_{i=1}^n K_h (-r_i(\hat\bbeta_h))x_{i,k}\bx_{i}^{\T}{\bv}_{(j)}- E\{ f_{\varepsilon_i|\bx_i}(0)x_{i,k}\bx_{i}^{\T}{\bv}_{(j)}  \} \bigg |\nonumber\\
 &~~~~ \leq  \underbrace{\max_{k\in[p]}\bigg| \frac{1}{n}\sum_{i=1}^n \{K_h (-r_i(\hat\bbeta_h)) -K_h (-\varepsilon_i)\}x_{i,k}\bx_{i}^{\T}{\bv}_{(j)} \bigg|}_{{\rm I}_1} \nonumber\\
 &~~~~~~~+ \underbrace{\max_{k\in[p]}\bigg| \frac{1}{n}\sum_{i=1}^n K_h (-\varepsilon_i)x_{i,k}\bx_{i}^{\T}{\bv}_{(j)} - E\{ f_{\varepsilon|\bx}(0)x_{i,k}\bx_{i}^{\T}{\bv}_{(j)}   \} \bigg | }_{{\rm I}_2}\,.
\end{align}
Then, by \eqref{tilde.w}, we have
\begin{align}\label{tilde.w.2}
    \big\|\tilde{\bw}_{(j)}^{\T}\{\nabla^2\hat{Q}_h({\hat\bbeta}_h)\}_{-j,-j}-  \{\nabla^2\hat{Q}_h(\hat{\bbeta}_h)\}_{j,-j}  \big\|_{\infty} \leq {\rm I}_1+{\rm I}_2\,.
\end{align}
We bound ${\rm I}_1$ and ${\rm I}_2$ in turn.

Since  $\varepsilon_i = y_i-\bx_i^{\T}\bbeta^*$ and $K_h(u)=h^{-1}K(u/h)$, we have
\begin{align*}
   K_h (-r_i(\hat\bbeta_h)) -K_h (-\varepsilon_i) =&~  \frac{1}{h^2} K'\bigg(-\frac{\varepsilon_i}{h} \bigg)\bx_i^{\T}(\hat{\bbeta}_h-\bbeta^*) \\
   &~+ \frac{1}{2h^3}K''\bigg(-\frac{r_i(\bar{\bbeta}_i)}{h} \bigg)\{\bx_i^{\T}(\hat{\bbeta}_h-\bbeta^*)\}^2\,,
\end{align*}
where $\bar{\bbeta}_i = (1-t_i)\hat{\bbeta}_h +t_i \bbeta^*$ for some $t_i\in[0,1]$.  Therefore, by the triangle inequality, we have
\begin{align*}
{\rm I}_1 \leq &~   \underbrace{\max_{k\in[p]}\bigg| \frac{1}{nh^2}\sum_{i=1}^n x_{i,k}\bx_{i}^{\T}{\bv}_{(j)}  K'\bigg(-\frac{\varepsilon_i}{h} \bigg)  \bx_i^{\T}(\hat{\bbeta}_h-\bbeta^*)    \bigg|}_{{\rm I}_{1,1}}\\
&~+ \underbrace{\max_{k\in[p]}\bigg| \frac{1}{2nh^3}\sum_{i=1}^n  x_{i,k}\bx_{i}^{\T}{\bv}_{(j)} K''\bigg(-\frac{r_i(\bar{\bbeta}_i)}{h} \bigg)\{\bx_i^{\T}(\hat{\bbeta}_h-\bbeta^*)\}^2  \bigg|}_{{\rm I}_{1,2}}\,.
\end{align*}
Let $\be_j \in \mathbb{R}^p$ denote the $j$th canonical basis vector, whose $j$th entry is one and all other entries are zero.  By Condition 1(b), for each $i\in[n]$ and $j\in[p]$, we have 
\begin{align*}
    \pr(|x_{i,j}|>z ) = \pr(|\langle \bSigma^{-1/2}\bx_i, \bSigma^{1/2}\be_j \rangle|>z)\leq 2 \exp\bigg(-\frac{z}{ \upsilon_0 \|\be_j\|_{\bSigma} }  \bigg) \leq 2 \exp\bigg(-\frac{z}{ \upsilon_0 \gamma_1^{1/2}}  \bigg)
\end{align*}
for any $z\geq 0$, where the last inequality follows from $\|\be_j\|_{\bSigma} \leq \|\bSigma^{1/2}\|_2 \leq \gamma_1^{1/2}$.  This implies that 
\begin{align}\label{psinorm.x}
    \|x_{i,j}\|_{\psi_1}\lesssim \upsilon_0\gamma_1^{1/2}
\end{align}
for any $i\in[n]$ and $j\in[p]$. By Condition 1(b) again, 
\begin{align*}
    \pr\{|\bx_{i}^{\T}{\bv}_{(j)}|>z \} = \pr\{|\langle \bSigma^{-1/2}\bx_i, \bSigma^{1/2}{\bv}_{(j)}\rangle|>z\}\leq 2 \exp\bigg\{-\frac{z}{ \upsilon_0 \|{\bv}_{(j)}\|_{\bSigma} }  \bigg\}  
\end{align*}
for any $i\in[n]$ and $j\in[p]$. It follows from Condition 1(a) that  $ {\bv}_{(j)}^{\T}\tilde{\bH}{\bv}_{(j)} \geq f_l {\bv}_{(j)}^{\T}\bSigma{\bv}_{(j)} = f_l \|{\bv}_{(j)}\|_{\bSigma}^2$. Moreover, by the construction of ${\bv}_{(j)}$ and $\tilde{\bw}_{(j)}=(\tilde{\bH}_{-j,-j})^{-1}\tilde{\bH}_{-j,j}$, we have
\begin{align*}
    {\bv}_{(j)}^{\T}\tilde{\bH}{\bv}_{(j)} = \tilde{H}_{j,j}-2\tilde{\bH}_{j,-j}\tilde{\bw}_{(j)}+ \tilde{\bw}_{(j)}^{\T}\tilde{\bH}_{-j,-j}\tilde{\bw}_{(j)} =\tilde{H}_{j,j}- \tilde{\bH}_{j,-j} (\tilde{\bH}_{-j,-j})^{-1}\tilde{\bH}_{-j,j} \leq \tilde{H}_{j,j}\,,
\end{align*}
where the last inequality follows from the fact that $\tilde{\bH}_{-j,-j}$ is positive definite. In addition, by Conditions 1(a) and 1(b), we have $\tilde{H}_{j,j} = E\{f_{\varepsilon|\bx}(0)x_{i,j}^2\}\lesssim f_u  \gamma_1$, which implies that  $\|{\bv}_{(j)}\|_{\bSigma} \lesssim \sqrt{f_u/f_l} \gamma_1^{1/2} $. Thus, for any $i\in[n]$ and $j\in [p]$,
\begin{align}\label{psinorm.xv}
    \|\bx_{i}^{\T}{\bv}_{(j)}\|_{\psi_1}\lesssim  \upsilon_0\gamma_1^{1/2}(f_u/f_l)^{1/2}\,.
\end{align}
Write $\xi_{i,(j,k,l)}=  x_{i,k}x_{i,l}\bx_{i}^{\T}{\bv}_{(j)}K'(-\varepsilon_i/h)$ for $i\in[n]$ and $k,l\in[p]$. By the triangle inequality,
\begin{align*}
    {\rm I}_{1,1} = &~ \max_{k\in[p]}\bigg | \frac{1}{nh^2}\sum_{i=1}^n  x_{i,k}\bx_{i}^{\T}{\bv}_{(j)} K'\bigg(-\frac{\varepsilon_i}{h} \bigg)  \bx_i^{\T}(\hat{\bbeta}_h-\bbeta^*)    \bigg | \\
    \leq  &~ \|\hat{\bbeta}_h-\bbeta^*\|_1 \cdot \frac{1}{h^2}\max_{k,l\in[p]} \bigg| \frac{1}{n }\sum_{i=1}^n   x_{i,k}x_{i,l}\bx_{i}^{\T}{\bv}_{(j)}K'\bigg(-\frac{\varepsilon_i}{h} \bigg) \bigg |\\
    \leq &~ \|\hat{\bbeta}_h-\bbeta^*\|_1 \cdot \frac{1}{h^2}\bigg[\max_{k,l\in[p]}\bigg| \frac{1}{n }\sum_{i=1}^n (1-E)\{\xi_{i,(j,k,l)}\}  \bigg | + \max_{ k,l\in[p]}\bigg| \frac{1}{n }\sum_{i=1}^n  E\{\xi_{i,(j,k,l)}\} \bigg |\bigg]\,.
\end{align*}
 Conditions 1(a) and 1(d) give 
\begin{align*}
     E\bigg\{\bigg|K'\bigg(-\frac{\varepsilon_i}{h} \bigg)\bigg|  \,\big|\, \bx_i\bigg\} = &~   \int_{-\infty}^{\infty} \bigg| K'\bigg(-\frac{u}{h} \bigg) \bigg|  f_{\varepsilon_i|\bx_i}(u)\,\mbox{d}u  \\
     = &~ h \int_{-\infty}^{\infty} | K'(v)|  f_{\varepsilon_i|\bx_i}(-hv)\,\mbox{d}v  \leq f_u C_K  h\,.
\end{align*}
By \eqref{psinorm.x} and \eqref{psinorm.xv}, we have $E\{| x_{i,k}x_{i,l}\bx_{i}^{\T}{\bv}_{(j)} |\}\lesssim \upsilon_0^3\gamma_1^{3/2} (f_u/f_l)^{1/2}$, which implies, for any $k,l\in[p]$, $|E\{\xi_{i,(j,k,l)}\}|\lesssim  \upsilon_0^3\gamma_1^{3/2} (f_u^3/f_l)^{1/2} C_K h $.  Hence,
\begin{align}\label{expe.bound}
    \max_{ k,l\in[p]}\bigg| \frac{1}{n }\sum_{i=1}^n E\{\xi_{i,(j,k,l)}\}      \bigg | \lesssim  \upsilon_0^3\gamma_1^{3/2} (f_u^3/f_l)^{1/2} C_K h \,.
\end{align}
By \eqref{psinorm.x} and \eqref{psinorm.xv}, $\|x_{i,k}x_{i,l}\bx_{i}^{\T}{\bv}_{(j)}  \|_{\psi_{1/3}} \leq \|x_{i,k}\|_{\psi_{1}}\|x_{i,l}\|_{\psi_{1}}\|\bx_{i}^{\T}{\bv}_{(j)}\|_{\psi_{1}}  \lesssim \upsilon_0^3\gamma_1^{3/2}(f_u/f_l)^{1/2} $. Since $|K'(-\varepsilon_i/h)|\leq \kappa_u'$, we have  $\|(1-E)\{\xi_{i,(j,k,l)}\}\|_{\psi_{1/3}}   \lesssim \kappa_u' \upsilon_0^3\gamma_1^{3/2}(f_u/f_l)^{1/2} $ for any $i\in[n]$ and $k,l\in[p]$.
 Then, by Theorem 3.1 of \citeS{Kuchibhotla2022S} with $\alpha=1/3$ and $\ba= (n^{-1},\ldots,n^{-1})^{\T}\in \RR^n$, we have
\begin{align*}
     \pr\bigg[\max_{k,l\in[p]}\bigg| \frac{1}{n }\sum_{i=1}^n (1-E)\{\xi_{i,(j,k,l)}\}  \bigg | \geq    \kappa_u' \upsilon_0^3\gamma_1^{3/2} (f_u/f_l)^{1/2}\bigg(\sqrt{\frac{C  z}{n}} + \frac{C  z^3}{n} \bigg)\bigg] \lesssim  p^2\exp(-z) 
\end{align*}
for any $z\geq 0$, which implies that, with probability at least $1-O(p^{-c_0})$, 
\begin{align} \label{psi.tail}
    \max_{k,l\in[p]}\bigg| \frac{1}{n }\sum_{i=1}^n (1-E)\{\xi_{i,(j,k,l)}\}  \bigg |  \lesssim \kappa_u' \upsilon_0^3\gamma_1^{3/2}(f_u/f_l)^{1/2} \bigg\{  \sqrt{\frac{\log p}{n}} + \frac{(\log p)^3}{n}\bigg\}\,.
\end{align} 
This, together with \eqref{expe.bound}, yields 
\begin{align} \label{I11.pre}
    {\rm I}_{1,1} \lesssim \upsilon_0^3\gamma_1^{3/2}(f_u/f_l)^{1/2} \bigg\{  \kappa_u'\sqrt{\frac{\log p}{nh^4}} + \frac{\kappa_u'(\log p)^3}{nh^2}+ \frac{f_u C_K}{h} \bigg\}\cdot \|\hat{\bbeta}_h-\bbeta^*\|_1
\end{align}
with probability at least $1-O(p^{-c_0})$.

Since $\ba^{\T}\bb \leq \|\ba\|_{\infty}\|\bb\|_1$ for any $\ba,\bb\in\RR^q$, by Condition 1(d), we have
\begin{align*}
    {\rm I}_{1,2} = &~ \frac{1}{2h^3} \max_{k\in[p]}\bigg| \frac{1}{n}\sum_{i=1}^n   x_{i,k}\bx_{i}^{\T}{\bv}_{(j)} K''\bigg(-\frac{r_i(\bar{\bbeta}_i)}{h} \bigg)\{\bx_i^{\T}(\hat{\bbeta}_h-\bbeta^*)\}^2 \bigg|\\
    \leq &~ \frac{1}{2h^3} \cdot \|\hat{\bbeta}_h-\bbeta^*\|_1^2\cdot \max_{ k\in[p]} \bigg\| \frac{1}{n}\sum_{i=1}^n   x_{i,k}\bx_{i}^{\T}{\bv}_{(j)}  K''\bigg(-\frac{r_i(\bar{\bbeta}_i)}{h} \bigg) \bx_i \bx_i^{\T}\bigg\|_{\infty} \\
    \leq &~  \frac{\kappa_u''}{2h^3}  \cdot \|\hat{\bbeta}_h-\bbeta^*\|_1^2\cdot   \max_{k,l,r\in[p]} \frac{1}{n}\sum_{i=1}^n | x_{i,k}x_{i,l}x_{i,r}\bx_{i}^{\T}{\bv}_{(j)}  | \,.
\end{align*}
By \eqref{psinorm.x} and \eqref{psinorm.xv}, $\|x_{i,k}x_{i,l}x_{i,r}\bx_{i}^{\T}{\bv}_{(j)}  \|_{\psi_{1/4}}  \lesssim \upsilon_0^4\gamma_1^{2}(f_u/f_l)^{1/2} $ for any $i\in[n]$. By Lemma \ref{lemma.triangle}, we have $\|n^{-1}\sum_{i=1}^n | x_{i,k}x_{i,l}x_{i,r}\bx_{i}^{\T}{\bv}_{(j)}  |\|_{\psi_{1/4}} \lesssim  \upsilon_0^4\gamma_1^2(f_u/f_l)^{1/2} $, which implies that 
\begin{align*}
    \pr\bigg\{ \max_{k,l,r\in[p]} \frac{1}{n}\sum_{i=1}^n | x_{i,k}x_{i,l}x_{i,r}\bx_{i}^{\T}{\bv}_{(j)}  | >z  \bigg\}  \lesssim p^3\exp\bigg[ -C\bigg\{\frac{z}{\upsilon_0^4\gamma_1^2(f_u/f_l)^{1/2}}\bigg\}^{1/4} \bigg]\,.
\end{align*}
Then, with probability at least $1-O(p^{-c_0})$, we have
\begin{align}\label{abs.moment.x}
   \max_{k,l,r\in[p]} \frac{1}{n}\sum_{i=1}^n | x_{i,k}x_{i,l}x_{i,r}\bx_{i}^{\T}{\bv}_{(j)}  |  \lesssim \upsilon_0^4\gamma_1^2(f_u/f_l)^{1/2} (\log p)^4\,.
\end{align}
Therefore, we have
\begin{align*} 
    {\rm I}_{1,2} \lesssim \frac{\kappa_u''\upsilon_0^4\gamma_1^2(f_u/f_l)^{1/2}(\log p)^4}{h^3}\cdot \|\hat{\bbeta}_h-\bbeta^*\|_1^2
\end{align*}
with probability at least $1-O(p^{-c_0})$. By Lemma \ref{Theorem.tan2022} and \eqref{I11.pre}, we obtain 
\begin{align}\label{Bound.I1}
     {\rm I}_{1} \leq C_{I_1} \sqrt{ \frac{f_u}{f_l}} \bigg\{    \frac{f_u}{f_l}   \sqrt{\frac{ s^2\log p}{nh^2}}+  \frac{1}{f_l^2}  \frac{s^2(\log p)^5 }{   nh^3} \bigg\}
\end{align}
with probability at least $1-O(p^{-c_0})$, where $C_{I_1}>0$ depends only on $(l_0,\upsilon_0,\gamma_p,\underline{\kappa},\kappa_2,\gamma_1,C_K,\kappa_u',\kappa_u'')$.

For ${\rm I}_2$, note that
\begin{align*}
    {\rm I}_2 =&~\max_{k\in[p]}\bigg| \frac{1}{n}\sum_{i=1}^n K_h (-\varepsilon_i)x_{i,k}\bx_{i}^{\T}{\bv}_{(j)} - E\{ f_{\varepsilon|\bx}(0)x_{i,k}\bx_{i}^{\T}{\bv}_{(j)}   \} \bigg |\\
    \leq &~ \underbrace{\max_{k\in[p]}\bigg| \frac{1}{n}\sum_{i=1}^n (1-E)\{K_h (-\varepsilon_i)x_{i,k}\bx_{i}^{\T}{\bv}_{(j)}  \} \bigg |}_{{\rm I}_{2,1}} \\
    &~+ \underbrace{\max_{k\in[p]}\bigg| \frac{1}{n}\sum_{i=1}^n  E\{K_h (-\varepsilon_i)x_{i,k}\bx_{i}^{\T}{\bv}_{(j)} \} - E\{ f_{\varepsilon|\bx}(0)x_{i,k}\bx_{i}^{\T}{\bv}_{(j)}   \} \bigg| }_{{\rm I}_{2,2}}\,.
\end{align*}
 Condition 1(d) yields $|K_h (-\varepsilon_i)|\leq h^{-1}\kappa_u$, which implies  $\||(1-E)\{K_h (-\varepsilon_i)x_{i,k}\bx_{i}^{\T}{\bv}_{(j)}\}|\|_{\psi_{1/2}}   \lesssim h^{-1}\kappa_u  \upsilon_0^2\gamma_1 (f_u/f_l)^{1/2} $ for any $i\in[n]$ and $k\in[p]$.  Then, by Theorem 3.1 of \citeS{Kuchibhotla2022S} and the argument for \eqref{psi.tail}, we obtain, with probability at least $1-O(p^{-c_0})$,
\begin{align*}
    {\rm I}_{2,1} \lesssim \kappa_u  \upsilon_0^2\gamma_1 (f_u/f_l)^{1/2} \bigg\{  \sqrt{\frac{\log p}{nh^2}} + \frac{(\log p)^2}{nh }\bigg\}\,.
\end{align*}
By Conditions 1(a) and 1(d), we have
\begin{align*}
     |E\{K_h (-\varepsilon_i)\,|\,\bx_i\} - f_{\varepsilon|\bx}(0) |= &~ \bigg| \int_{-\infty}^{\infty} \frac{1}{h}K\bigg( -\frac{u}{h}\bigg) f_{\varepsilon_i|\bx_i}(u)\,\mbox{d}u - f_{\varepsilon|\bx}(0)\bigg|\\
    =&~ \bigg| \int_{-\infty}^{\infty} K(v) \{f_{\varepsilon_i|\bx_i}(-hv)-f_{\varepsilon|\bx}(0)\}\,\mbox{d}v\bigg|\leq l_0 \kappa_1 h\,.
\end{align*}
Since $\max_{i\in[n],k\in[p]}E\{|x_{i,k}\bx_{i}^{\T}{\bv}_{(j)}|\}\lesssim \upsilon_0^2\gamma_1(f_u/f_l)^{1/2}$, we have ${\rm I}_{2,2} \lesssim \upsilon_0^2\gamma_1 (f_u/f_l)^{1/2}l_0 \kappa_1 h$. Thus, 
\begin{align*}
     {\rm I}_{2} \leq C_{I_2}\sqrt{ \frac{f_u}{f_l}}  \bigg\{   \sqrt{\frac{\log p}{nh^2}} + \frac{(\log p)^2}{nh}+h\bigg\}
\end{align*}
with probability at least $1-O(p^{-c_0})$, where $C_{I_2}>0$ depends only on $(l_0,\upsilon_0,\gamma_1,\kappa_1,\kappa_u)$.
 Hence, by \eqref{h.req}, \eqref{tilde.w.2}, and \eqref{Bound.I1}, we have
\begin{align}
   &~\big\|\tilde{\bw}_{(j)}^{\T}\{\nabla^2\hat{Q}_h({\hat\bbeta}_h)\}_{-j,-j}-  \{\nabla^2\hat{Q}_h(\hat{\bbeta}_h)\}_{j,-j}  \big\|_{\infty}\nonumber\\
  &~~~~ \leq  \tilde{C} \sqrt{ \frac{f_u}{f_l}} \bigg\{      \frac{f_u}{f_l}   \sqrt{\frac{ s^2\log p}{nh^2}}+  \frac{1}{f_l^2}  \frac{s^2(\log p)^5 }{   nh^3}  \bigg\}\leq \Delta \label{w.solution}
\end{align}
with probability at least $1-O(p^{-c_0})$, where $\tilde{C}>0$ depends only on $(l_0,\upsilon_0,\gamma_1,\gamma_p,\underline{\kappa},\kappa_1, \kappa_2,C_K,\kappa_u,\kappa_u',\kappa_u'')$. Thus, with probability at least $1-O(p^{-c_0})$, $\tilde{\bw}_{(j)}$ is a feasible solution to the optimization problem \eqref{Ch3dantzig}.

We next bound $\|\hat{\bw}_{(j)}- \tilde{\bw}_{(j)}\|_1$. Let $\hat{\bw}_{(j)}= \{\hat{w}_{(j),1},\ldots, \hat{w}_{(j),p-1}\}^{\T}$ and $\hat{\bdelta}_{(j)}=\hat{\bw}_{(j)}- \tilde{\bw}_{(j)}=\{\hat{\delta}_{(j),1},\ldots, \hat{\delta}_{(j),p-1}\}^{\T}$. It suffices to consider $\hat{\bdelta}_{(j)} \neq {\mathbf 0}$ for any $j\in [p]$, since the bound is immediate otherwise. Feasibility and the minimizing property give 
\begin{align*}
    \|\tilde{\bw}_{(j)}\|_1 \geq \|\hat{\bw}_{(j)}\|_1 = \sum_{k\in \tilde{S}_j}|\hat{w}_{(j),k}|+\sum_{k\in \tilde{S}_j^{{\rm c}}}|\hat{w}_{(j),k}|\geq \|\tilde{\bw}_{(j)}\|_1-\sum_{k\in \tilde{S}_j}|{\hat{\delta}_{(j),k}}|+\sum_{k\in \tilde{S}_j^{{\rm c}}}|\hat{\delta}_{(j),k}| 
\end{align*}
with probability at least $1-O(p^{-c_0})$, where $\tilde{S}_j={\rm supp}\{\tilde{\bw}_{(j)}\}$ and $\tilde{S}_j^{{\rm c}}=[p-1]\!\setminus \!\tilde{S}_j$. This implies $\sum_{k\in \tilde{S}_j}|{\hat{\delta}_{(j),k}}| \geq \sum_{k\in \tilde{S}_j^{{\rm c}}}|\hat{\delta}_{(j),k}|$. By Conditions 1(a) and 1(b), $\tilde{\bH}$ is positive definite. Since $\tilde{\bH}\tilde{\bH}^{-1}=\bI_p$, we have $\tilde{\bH}_{-j,j}(\tilde{\bH}^{-1})_{j,j}+\tilde{\bH}_{-j,-j}(\tilde{\bH}^{-1})_{-j,j}={\mathbf 0}$, which implies
\begin{align*}
    \tilde{\bw}_{(j)} = (\tilde{\bH}_{-j,-j})^{-1}\tilde{\bH}_{-j,j}= -\frac{(\tilde{\bH}^{-1})_{-j,j}}{(\tilde{\bH}^{-1})_{j,j}}\,.
\end{align*}
Since $(\tilde{\bH}^{-1})_{j,j}>0$, we have $\tilde{S}_j = {\rm supp}\{(\tilde{\bH}^{-1})_{-j,j}\}$. Since $\tilde{\bH}^{-1}$ is symmetric, it follows from Condition 1(c) that $|\tilde{S}_j |\leq \tilde{s}$ for any $j\in [p]$. Then, we obtain 
\begin{align}\label{Delta.norm12}
    \|\hat{\bdelta}_{(j)}\|_1 = \sum_{k\in \tilde{S}_j}|{\hat{\delta}_{(j),k}}| + \sum_{k\in \tilde{S}_j^{{\rm c}}}|\hat{\delta}_{(j),k}| \leq 2 \sum_{k\in \tilde{S}_j}|{\hat{\delta}_{(j),k}}| \leq 2 \sqrt{\tilde{s} \sum_{k\in \tilde{S}_j}|{\hat{\delta}_{(j),k}}|^2 } \leq 2\sqrt{ \tilde{s} }\|\hat{\bdelta}_{(j)}\|_2\,.
\end{align}
For an upper bound, observe that  
\begin{align}\label{upper.delta}
   &~ \hat{\bdelta}_{(j)}^{\T}\{\nabla^2\hat{Q}_h({\hat\bbeta}_h)\}_{-j,-j}\hat{\bdelta}_{(j)}  
    \leq   \big\|\hat{\bw}_{(j)}^{\T}\{\nabla^2\hat{Q}_h({\hat\bbeta}_h)\}_{-j,-j}-  \tilde{\bw}_{(j)}^{\T}\{\nabla^2\hat{Q}_h({\hat\bbeta}_h)\}_{-j,-j}   \big\|_{\infty}\|\hat{\bdelta}_{(j)}\|_1\nonumber\\
   &~~~~~~~~~~~~~~~~~~ \leq \big\|\hat{\bw}_{(j)}^{\T}\{\nabla^2\hat{Q}_h({\hat\bbeta}_h)\}_{-j,-j}-  \{\nabla^2\hat{Q}_h(\hat{\bbeta}_h)\}_{j,-j}  \big\|_{\infty} \|\hat{\bdelta}_{(j)}\|_1\nonumber\\
   &~~~~~~~~~~~~~~~~~~~~~ + \big\|\tilde{\bw}_{(j)}^{\T}\{\nabla^2\hat{Q}_h({\hat\bbeta}_h)\}_{-j,-j}-  \{\nabla^2\hat{Q}_h(\hat{\bbeta}_h)\}_{j,-j}  \big\|_{\infty}\|\hat{\bdelta}_{(j)}\|_1 
   \leq  2\Delta \|\hat{\bdelta}_{(j)}\|_1 
\end{align}
with probability at least $1-O(p^{-c_0})$, where the last inequality follows from \eqref{Ch3dantzig} and \eqref{w.solution}.  For a lower bound,  
\begin{align}\label{lower.delta.pre}
    \frac{\hat{\bdelta}_{(j)}^{\T}\{\nabla^2\hat{Q}_h({\hat\bbeta}_h)\}_{-j,-j}\hat{\bdelta}_{(j)}}{\|\hat{\bdelta}_{(j)}\|_2^2} \geq  \frac{\hat{\bdelta}_{(j)}^{\T}\tilde{\bH}_{-j,-j}\hat{\bdelta}_{(j)}}{\|\hat{\bdelta}_{(j)}\|_2^2}-\bigg| \frac{\hat{\bdelta}_{(j)}^{\T}[\{\nabla^2\hat{Q}_h({\hat\bbeta}_h)\}_{-j,-j}-\tilde{\bH}_{-j,-j}]\hat{\bdelta}_{(j)}}{\|\hat{\bdelta}_{(j)}\|_2^2} \bigg| \,.
\end{align}
 By Conditions 1(a) and 1(b), we have 
\begin{align}\label{lower1}
    \frac{\hat{\bdelta}_{(j)}^{\T}\tilde{\bH}_{-j,-j}\hat{\bdelta}_{(j)}}{\|\hat{\bdelta}_{(j)}\|_2^2} \geq  f_l \frac{\hat{\bdelta}_{(j)}^{\T}E(  \bx_{-j}\bx_{-j}^{\T}  )\hat{\bdelta}_{(j)}}{\|\hat{\bdelta}_{(j)}\|_2^2} \geq f_l \gamma_p\,.
\end{align}
Note that 
\begin{align*}
    &\|\nabla^2\hat{Q}_h({\hat\bbeta}_h) -\tilde{\bH} \|_{\infty} = \max_{j,k\in[p]}\bigg| \frac{1}{n}\sum_{i=1}^n K_h (-r_i(\hat\bbeta_h))x_{i,j}x_{i,k}- E\{ f_{\varepsilon_i|\bx_i}(0)x_{i,j}x_{i,k} \} \bigg|\,.
\end{align*}
Since $f_l < f_u$, the argument for \eqref{w.solution}, with $x_{i,k}\bx_{i}^{\T}{\bv}_{(j)}$ replaced by $x_{i,j}x_{i,k}$, gives 
\begin{align}\label{cons.hatQ}
    \|\nabla^2\hat{Q}_h({\hat\bbeta}_h) -\tilde{\bH} \|_{\infty} \leq \Delta
\end{align}
with probability at least $1-O(p^{-c_0})$. Since $|\ba^{\T} \bB \ba|\leq \|\ba\|_1^2 \|\bB\|_{\infty}$ for any $\ba\in\RR^q$ and $\bB\in\RR^{q\times q}$, by \eqref{Delta.norm12}, we have
\begin{align*} 
   \bigg| \frac{\hat{\bdelta}_{(j)}^{\T}[\{\nabla^2\hat{Q}_h({\hat\bbeta}_h)\}_{-j,-j}-\tilde{\bH}_{-j,-j}]\hat{\bdelta}_{(j)}}{\|\hat{\bdelta}_{(j)}\|_2^2} \bigg| \leq \frac{\|\hat{\bdelta}_{(j)}\|_1^2}{\|\hat{\bdelta}_{(j)}\|_2^2}\cdot \|\nabla^2\hat{Q}_h({\hat\bbeta}_h) -\tilde{\bH} \|_{\infty}\leq  4 \tilde{s} \Delta 
\end{align*}
with probability at least $1-O(p^{-c_0})$. Then, provided that $8 \tilde{s} \Delta  \leq f_l \gamma_p$,   \eqref{lower.delta.pre} and \eqref{lower1} yield
\begin{align*}
    \hat{\bdelta}_{(j)}^{\T}\{\nabla^2\hat{Q}_h({\hat\bbeta}_h)\}_{-j,-j}\hat{\bdelta}_{(j)} \geq \frac{f_l \gamma_p  \|\hat{\bdelta}_{(j)}\|_2^2 }{2}\,.
\end{align*}
This, together with \eqref{Delta.norm12} and \eqref{upper.delta}, yields
\begin{align*}
    \|\hat{\bdelta}_{(j)}\|_1^2 \leq 4\tilde{s}\|\hat{\bdelta}_{(j)}\|_2^2 \lesssim \frac{\tilde{s}\Delta}{f_l }\|\hat{\bdelta}_{(j)}\|_1
\end{align*}
with probability at least $1-O(p^{-c_0})$.  It follows that, with probability at least $1-O(p^{-c_0})$,
\begin{align*}
\|\hat{\bw}_{(j)}- \tilde{\bw}_{(j)}\|_1=\|\hat{\bdelta}_{(j)}\|_1 \lesssim \frac{\tilde{s}\Delta}{f_l} 
\end{align*}
for any $j\in [p]$. This completes the proof of Lemma \ref{lemma.w.cons}. \hfill$\Box$

\subsection{Proof of Lemma \ref{lemma.sigma.cons}}\label{proof.lemma.sigma.cons}

Note that $\hat{\sigma}_{j}^2 =  \hat{\bv}_{(j)}^{\T} \hat{\bJ} \hat{\bv}_{(j)} $ and ${\sigma}_{j}^2 =   {\bv}_{(j)}^{\T} {\bJ} {\bv}_{(j)} $, where  
\begin{align*}
    \hat{\bJ}= \frac{1}{n}\sn \{   \tau-\bar{K}_h(  -r_i(\hat{\bbeta}_h))  \}^2\bx_i\bx_i^{\T} ~~\mbox{and}~~{\bJ}= \frac{1}{n}\sn {\rm var}[\{   \tau-\bar{K}_h(  -\varepsilon_i) \} \bx_i]\,.
\end{align*}
Since $|\ba^{\T} \bB \ba|\leq \|\ba\|_1^2 \|\bB\|_{\infty}$ and $|\ba^{\T} \bc|\leq \|\ba\|_1\|\bc\|_{\infty}$ for any $\ba,\bc\in\RR^q$ and $\bB\in\RR^{q\times q}$, we have 
\begin{align*}
     |\hat{\sigma}_{j}^2- {\sigma}_{j}^2| = &~   \big| \hat{\bv}_{(j)}^{\T} \hat{\bJ} \hat{\bv}_{(j)} - {\bv}_{(j)}^{\T} {\bJ} {\bv}_{(j)} \big|\\
    \lesssim &~  \|\hat{\bv}_{(j)} - {\bv}_{(j)}\|_1^2 \|\hat{\bJ}-{\bJ}  \|_{\infty} + \|\hat{\bv}_{(j)} - {\bv}_{(j)}\|_1 ^2 \|{\bJ}  \|_{\infty}+ \|\hat{\bv}_{(j)} - {\bv}_{(j)}\|_1  \|{\bJ} {\bv}_{(j)} \|_{\infty} \\
    &~+   \|\hat{\bv}_{(j)} - {\bv}_{(j)}\|_1 \|\hat{\bJ}{\bv}_{(j)}-{\bJ}{\bv}_{(j)}  \|_{\infty}  +   \big| {\bv}_{(j)}^{\T} \hat{\bJ} {\bv}_{(j)} - {\bv}_{(j)}^{\T} {\bJ} {\bv}_{(j)} \big|  \,.
\end{align*}
By Lemma \ref{lemma.w.cons}, we have $\|\hat{\bv}_{(j)} - {\bv}_{(j)}\|_1  = \|\hat{\bw}_{(j)} - \tilde{\bw}_{(j)}\|_1\lesssim \tilde{s}\Delta/f_l$ with probability at least $1-O(p^{-c_0})$. Since $|\tau-\bar{K}_h(  -\varepsilon_i)|\leq 1$, it follows from \eqref{psinorm.x} and \eqref{psinorm.xv} that $\|{\bJ}  \|_{\infty} \lesssim \upsilon_0^2\gamma_1 $ and $\|{\bJ} {\bv}_{(j)} \|_{\infty} \lesssim \upsilon_0^2\gamma_1 (f_u/f_l)^{1/2}$. Then, provided that $\tilde{s}\Delta \lesssim f_l$,  we obtain
\begin{align}\label{dis.sigma2}
    |\hat{\sigma}_{j}^2- {\sigma}_{j}^2| \lesssim  \frac{\tilde{s}\Delta}{f_l}\bigg\{ \sqrt{\frac{f_u}{f_l}}+\frac{\tilde{s}\Delta}{f_l}\|\hat{\bJ}-{\bJ}  \|_{\infty}+  \|\hat{\bJ}{\bv}_{(j)}-{\bJ}{\bv}_{(j)}  \|_{\infty}\bigg\}+\big| {\bv}_{(j)}^{\T} \hat{\bJ} {\bv}_{(j)} - {\bv}_{(j)}^{\T} {\bJ} {\bv}_{(j)} \big| 
\end{align}
with probability at least $1-O(p^{-c_0})$. Write $\psi_{i}(\bbeta) = \tau -\bar{K}_h(  -r_i({\bbeta}) )$ and $Q_{i,j}= \bx_{i}^{\T}{\bv}_{(j)}$.  We have
\begin{align*}
    \big| {\bv}_{(j)}^{\T} \hat{\bJ} {\bv}_{(j)} - {\bv}_{(j)}^{\T} {\bJ} {\bv}_{(j)} \big| =&~   \bigg| \frac{1}{n}\sn \psi_i^2(\hat{\bbeta}_h)Q_{i,j}^2 -  \frac{1}{n}\sn \big(E \{ \psi_i^2({\bbeta}^*)Q_{i,j}^2 \} - [E \{ \psi_i({\bbeta}^*)Q_{i,j}\}]^2 \big) \bigg|\\
    \leq &~ \underbrace{ \bigg| \frac{1}{n}\sn \{\psi_i^2(\hat{\bbeta}_h)-\psi_i^2({\bbeta}^*)\}Q_{i,j}^2\bigg|}_{{\rm II}_1} 
      + \underbrace{ \bigg| \frac{1}{n}\sn  (1-E)\{ \psi_i^2({\bbeta}^*)Q_{i,j}^2\}\bigg|}_{{\rm II}_2}\\
      &~+ \underbrace{ \bigg|  \frac{1}{n}\sn   [E \{ \psi_i({\bbeta}^*)Q_{i,j}\}]^2  \bigg|}_{{\rm II}_3}\,.
\end{align*}
Since $\bar{K}_h(u)=\int_{-\infty}^{u/h}K(v)\,\mbox{d}v$ and  $|\psi_{i}(\bbeta)|\leq 1$ for any $i\in[n]$ and $\bbeta\in\RR^p$,  by Condition 1(d),
\begin{align*}
    |\psi_i^2(\hat{\bbeta}_h)-\psi_i^2({\bbeta}^*)| \leq 2|\psi_i(\hat{\bbeta}_h)-\psi_i({\bbeta}^*)| =2  \bigg|\int_{-r_i(\bbeta^*)/h}^{-r_i(\hat\bbeta_h)/h} K(u)\,\mbox{d}u\bigg| \leq \frac{2\kappa_u}{h}|\bx_i^{\T}(\hat{\bbeta}_h-\bbeta^*) |\,,
\end{align*}
which implies  ${\rm II}_1 \leq 2\kappa_u {h}^{-1}\| \hat{\bbeta}_h-\bbeta^*\|_{1}\cdot \max_{k\in[p]} {n}^{-1}\sn|x_{i,k}Q_{i,j}^2| $. By \eqref{psinorm.x} and \eqref{psinorm.xv}, we have  $\||x_{i,k}Q_{i,j}^2|\|_{\psi_{1/3}} \leq \||x_{i,k} |\|_{\psi_1}\|| Q_{i,j}|\|_{\psi_{1}}^2 \lesssim \upsilon_0^3\gamma_1^{3/2}(f_u/f_l)$. The argument for \eqref{abs.moment.x} gives $\max_{k\in[p]} {n}^{-1}\sn|x_{i,k}Q_{i,j}^2| \lesssim \upsilon_0^3\gamma_1^{3/2}(f_u/f_l)(\log p)^3$ with probability at least $1-O(p^{-c_0})$. Then,  by Lemma \ref{Theorem.tan2022}, we obtain
\begin{align*}
    {\rm II}_1 \leq C_1    \frac{f_u}{f_l^2}    \sqrt{\frac{ s^2(\log p)^7}{nh^2}}
\end{align*}
with probability at least $1-O(p^{-c_0})$, where $C_1>0$ depends on $(l_0,\upsilon_0,\gamma_1,\gamma_p,\underline{\kappa},\kappa_u,\kappa_2)$. Note that $\|\psi_i^2({\bbeta}^*)Q_{i,j}^2\|_{\psi_{1/2}}\lesssim \upsilon_0^2\gamma_1(f_u/f_l)$ for any $i\in[n]$.  Then, by Theorem 3.1 of \citeS{Kuchibhotla2022S} with $\alpha=1/2$ and $\ba= (n^{-1},\ldots,n^{-1})^{\T}\in \RR^n$, we have
\begin{align*}
     \pr\bigg\{{\rm II}_2 \geq \upsilon_0^2\gamma_1(f_u/f_l)\bigg(\sqrt{\frac{C   z}{n}} + \frac{C  z^2}{n}\bigg)\bigg\} \lesssim  \exp(-z) 
\end{align*}
for any $z\geq 0$, which implies that, with probability at least $1-O(p^{-c_0})$, 
\begin{align*}
    {\rm II}_2 \lesssim \upsilon_0^2 \gamma_1 \frac{f_u}{f_l}  \bigg\{  \sqrt{\frac{\log p}{n}} + \frac{(\log p)^2}{n}\bigg\}\,.
\end{align*}

For ${\rm II}_3$, note that $\psi_i({\bbeta}^*)=  \tau -\bar{K}_h(  -\varepsilon_i )$. Fubini's theorem gives
\begin{align*}
    E\{ \bar{K}_h(-\varepsilon_i)\,|\,\bx_i \} = &~ \int_{-\infty}^{\infty} \bigg\{\int_{-\infty}^{-u/h}  {K}(v) \,\mbox{d} v\bigg\}f_{\varepsilon_i|\bx_i}(u) \,\mbox{d}u\\
    =&~ \int_{-\infty}^{\infty}{K}(v) \bigg\{\int_{-\infty}^{- hv} f_{\varepsilon_i|\bx_i}(u) \,\mbox{d} u\bigg\}\,\mbox{d} v = \int_{-\infty}^{\infty}{K}(v) F_{\varepsilon_i|\bx_i}(- hv)\,\mbox{d} v \,.
\end{align*}
Since $\tau = F_{\varepsilon_i|\bx_i}(0)=  \int_{-\infty}^{\infty}{K}(v) F_{\varepsilon_i|\bx_i}(0)\,\mbox{d} v$, it follows from  Conditions 1(a) and 1(d) that 
\begin{align}\label{psi.beta}
    &~|E \{ \psi_i({\bbeta}^*)\,|\,\bx_i\}| =  \bigg| \int_{-\infty}^{\infty}{K}(v) \{F_{\varepsilon_i|\bx_i}(- hv) -F_{\varepsilon_i|\bx_i}(0)\}\,\mbox{d} v  \bigg| \\
     &~~~~=  \bigg| \int_{-\infty}^{\infty}{K}(v) \int_{0}^{-hv} f_{\varepsilon_i|\bx_i}(u)\,\mbox{d} u\,\mbox{d} v  \bigg|\nonumber\\
   &~~~~\leq   \underbrace{\bigg| \int_{-\infty}^{\infty}{K}(v) (-hv) f_{\varepsilon_i|\bx_i}(0) \,\mbox{d} v  \bigg|}_{=0} + \int_{-\infty}^{\infty}{K}(v) \bigg| \int_{0}^{-hv}\{ f_{\varepsilon_i|\bx_i}(u) -f_{\varepsilon_i|\bx_i}(0)\}\,\mbox{d} u\bigg|\,\mbox{d} v\nonumber  \\
  &~~~~\leq_{(i)}    \frac{l_0 h^2}{2} \int_{-\infty}^{\infty}v^2 {K}(v) \mbox{d} v \leq \frac{l_0 h^2 \kappa_2}{2} 
\end{align}
for any $i\in[n]$, where $(i)$ follows from $|\int_{0}^{-hv}\{ f_{\varepsilon_i|\bx_i}(u) -f_{\varepsilon_i|\bx_i}(0)\}\,\mbox{d} u |\leq l_0\int_{0}^{ h|v|}  u  \,\mbox{d} u = l_0 h^2 v^2/2$. Since $E(|Q_{i,j}|)\lesssim \upsilon_0\gamma_1^{1/2}(f_u/f_l)^{1/2}$, we obtain  
\begin{align*}
    {\rm II}_3 \leq \max_{i\in[n]}  |E \{ \psi_i({\bbeta}^*) Q_{i,j} \}   |^2 \lesssim l_0^2\kappa_2^2 \upsilon_0^2\gamma_1 (f_u/f_l)h^4  \,.
\end{align*}
Therefore, with probability at least $1-O(p^{-c_0})$,
\begin{align*}
    \big| {\bv}_{(j)}^{\T} \hat{\bJ} {\bv}_{(j)} - {\bv}_{(j)}^{\T} {\bJ} {\bv}_{(j)} \big| \leq C_2 \frac{f_u}{f_l} \bigg\{ \frac{1}{f_l} \sqrt{\frac{ s^2(\log p)^7}{nh^2}} +    h^4  \bigg\} \,,
\end{align*}
where $C_2>0$ depends on $(l_0,\upsilon_0,\gamma_1,\gamma_p,\underline{\kappa},\kappa_u,\kappa_2)$. Since $f_l < f_u$, a similar argument, with larger concentration constants if needed, gives the following bound simultaneously with the preceding one:
\begin{align*}
    \|\hat{\bJ}-{\bJ}  \|_{\infty}+  \|\hat{\bJ}{\bv}_{(j)}-{\bJ}{\bv}_{(j)}  \|_{\infty} \leq C_2 \frac{f_u}{f_l} \bigg\{ \frac{1}{f_l} \sqrt{\frac{ s^2(\log p)^7}{nh^2}} +    h^4  \bigg\} 
\end{align*}
with probability at least $1-O(p^{-c_0})$. Write 
\begin{align*}
    \tilde\Delta  =  \frac{\tilde{s}f_u^{1/2}}{f_l^{3/2}}\Delta + C_2 \frac{f_u}{f_l^2}   \sqrt{\frac{ s^2(\log p)^7}{nh^2}}  \,.
\end{align*}
By \eqref{h.req} and  \eqref{dis.sigma2}, if $\tilde\Delta \leq C$, we have
\begin{align*}
    |\hat{\sigma}_{j}^2- {\sigma}_{j}^2| \lesssim \tilde\Delta
\end{align*}
with probability at least $1-O(p^{-c_0})$.

By the law of total variance, for each $i\in[n]$,
\begin{align*}
{\rm var}\{\psi_i(\bbeta^*)Q_{i,j}\}
=&~ E[{\rm var}\{\psi_i(\bbeta^*)Q_{i,j}\,|\,\bx_i\}]
+{\rm var}[E\{\psi_i(\bbeta^*)Q_{i,j}\,|\,\bx_i\}]\\
\geq&~ E[Q_{i,j}^2{\rm var}\{\psi_i(\bbeta^*)\,|\,\bx_i\}]\,,
\end{align*} 
where the inequality follows since $Q_{i,j}=\bx_i^{\T}\bv_{(j)}$ is fixed conditional on $\bx_i$. Hence,  
\begin{align*}
    \sigma_j^2 \geq   \frac{1}{n} \sn E[ Q_{i,j}^2 {\rm var}\{\psi_i({\bbeta}^*) \,|\, \bx_i\} ] = \frac{1}{n} \sn E\big\{ Q_{i,j}^2 \big( E\{ |\psi_i({\bbeta}^*)|^2\,|\, \bx_i \} - [E \{ \psi_i({\bbeta}^*)\,|\,\bx_i \}]^2\big) \big\}\,.
\end{align*} 
 Note that 
  \begin{align*} 
    E\{ |\psi_i({\bbeta}^*)|^2 \,|\,\bx_i\}  = \tau^2+E\{ | \bar{K}_h(  -\varepsilon_i)|^2 \,|\,\bx_i\}-2\tau [E\{  \bar{K}_h(  -\varepsilon_i)  \,|\,\bx_i\}-\tau]-2\tau^2\,. 
\end{align*}
Under Conditions 1(a) and 1(d), equation~(G.5) in the supplementary material of \citeS{tan2022highS} yields   $|E\{  \bar{K}_h(  -\varepsilon_i)  \,|\,\bx_i\} -\tau | \leq l_0\kappa_2 h^2/2$, which implies $ E\{ |\tau-\bar{K}_h(  -\varepsilon_i)|^2 \,|\,\bx_i\} \geq  E\{ | \bar{K}_h(  -\varepsilon_i)|^2 \,|\,\bx_i\}  -\tau^2- \tau l_0\kappa_2 h^2$. Since $\tau = F_{\varepsilon_i|\bx_i}(0)$, we have 
 \begin{align*}
     &E\{  \bar{K}_{h}^2(- \varepsilon_i)  \,|\, \bx_i \} = \int_{-\infty}^{\infty} \bar{K}^2(-u/h)\,\mbox{d}F_{\varepsilon_i|\bx_i}(u) \\
	&~~~ \overset{{\rm (i)}}{=} \int_{-\infty}^{\infty} F_{\varepsilon_i|\bx_i}(-uh)\,\mbox{d}\bar{K}^2(u) 
	=   2 \int_{-\infty}^{\infty} K(u)\bar{K}(u)F_{\varepsilon_i|\bx_i}(-uh)\,\mbox{d}u \\
	&~~~ \geq 2 \int_{-\infty}^{\infty} K(u)\bar{K}(u)F_{\varepsilon_i|\bx_i}(0)\,\mbox{d}u -\bigg|2 \int_{-\infty}^{\infty} K(u)\bar{K}(u)\{F_{\varepsilon_i|\bx_i}(-uh)-F_{\varepsilon_i|\bx_i}(0)\}\,\mbox{d}u \bigg| \\
	&~~~ \overset{{\rm (ii)}}{\geq}  2\tau\int_{-\infty}^{\infty} K(u)\bar{K}(u) \,\mbox{d}u -2f_uh  \int_{-\infty}^{\infty} |u|K(u)\bar{K}(u) \,\mbox{d}u  \overset{{\rm (iii)}}{\geq} \tau - 2 f_u\kappa_1 h\,,
 \end{align*}  
where (i) follows from a change of variables and integration by parts, (ii) follows from Condition 1(a), and (iii) follows from
 $\int_{-\infty}^{\infty} K(u)\bar{K}(u) \,\mbox{d}u = 1/2$ and $\sup_{u\in\RR}\bar{K}(u)\leq 1$. Provided that $2 f_u\kappa_1 h + \tau l_0\kappa_2 h^2 \leq \tau(1-\tau)/2$, we have $ E\{ |\psi_i({\bbeta}^*)|^2 \,|\,\bx_i\} \geq  \tau-\tau^2 -2 f_u\kappa_1 h - \tau l_0\kappa_2 h^2 \geq \tau(1-\tau)/2$. Moreover, \eqref{psi.beta} implies $[E \{ \psi_i({\bbeta}^*)\,|\,\bx_i \}]^2 \leq {l_0^2 h^4 \kappa_2^2}/{4}$. Provided that $l_0^2 h^4 \kappa_2^2 \leq \tau(1-\tau)$, we obtain  $\sigma_j^2 \gtrsim  E(|\bx_i^{\T} {\bv}_{(j)}|^2)= \|\bv_{(j)}\|_{\bSigma}^2$. Since $\|\bv_{(j)}\|_2\geq 1$,   we have $ \|\bv_{(j)}\|_{\bSigma}^2 \geq \|\bv_{(j)}\|_2^2 \gamma_p \geq \gamma_p$. Thus,  for any $j\in [p]$,  we have 
\begin{align*} 
     |\hat{\sigma}_{j} - {\sigma}_{j}|  = \frac{ |\hat{\sigma}_{j}^2- {\sigma}_{j}^2| }{\hat{\sigma}_{j} + {\sigma}_{j}} \lesssim {\tilde{\Delta}}  
 \end{align*}
with probability at least $1-O(p^{-c_0})$. This completes the proof of Lemma \ref{lemma.sigma.cons}.\hfill$\Box$

{\small{
		\setlength{\bibsep}{3pt}
		\bibliographystyleS{apalike}
		\bibliographyS{refs}
}}

\end{document}